\documentclass{aa}

\usepackage{graphicx}
\usepackage{txfonts}
\usepackage[margincaption]{sidecap}
\usepackage{xcolor}
\usepackage{hyperref}
\hypersetup{colorlinks = true,
            linkcolor = blue,
            urlcolor = cyan,
            citecolor = blue,
            anchorcolor = blue}
\usepackage{orcidlink}
\usepackage[nice]{nicefrac}
\usepackage[varvw]{newtxmath}
\newcommand{\teff}{$T_\mathrm{eff}$}
\newcommand{\logg}{$\log{g}$}
\newcommand{\feh}{[Fe/H]}

\newcommand{\Mstar}{$M_*$}

\newcommand{\kms}{km\,s$^{-1}$}
\newcommand{\vbroad}{$V_{\rm broad}$}
\newcommand{\sweet}{SWEET-Cat}

\begin{document} 

   \title{Ariel stellar characterisation
   \thanks{Based on data from public telescope archives and from observations collected at the European Southern Observatory (ESO) under programmes 105.20P2.001 and 106.21QS.001, at the Large Binocular Telescope (LBT) Observatory under programmes 2021$\_$2022$\_$25 and IT-2022-024, with the Italian Telescopio Nazionale Galileo (TNG) under programmes AOT41$\_$TAC25 and AOT42$\_$TAC20, and with the Southern African Large Telescope (SALT) under programme 2023-1-SCI-005.}$^,$
   \thanks{Table~\ref{table:cno_abundances} is available in electronic form at the CDS via anonymous ftp to cdsarc.u-strasbg.fr (130.79.128.5) or via http://cdsweb.u-strasbg.fr/cgi-bin/qcat?J/A+A/.}
   }

   \subtitle{II. Chemical abundances of carbon, nitrogen, and oxygen for 181 planet-host FGK dwarf stars}
   \titlerunning{Ariel homogeneous stellar C, N, and O abundances}
   \authorrunning{da Silva et al.}

   \author{
        da Silva, R. \inst{\ref{oar},\ref{asi}}\and
        Danielski, C. \inst{\ref{oaa},\ref{iaa}} \and
        Delgado Mena, E. \inst{\ref{porto}}\and
        Magrini, L. \inst{\ref{oaa}} \and
        Turrini, D. \inst{\ref{oato}} 
        \and
        Biazzo, K. \inst{\ref{oar}} \and
        Tsantaki, M. \inst{\ref{oaa}}
        \and \\
        Rainer, M. \inst{\ref{oabr}} \and
        Helminiak, K.G. \inst{\ref{nicolaus}} \and
        Benatti, S. \inst{\ref{oapa}}\and
        Adibekyan, V. \inst{\ref{porto}} \and
        Sanna, N. \inst{\ref{oaa}} \and
        Sousa, S. \inst{\ref{porto}} \and
        Casali, G. \inst{\ref{aust1},\ref{aust2},\ref{oabo}} \and \\
        Van der Swaelmen, M. \inst{\ref{oaa}}
        }


   \institute{
   INAF - Osservatorio Astronomico di Roma, via Frascati 33, 00078 Monte Porzio Catone, Italy. \label{oar}
   \and
   Agenzia Spaziale Italiana, Space Science Data Center, via del Politecnico snc, 00133 Rome, Italy. \label{asi}
   \and
   INAF - Osservatorio Astrofisico di Arcetri, Largo E. Fermi 5, 50125, Firenze, Italy. \label{oaa}
   \and
   Instituto de Astrof\'isica de Andaluc\'ia, CSIC, Glorieta de la Astronom\'ia s/n, 18008, Granada, Spain. \label{iaa}
   \and
   Instituto de Astrofísica e Ciências do Espaço, Universidade do Porto, CAUP, Rua das Estrelas, P-4150-762 Porto, Portugal. \label{porto}
   \and
   INAF - Osservatorio Astrofisico di Torino, via Osservatorio 20, 10025, Pino Torinese \label{oato}
   \and
   INAF - Osservatorio Astronomico di Brera, Via E. Bianchi 46, 23807 Merate (LC), Italy. \label{oabr}
   \and
   Nicolaus Copernicus Astronomical Center, Polish Academy of Sciences, ul. Rabiańska 8, 87-100 Toruń, Poland. \label{nicolaus}
   \and
   INAF - Osservatorio Astronomico di Palermo, Piazza del Parlamento 1, 90134, Palermo, Italy. \label{oapa}
   \and
   Research School of Astronomy \& Astrophysics, Australian National University, Cotter Rd., Weston, ACT 2611, Australia. \label{aust1}
   \and
   ARC Centre of Excellence for All Sky Astrophysics in 3 Dimensions (ASTRO 3D), Stromlo, Australia. \label{aust2}
   \and
   INAF-Osservatorio di Astrofisica e Scienza dello Spazio di Bologna, via P. Gobetti 93/3, 40129, Bologna, Italy. \label{oabo}
   }

   \date{Received ...; accepted ...}

 
  \abstract
   {One of the ultimate goals of the ESA Ariel space mission is to shed light on the formation pathways and evolution of planetary systems in the Solar neighbourhood. Such an endeavour is only possible by performing a large chemical survey of not only the planets, but also their host stars, inasmuch as stellar elemental abundances are the cipher key to decode the planetary compositional signatures.}
   {This work aims at providing homogeneous chemical abundances of C, N, and O of a sample of 181 stars belonging to the Tier 1 of the Ariel Mission Candidate Sample.}
   {We applied the spectral synthesis and the equivalent width methods to a variety of atomic and molecular indicators (the \ion{C}{i} lines at 5052 and 5380.3~\AA, the [\ion{O}{i}] forbidden line at 6300.3~\AA, the ${\rm C}_2$ bands at 5128 and 5165~\AA, and the CN band at 4215~\AA) using high-resolution and high signal-to-noise spectra collected with several spectrographs.}
   {We provide carbon abundances for 180 stars, nitrogen abundances for 105 stars, and oxygen abundances for 89 stars. We analyse the results in the light of the Galactic chemical evolution, and in terms of the planetary companions properties. Our sample basically follows the typical trends with metallicity expected for the [C/Fe], [N/Fe], and [O/Fe] abundance ratios. The fraction between carbon and oxygen abundances, both yields of primary production, is consistent with a constant ratio as [O/H] increases, whereas the abundance of nitrogen tends to increase with the increasing of the oxygen abundance, supporting the theoretical assumption of a secondary production of nitrogen. The [C/N], [C/O], and [N/O] abundance ratios are also correlated with [Fe/H], which might introduce biases in the interpretation of the planetary compositions and formation histories if host stars of different metallicity are compared. We provide relations that can be used to qualitatively estimate whether the atmospheric composition of planets is enriched or not with respect to the host stars.
   }
   {}


   \maketitle

\section{Introduction}

\subsection{Stellar abundances and planetary formation}

In the last years the developments seen in the field of exoplanet have set the framework for a joint analysis of the star and its planets as a whole. Mostly, it has been understood that for reaching a precise characterisation of the planet, we need a precise characterisation of the host star \citep{Thorngrenetal2016,BitschBattistini2020,Adibekyanetal2021,Barrosetal2023,Hawthornetal2023,Psaridietal2024}. Besides, the \textit{discourse} grows wider when more and more of the discovered systems are found to host multiple planets \citep{Cabreraetal2014,Campanteetal2015,Gillonetal2017,Leleuetal2021,Luqueetal2023}. In this situation, the star arises as the common denominator to the formation and evolution of the planetary bodies within the same system. Understanding the origins of planets becomes hence an interplay between delineating the host star features (its fundamental parameters and its chemical traits), the planetary features (the atmospheric and bulk compositions, radius and mass), and finally the circumstellar discs and interstellar medium features, from which these systems were born. The temporal dimension to the problem is eventually given by the age of the system, i.e., the age of the star.

To date, the Solar neighbourhood has been found hosting more than 5600 exoplanets (NASA exoplanet archive\footnote{\url{https://exoplanetarchive.ipac.caltech.edu}}) for a total of 4179 single or multiple stars. Throughout the years, various teams dedicated time to characterise the host stars of each system case by case, but only more recently the analysis was up-scaled and a population approach has been applied to a larger sample of stars. The goal, within the field, being the identification of possible correlations between homogeneously determined stellar parameters (e.g., \sweet\footnote{\url{http://sweetcat.iastro.pt/}}: \citealt{Santosetal2013,sousa21}, GIARPS: \citealt{Baratellaetal2020,Biazzoetal2022}, Ariel: \citealt{Danielskietal2022,Brucalassietal2022,Magrinietal2022}, CARMENES: \citealt{Passeggeretal2018,Passeggeretal2019,Marfiletal2020,Marfiletal2021}, ExoFOP\footnote{\url{https://www.ipac.caltech.edu/project/exofop}}), and exoplanet parameters. The larger and more diverse is the sample analysed, the better the chances to identify relevant trends which allows shedding light on the original location of the planetary embryo in the protoplanetary disc, and the primordial formation pathways. It is established that, as soon as the protostellar cloud collapses, the protoplanetary disc cools down and the snowline of the various volatile species (e.g., H$_2$O, CO$_2$, CO, CH$_4$, N$_2$) move towards the central star, causing the chemical composition of both solids and gas to evolve as function of time and their location in the disc \citep{Eistrupetal2016,Eistrupetal2018,BoothIlee2019,Madhusudhan2019}. While in the inner part of the snowline the various species contribute to the gas composition, beyond the snowline they contribute to the solids composition. For such, assuming that the planet migrates inside the disc during its formation, the net planetary composition is provided in first approximation (i.e., without accounting for chemical evolution of the disc) by the cumulative accretion history over the migration trail across both gas and solids. As a consequence, the present-day atmospheric chemical composition of exoplanets can be used to retrieve key information on the planet's birthplace and on the time the planet migrated to its present orbit \citep{Cridlandetal2019,Cridlandetal2020,Schneideretal2021a,Schneideretal2021b,Turrinietal2021,Pacettietal2022,Khorshidetal2022}.

In the light of the above, one of the main focus of the last decade of the exoplanet community has been the determination of the carbon-to-oxygen ratio (C/O\footnote{${\rm X1/X2} \equiv 10^{\log{\epsilon({\rm X1})}} / 10^{\log{\epsilon({\rm X2})}}$, where $\log{\epsilon({\rm X1})}$ and $\log{\epsilon({\rm X2})}$ are absolute abundances. It should not be confused with the solar-normalised ratio ${\rm [X1/X2]} \equiv \log{\epsilon({\rm X1/X2})}_{\rm star} - \log{\epsilon({\rm X1/X2})}_{\rm Sun}$.}, e.g., \citealt{Mosesetal2013a,Mosesetal2013b,Notsuetal2020}), used as a formation location proxy of giant planets (see \citealt{Obergetal2011,Madhusudhanetal2016,Madhusudhan2019} and references therein). When a super-solar C/O ratio is detected, the planet is theorised to have formed in a low-metallicity environment and hence to have accreted most of C and O in gas state. On the other hand, when a sub-solar C/O ratio is found, the planet is expected to have formed in a high metallicity environment where C and O were dominated by the accretion of solids. However, recent theoretical developments showed that the C/O ratio alone provides limited information with regard to the formation region, and that the inclusion of both carbon-to-nitrogen (C/N, \citealt{Turrinietal2021,Pacettietal2022}) and nitrogen-to-oxygen (N/O, e.g., \citealt{Pisoetal2016,Turrinietal2021,Ohnoetal2023}) ratios enable breaking the degeneracy in the information provided by C/O \citep{Turrinietal2021,Fonteetal2023}.

An important aspect about the various planetary elemental ratios though is the fact that they cannot be directly used for comparison purposes between objects in different planetary systems, due to the variety of scale range they present. Consequently, when performing population studies and/or comparison planetology, it is extremely important to normalise these ratios to the elemental abundance ratios of their own star \citep{Turrinietal2021}. Normalised elemental ratios open up the possibility to directly compare formation and migration histories of giant planets that reside in different planetary systems, always in the event that stellar abundances are homogeneously derived among the stellar sample \citep{Turrinietal2022,KoleckiWang2022}. Furthermore, \cite{Turrinietal2021} showed that when the abundance pattern of the stellar normalised ratios ${\rm X1/X2}^\ast$\footnote{${\rm X1/X2}^\ast \equiv {\rm (X1/X2)}_{\rm planet} / {{\rm (X1/X2)}_{\rm star}}$ follows the syntax by \citealt{Turrinietal2021} and refers to the planetary elemental ratio over the stellar elemental ratio.} is C/N$^*$ > C/O$^*$ > N/O$^*$, then giant planets are solid-dominated and have high metallicity; with the opposite pattern N/O$^*$ > C/O$^*$ > C/N$^*$, planets are gas-dominated and have low metallicity. The first data application of normalised ratios has been presented by \cite{Biazzoetal2022} for a sample of $\sim$30 transiting planets whose stellar abundances have been homogeneously determined. As a result of their analysis the authors identified planets that were originally formed outside the CO$_2$ snowline, others between the CO$_2$ and CH$_4$ snowlines, and others between the N$_2$ and CO$_2$ snowlines. We refer to \cite{Biazzoetal2022} for more details.

Before proceeding, it is important to point out that, while analogous relations between normalised ratios are yet to be defined for the secondary or mixed atmospheres of solid planets, the knowledge of the stellar abundances is of paramount importance to probe their interior state by means of their density \citep[see, e.g.,][]{Bondetal2010,Thiabaudetal2014,Adibekyanetal2021} and properly constrain their migration history/formation region \citep{BitschBattistini2020,Adibekyanetal2021}. Specifically, \citet{BitschBattistini2020} argues that the variations in the abundance of oxygen among stars of different metallicity can shift the disc region characterised by peak pebble accretion efficiency from the snowline of H$_2$O to the outer one of CO. Such a change would impact the budget of volatile elements in the forming solid planets and, consequently, affect their mass-radius relations, meaning that the stellar composition is a key parameter to account for reliable population studies.

To summarise, the ingredients needed to start solving the puzzle of planet formation on a large scale are the following: a statistical broad sample of planetary systems, a homogeneous determination of the age of the system, and homogeneous elemental abundances of C, N, and O for all stars and planets. Homogeneous refractory elements are needed too, for instance sulphur, which has been shown to be mainly incorporated in refractory solids in planetary formation processes \citep{Palmeetal2014,Kamaetal2019,Keyteetal2024}, and its sulphur-to-nitrogen normalised ratio. We refer to Delgado Mena (in prep.) for a thorough discussion on the subject of refractories.

\begin{figure*}
\centering
\begin{minipage}[t]{0.33\textwidth}
\centering
\resizebox{\hsize}{!}{\includegraphics{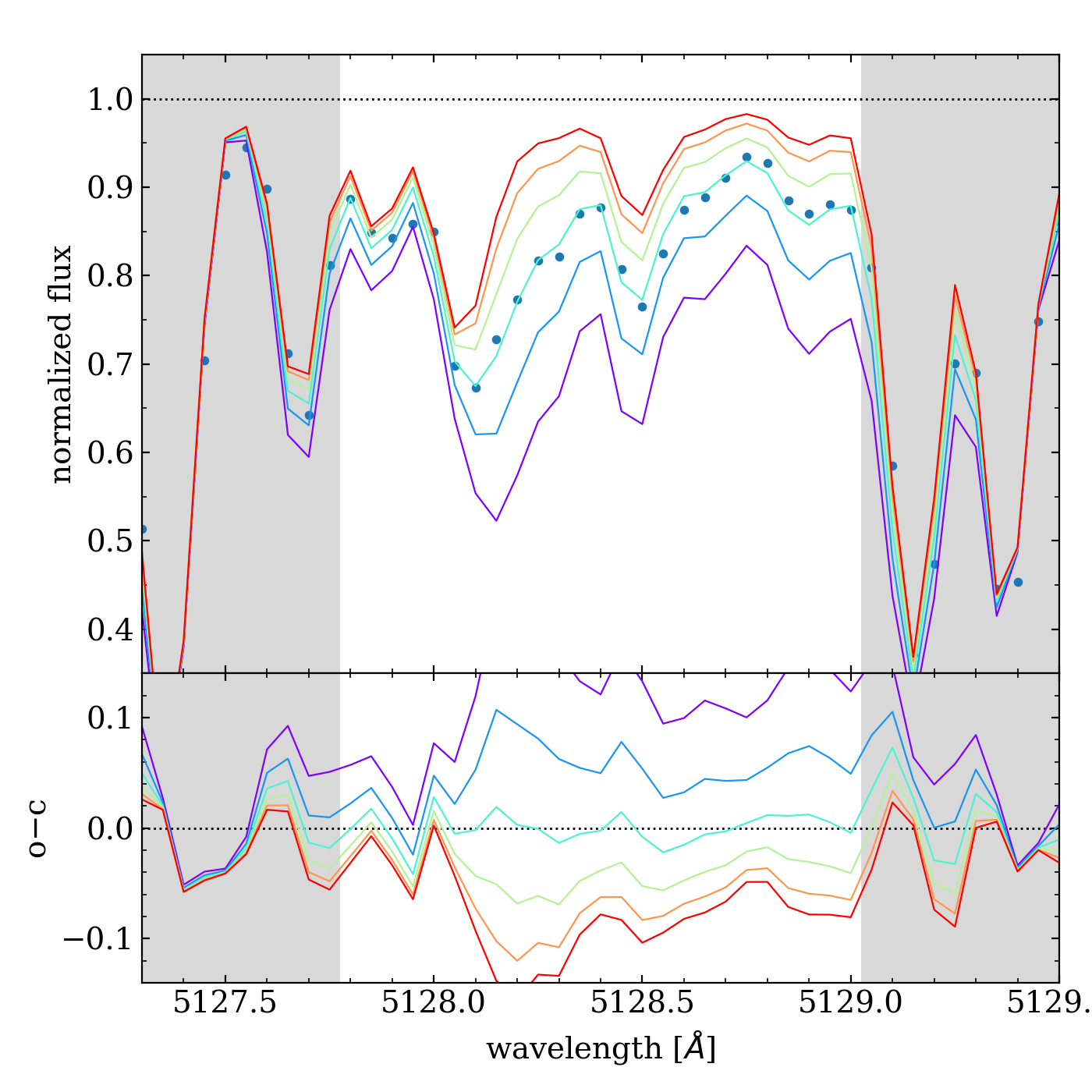}}
\end{minipage}
\begin{minipage}[t]{0.33\textwidth}
\centering
\resizebox{\hsize}{!}{\includegraphics{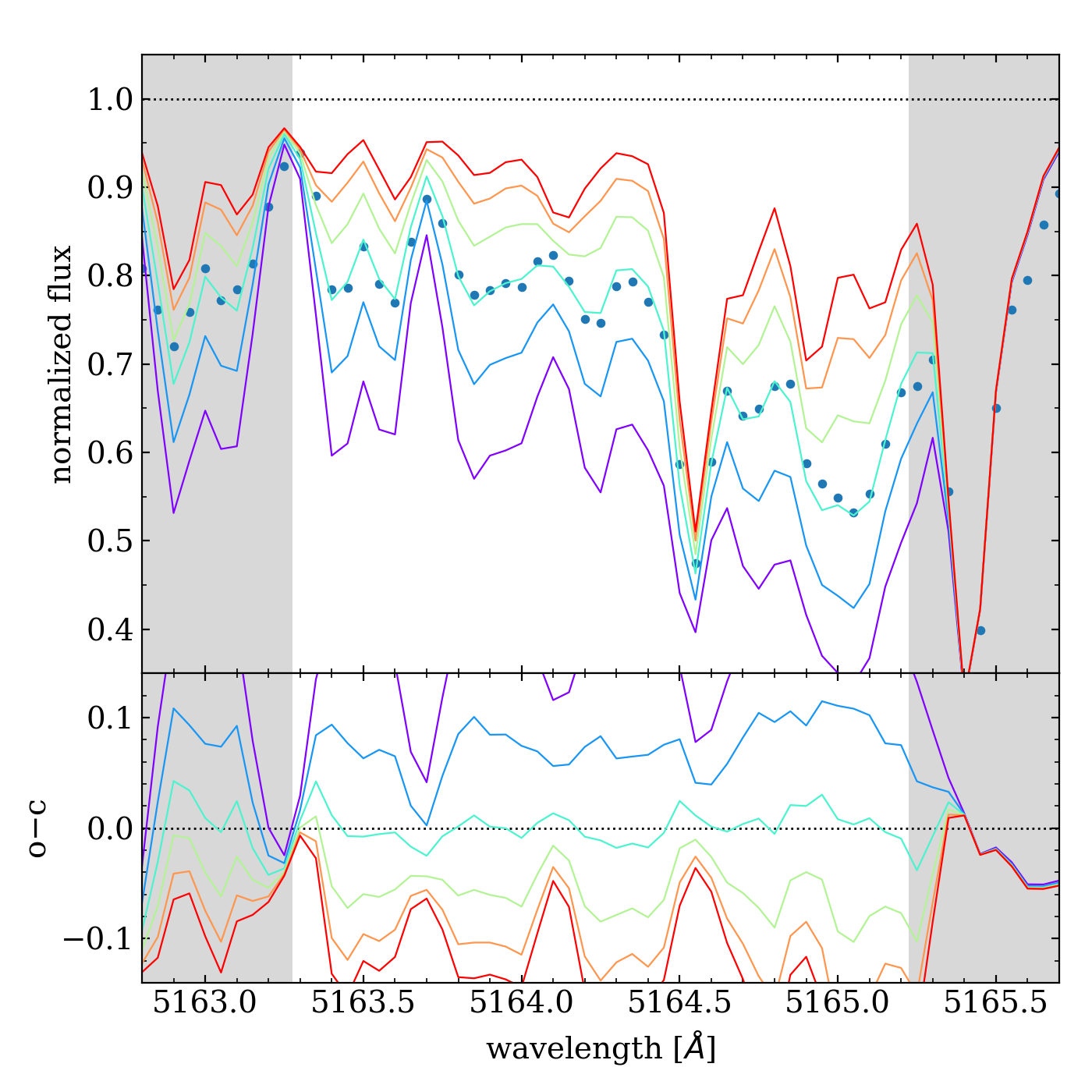}}
\end{minipage}
\begin{minipage}[t]{0.33\textwidth}
\centering
\resizebox{\hsize}{!}{\includegraphics{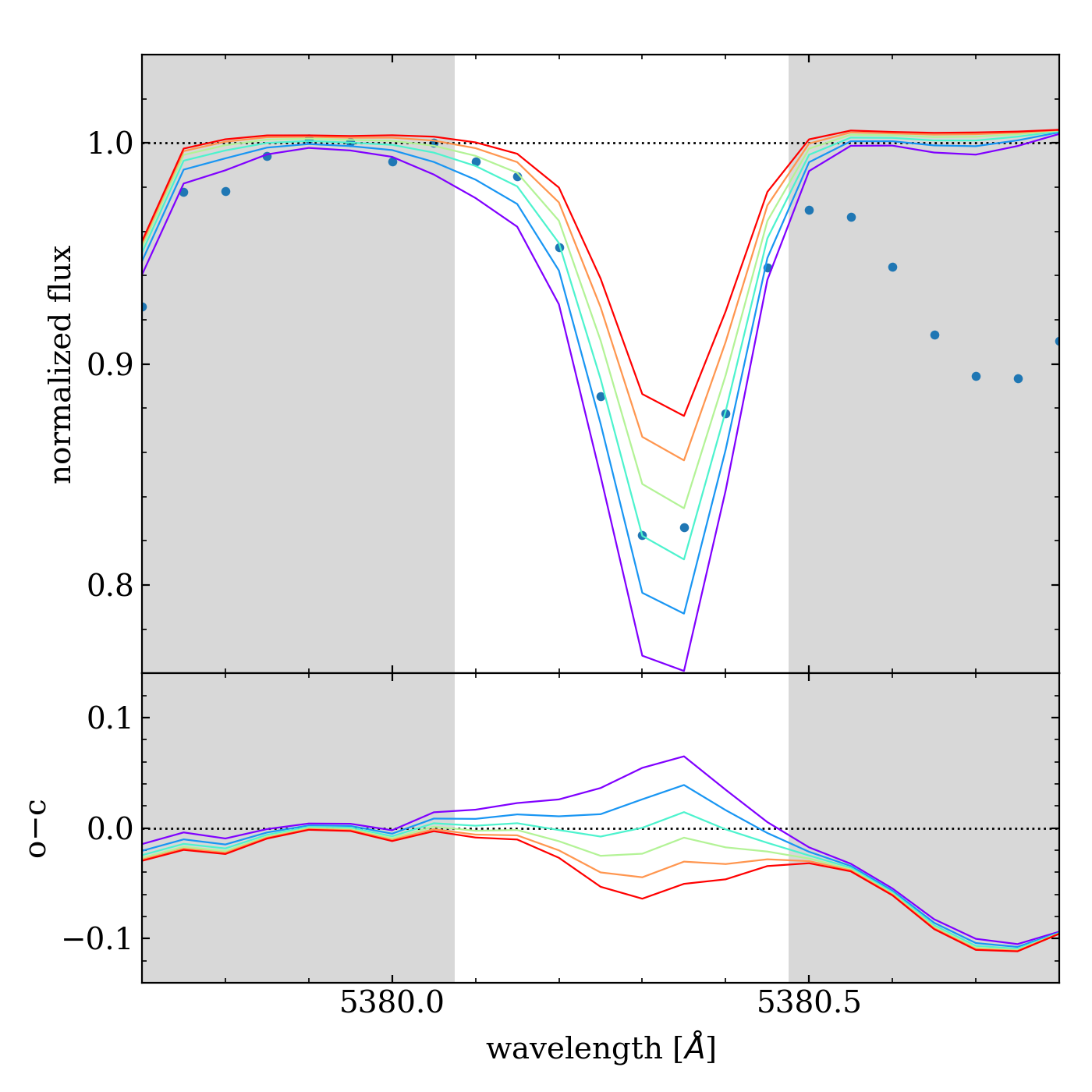}}
\end{minipage}
\caption{Spectral synthesis examples. The panels show the ${\rm C}_2$ molecular band regions at 5128~\AA\ (left) and 5165~\AA\ (middle), and the \ion{C}{i} atomic line at 5380.3~\AA\ (right) for the HARPS spectrum of \object{K2-24} (S/N$\sim$50 at 5000~\AA\, \teff\ = 5768~K, [Fe/H] = 0.41). The blue dots represent the observed spectrum, and the coloured lines are the synthetic spectra fitted to the data for models with different C abundances, separated by 0.1~dex. The bottom panels show the differences between observed and computed spectra. The shaded regions indicate wavelengths not used in the fit.}
\label{figure:c_synthesis_ex}
\end{figure*}

The large scale approach will soon be employed by the ESA Ariel space mission \citep{Tinettietal2018,Tinettietal2021}, which is a mission fully dedicated to the study of exoplanets, whose launch in L2 is foreseen for 2029. Ariel will have the sensitivity to study the atmospheres of diverse range of worlds, from Earth-mass planets to brown dwarfs \citep{Zingalesetal2018,Edwardsetal2019,Tinettietal2021}, with the goal to determine the chemical composition of their atmospheres (e.g., \citealt{Changeatetal2022,Wangetal2023}) and use such information to eventually constrain their formation and evolution. The Tier 1 planets orbit stars with effective temperatures ranging between 2500~K and 7500~K, and with spectral range coverage from M-dwarfs to A type stars, approximately. To meet Ariel scientific goals, all hosting-stars need to be precisely characterised in a uniform way throughout the whole spectral and evolutionary range, endeavour that has been undertaken by the ``Stellar Characterisation'' working group of the Ariel consortium since the mission has been adopted by ESA in 2020 \citep{Danielskietal2022}. The first release of the Ariel stellar catalogue of homogeneous atmospheric parameters and kinematics for 187 FGK dwarf stars has been issued by \cite{Magrinietal2022}. These data has been then used as input to determine homogeneous ages, masses, and radii (Bossini, in prep.) and homogeneous abundances of refractory elements (Delgado Mena, in prep.) for a sample based on the same 187 stars. Similarly, the determination of fundamental parameters for the hot stars and fast rotators within the Tier 1 is being tackled (Tsantaki, in prep.; Ramler, in prep.), as well as for the cool M-dwarfs end of the Tier-1 (Danielski, in prep.), the latter in collaboration with the CARMENES consortium.

\subsection{Stellar C, N, and O abundances}

Carbon, nitrogen, and oxygen are among the most abundant elements in our Solar System, in the Milky Way \citep{Houetal2000,Asplundetal2021}, and in the Universe \citep[e.g.][]{MaiolinoMannucci2019}, playing a key role in the several chain of reactions that occur in stellar nucleosynthesis processes. Special attention has been given to these elements also in the context of exoplanetary systems when comparing their abundances with stars for which no planet has been detected. The well known correlation between the stellar iron abundance and the presence of giant planets \citep[see, e.g.,][and references therein]{Santosetal2004,FischerValenti2005} was afterwards extended to other heavy elements \citep[e.g.,][]{Adibekyanetal2012,daSilvaetal2015}. Besides, a correlation between the stellar metallicity and the presence of low-mass planets was also identified \citep[e.g.,][]{WangFischer2015}. Several studies tried to verify whether such correlations are also valid for light elements, specially C, N, and O, but the results are quite variable. The utilisation of different abundance indicators (atomic or molecular), the 1D local thermodynamic equilibrium (LTE) versus 3D non-LTE assumptions, and the inhomogeneity of the different datasets are among the main reasons for these discrepancies.

To date, various chemical surveys have been dedicated to overcome these problems, specially in what concerns the homogeneity of the sample. \citet{Adibekyanetal2012} measured 12 refractory elements for 135 planet-host stars within a larger sample of HARPS-GTO 1111 FGK dwarfs stars. \citet{daSilvaetal2015} performed a homogeneous abundance determination of C, N, and O plus 11 refractory elements for a sample of 309 stars in the solar neighbourhood, with and without detected planets, including 140 dwarfs, 29 subgiants, and 140 giants. \citet{DelgadoMenaetal2017,DelgadoMenaetal2021} measured neutron-capture and C abundances within the HARPS-GTO sample for 136 and 152 planet-host stars, respectively. This sample has been also used to determine abundances of oxygen \citep{Bertrandelisetal2015}, nitrogen \citep{SuarezAndresetal2016}, and carbon \citep{SuarezAndresetal2017}. The abundances of sulphur for the same sample were presented in \citet{CostaSilvaetal2020}. More recently, \citet{Tautvaisienetal2022} measured 24 chemical species for 25 host stars within a total sample of 848 stars in the northern hemisphere. \citet{Polanskietal2022} measured 15 elements (among which C, N, and O) for a JWST sample of 25 exoplanet host stars using a data driven machine learning tool.

In this manuscript, we focus on the determination of carbon, nitrogen, and oxygen for a subset of stars from \citet{Magrinietal2022} belonging to the Ariel Mission Reference Sample. This work presents the largest homogeneous chemical C, N, and O compilation of exoplanet stellar hosts currently available and for which we can provide stellar C/O, N/O, and C/N elemental ratios to be used in the study of planetary formation of the relative planetary companions. We note that the X1/X2 notation is normally used when referring to the abundance ratios of planetary atmospheres whereas the [X1/X2] notation, which is normalised to the Sun, is normally used for stars. Here we derive the stellar [X1/X2] abundance ratios, we investigate possible correlations with other parameters, and we discuss both [X1/X2] and X1/X2 ratios in the context of stellar and planetary composition.

This paper in structure as follows. In Sect.~\ref{section:sample} we present the sample of stars and in Sect.~\ref{section:abundances} we describe how we perform the analysis of their spectra to obtain the abundances. In Sect.~\ref{section:results} we discuss our main results regarding the [X/Fe], [X1/X2], and X1/X2 abundance ratios and their relation with other stellar and planetary parameters. Finally, a brief summary, a few conclusions, and some final remarks are presented in Sect.~\ref{section:conclusions}.

\section{Observations and sample properties}
\label{section:sample}

The stars analysed in the current work is a subset of the 187 FGK planet-host dwarf stars studied by \citet{Magrinietal2022}. These authors provide a homogeneous catalogue containing stellar atmospheric parameters (metallicity [Fe/H], effective temperature \teff, surface gravity \logg, and microturbulent velocity $\xi$) derived using high-resolution and high signal-to-noise ratio (S/N) spectra collected with several instruments: HARPS@3.6m of ESO, HARPS-N@TNG, UVES@VLT, FEROS@2.2m of ESO, SOPHIE@1.93m of OHP, FIES@NOT, PEPSI@LBT, HIRES@Keck, and NARVAL@TBL spectrographs (for more details about these observations we refer to the aforementioned paper and references therein). Our final sample contains 181 stars for which we were able to derive reliable abundances for at least one of the three elements under investigation. For six out of the 187 stars in the original sample, none of the C, N, and O abundance indicators could be used for different reasons, which include too weak spectral lines in hot stars, too broad line profile in fast rotators, low S/N spectra, and out-of-coverage wavelength. Note that the overall sample by \cite{Magrinietal2022} is a subset of planetary hosting stars belonging to the Ariel mission Candidate Sample, which currently accounts for a total of $\sim$1000 potential planets whose host stars have spectral types ranging from A to M, with the majority being dwarfs \citep{Zingalesetal2018,Edwardsetal2019,EdwardsTinetti2021}. The selection by \citet{Magrinietal2022} was purely done on the data availability basis; more stellar parameters will be released in future works (e.g., Tsantaki, in prep.).

\begin{figure*}
\centering
\resizebox{0.8\hsize}{!}{\includegraphics{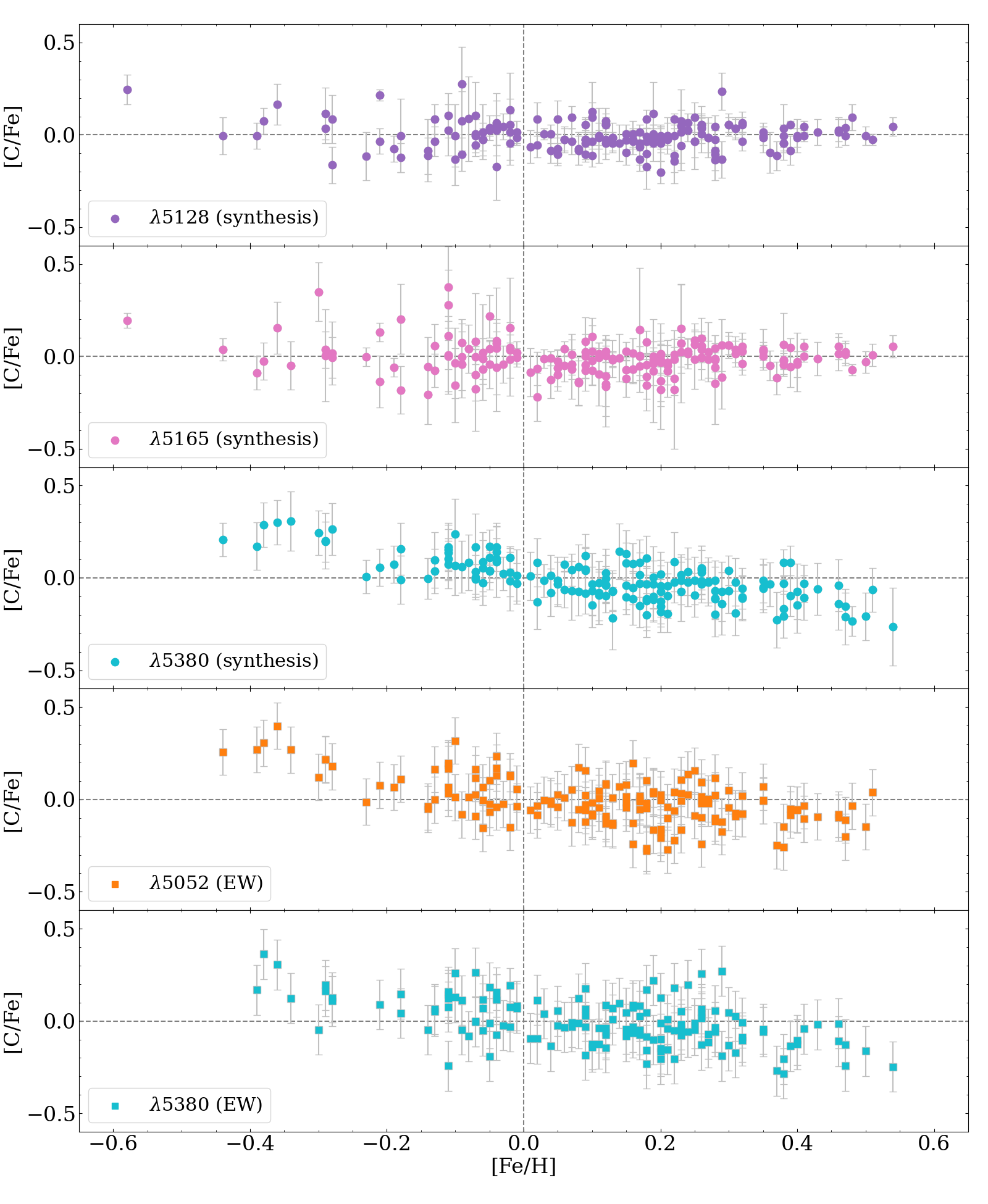}}
\caption{[C/Fe] abundance ratios as a function of the stellar metallicity. The panels show the carbon abundances derived from the spectral synthesis (circles) of two molecular bands (5128 and 5165~\AA) and one atomic line (5380.3~\AA) and from the equivalent widths (squares) of two atomic lines (5052.2 and 5380.3~\AA). Stars cooler than 5000~K for 5052.2~\AA\ (EW) and than 5200~K for 5380.3~\AA\ (EW) are not plotted (see discussion in the Appendix~\ref{appendix:atm_par_dependence}).}
\label{figure:cfe_feh_5}
\end{figure*}

Some of the abundance indicators that we adopt to derive the C, N, and O abundances (see next section) normally requires very high S/N ($\sim$200). Unfortunately, this could not be achieved for all the sample stars using the spectra already available. Therefore, additional spectra with S/N higher than the previous observations were recently collected for a selection of stars. The star \object{Kepler-5} already had two PEPSI spectra collected in 2021, and an additional spectrum was collected in 2023 with the same spectrograph. One new PEPSI spectrum was also collected last year for \object{WASP-90}, which had been previously observed with the VLT-UVES instrument. Finally, new spectra of \object{HATS-33}, \object{KELT-10}, \object{WASP-7}, \object{WASP-15}, \object{WASP-32}, \object{WASP-72}, \object{WASP-74}, \object{WASP-79}, \object{WASP-90}, \object{WASP-94A}, \object{WASP-100}, \object{WASP-120}, and \object{WASP-126} were collected using the high-resolution spectrograph (HRS) mounted on the Southern African Large Telescope (SALT) situated at the South African Astronomical Observatory (SAAO) at Sutherland (South Africa). The SALT HRS cover a wavelength range of about 370-890~nm with a spectral resolution of $\sim$65,000. The spectra were reduced using the SALT science pipeline\footnote{\url{http://pysalt.salt.ac.za/}} \citep{Crawfordetal2010}, then normalised and corrected for Doppler velocity shifts using the iSpec tool \citep{BlancoCuaresmaetal2014,BlancoCuaresma2019}.

Regarding the utilisation of different spectrographs and the claim of abundance homogeneity, we note that at least ten of the stars in our sample were observed with more than one instrument, allowing us to estimate the difference in the abundances derived from different spectra of the same star. For about 30 pairs of spectra (considering several abundance indicators), we estimate a mean abundance difference of $-$0.02 $\pm$ 0.06~dex, which is much smaller than the typical uncertainties derived for the C, N, and O abundances (see next section).

\section{Abundance determination}
\label{section:abundances}

\subsection{Carbon abundances}
\label{section:c_abundances}

The carbon abundances were derived using two methods, one based on the spectral synthesis of atomic and molecular features, and another one based on equivalent width (EW) measurements of atomic line profiles. These two approaches allow us to identify possible systematic problems when deriving the abundances, for instance, due to line blending (which makes it difficult to measure weak lines) or LTE departures (which might cause a dependence of the derived abundances on the effective temperature). The spectral synthesis method was applied to molecular lines of the electronic-vibrational band heads of the ${\rm C}_2$ Swan System at 5128 and 5165~\AA, and to the \ion{C}{i} atomic line centred at 5380.3~\AA. The equivalent width method was used for two \ion{C}{i} atomic lines, centred at 5052.2 and 5380.3~\AA. Other ${\rm C}_2$ features are also available in the spectra of FGK dwarf stars, such as the bands of the Swan System at 5135.6 and 5635.2~\AA. However, from our experience, the abundances derived from these features are more affected by the presence of atomic lines, resulting in a larger scatter.

The synthetic spectra were computed under the assumption of LTE and molecular equilibrium using pyMOOGi\footnote{\url{https://github.com/madamow/pymoogi}}, a Python wrapper of the MOOG radiative transfer code \citep[][version 2019]{Sneden1973}, and then fitted to the observed spectra. MOOG requires as inputs the model atmosphere of each star (which includes a list molecules used for molecular equilibrium calculations), a line list of atomic and molecular transitions, and a combination of spectral line broadening parameters. In order to minimise the impact that strong spectral features may have on the spectral synthesis computation, and to properly match synthetic and observed data, the collected spectra were carefully normalised based on continuum windows identified just before and just after the spectral regions used to measure the abundances. Moreover, the opacity contribution of strong atomic lines, with respect to weaker features, are independently taken into account by MOOG, providing a more appropriate representation of the wings of strong line profiles.

\begin{figure*}
\centering
\begin{minipage}[t]{0.565\textwidth}
\centering
\resizebox{\hsize}{!}{\includegraphics{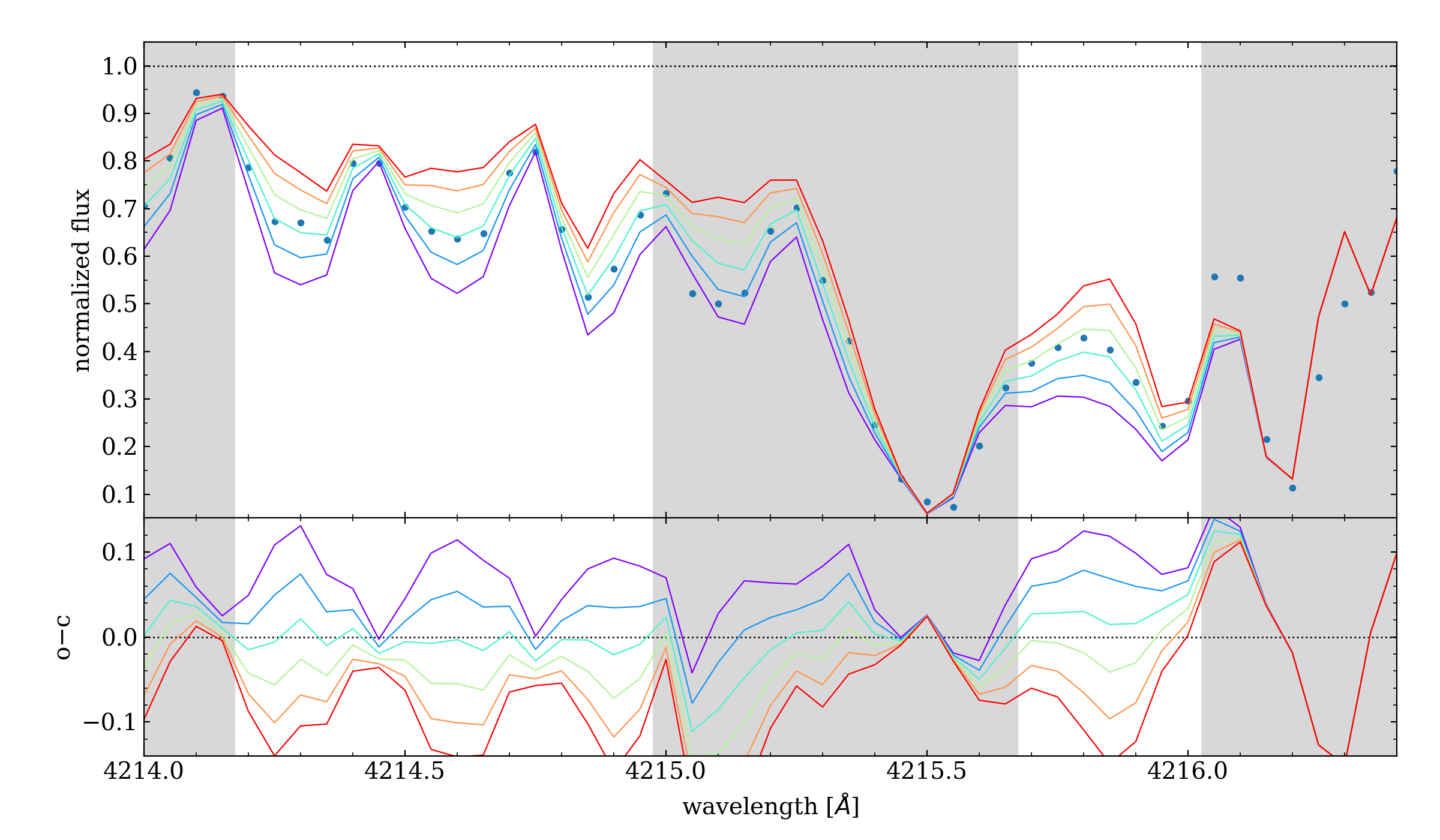}}
\end{minipage}
\begin{minipage}[t]{0.33\textwidth}
\centering
\resizebox{\hsize}{!}{\includegraphics{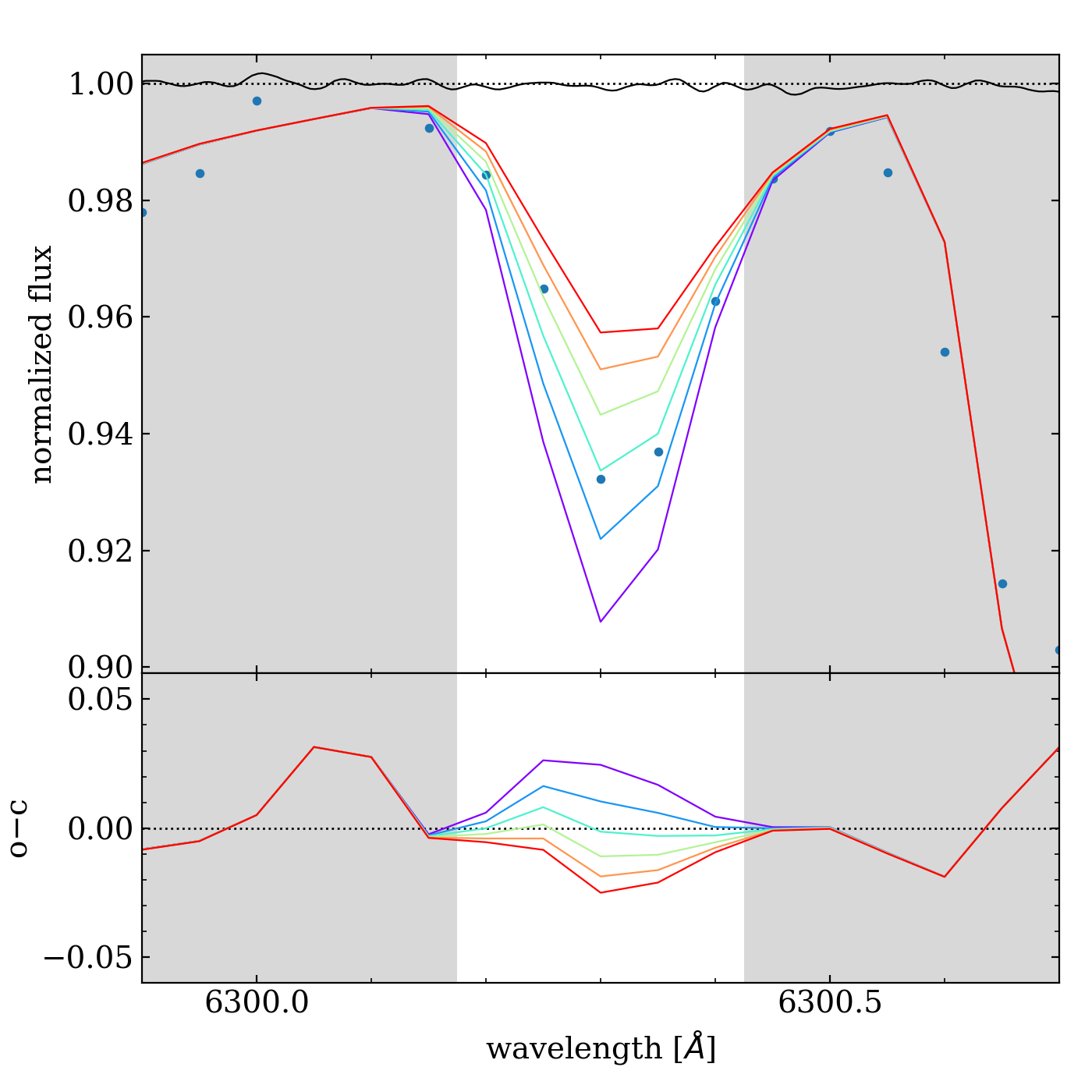}}
\end{minipage}
\caption{Same as in Fig.~\ref{figure:c_synthesis_ex} but showing the spectral synthesis of the CN molecular band at 4215~\AA\ (left panel) and of the O atomic line at 6300.3~\AA\ (right panel). The black solid line on the top of the right panel represents the spectrum of telluric lines.}
\label{figure:n_o_synthesis_ex}
\end{figure*}

We interpolated on a grid of MARCS model atmospheres \citep{Gustafssonetal2008} using the stellar atmospheric parameters derived by \citet{Magrinietal2022}, namely [Fe/H], \teff, \logg, and $\xi$, and including the elemental abundances of each star, whenever available, derived by Delgado Mena (in prep.). If not available, we used the standard solar abundances from \citet{Asplundetal2009} internally adopted by MOOG, except for iron, for which we adopted $\log{\epsilon_\odot}$(Fe) = 7.47, a logarithm scale in which $\log{\epsilon_\odot}$(H) = 12. We note that, for the molecular equilibrium calculations \citep[see, e.g.,][for a discussion about the chemical equilibrium in stellar atmospheres]{Tsuji1973,Lardoetal2012}, we tried to include in the model atmospheres the [O/Fe] values that we derived, but it increased the scatter of the [C/Fe] values. Therefore, in the current work, we preferred to adopt the solar oxygen value when deriving the carbon abundances. The atomic line data were taken from the Vienna Atomic Line Database \citep[VALD,][]{Piskunovetal1995,Ryabchikovaetal1997,Kupkaetal1999,Kupkaetal2000} whereas the molecular transitions are from \citet{Kurucz1992}, which adopts a dissociation energy ${\rm D}_0$ = 6.156~eV for the ${\rm C}_2$ transitions. Additional molecular features of MgH that may contribute to the synthetic spectrum in the studied regions were also included. The oscillator strengths of both atomic and molecular transitions were revised and changed wherever needed in order to match an observed solar spectrum (for this, we used a high-resolution and high S/N spectrum of Vesta collected with HARPS; for molecular features, we applied a global correction to the oscillator strengths, multiplying them by a constant factor, whereas single lines of both molecular and atomic transitions were slightly adjusted in some specific wavelengths). The spectral line broadening is a convolution of the following parameters: the instrumental broadening, which depends on the spectral resolution $R$ according to the relation $\lambda/R$; the stellar limb darkening, which we calculated from an interpolation of \teff\ and \logg\ on the Table~1 of \citet{DiazCordoves1995}; the stellar projected equatorial rotational velocity ($v\sin{i}$); and stellar velocity fields known as macroturbulence ($v_{\rm mac}$). Disentangling the latter two parameters is not an easy task. Therefore, in the parameter file passed as input to MOOG, we preferred to include a combined contribution of both $v\sin{i}$ and $v_{\rm mac}$ by defining a new parameter that we named as \vbroad. We estimated \vbroad\ by keeping fixed the instrumental broadening and the limb darkening and then fitting the profile of isolated and relatively strong (not saturated) atomic lines close to the spectral regions of interest.

Figure~\ref{figure:c_synthesis_ex} shows a few examples of spectral synthesis applied to the HARPS spectrum of the star \object{K2-24} for three of the carbon indicators investigated in the current work. For more details on the spectral synthesis method applied to atomic and molecular features, and on the input parameters used, we refer to \citet[][and references therein]{daSilvaetal2012,daSilvaetal2015}.

We estimated the uncertainties in the C abundances for each star in our sample by performing the spectral synthesis using modified versions of the model atmospheres. These are models perturbed by one standard deviation, i.e., by the errors estimated for the atmospheric parameters and for \vbroad\ (a value of 1~\kms\ was adopted). We used the perturbed models one at a time, and then calculated the total uncertainty as the quadratic sum of the single contributions.

The abundances of carbon from EWs were derived under LTE following the procedure by \citet{DelgadoMenaetal2021}. The EWs of the two atomic \ion{C}{i} lines were measured with ARES\footnote{The latest version, ARES v2, can be downloaded at \url{https://github.com/sousasag/ARES}.} \citep{Sousaetal2015} and then used as input together with the previously used MARCS model atmospheres in the MOOG code. When large discrepancies were found between the two lines, a visual inspection (and eventually measurement) was done with the \textit{splot} task of the Image Reduction and Analysis Facility (IRAF\,\footnote{The IRAF package is distributed by the National Optical Astronomy Observatories (NOAO), USA.}). All the [X/Fe] abundance ratios were obtained by doing a differential analysis with respect to the HARPS Vesta spectrum aforementioned.

Figure~\ref{figure:cfe_feh_5} shows the [C/Fe] abundance ratios as a function of the stellar metallicity for each of the aforementioned indicators. As explained in the Appendix~\ref{appendix:atm_par_dependence}, the EW method usually overestimates the abundances derived from \ion{C}{i} atomic lines in cool stars \citep[see, e.g.,][]{DelgadoMenaetal2021,Biazzoetal2022}. This is seen in our data plotted in Fig.~\ref{figure:xfe_teff_feh} for the line at 5052.2~\AA\ in stars cooler than 5000~K, and for the line at 5380.3~\AA\ in stars cooler than 5200~K. We do not see the same behaviour for the line at 5380.3~\AA\ if we apply the spectral synthesis method, which is more suited to deal with spectral line profiles in cool stars (see also Fig.~\ref{figure:delta_cfe5380_teff}, which shows the differences in abundance derived from this line using the two methods). Therefore, the two lower panels of Fig.~\ref{figure:cfe_feh_5} do not include such stars.

\begin{table*}
\centering
\caption{Excerpt from the list of C, N, and O abundances for our sample of 181 stars.}
\label{table:cno_abundances}
{\tiny
\begin{tabular}{l r@{}l r@{ }l r@{ }l r@{ }l r@{ }l r@{ }l r@{ }l r@{ }l r@{ }l}
\hline\hline\noalign{\smallskip}
ID &
\multicolumn{2}{c}{\parbox[c]{1.0cm}{\centering \vbroad\ [\kms]}} &
\multicolumn{2}{c}{\parbox[c]{0.9cm}{\centering [C/Fe] $\lambda$5128}} &
\multicolumn{2}{c}{\parbox[c]{0.9cm}{\centering [C/Fe] $\lambda$5165}} &
\multicolumn{2}{c}{\parbox[c]{0.9cm}{\centering [C/Fe] $\lambda$5380}} &
\multicolumn{2}{c}{\parbox[c]{1.1cm}{\centering [C/Fe] $\lambda5052_{\rm EW}$}} &
\multicolumn{2}{c}{\parbox[c]{1.1cm}{\centering [C/Fe] $\lambda5380_{\rm EW}$}} &
\multicolumn{2}{c}{\parbox[c]{0.9cm}{\centering [C/Fe]}} &
\multicolumn{2}{c}{\parbox[c]{0.9cm}{\centering [N/Fe] $\lambda4215$}} &
\multicolumn{2}{c}{\parbox[c]{0.9cm}{\centering [O/Fe] $\lambda6300$}} \\
\noalign{\smallskip}\hline\noalign{\smallskip}
\object{CoRoT-10} &  3&.1   &    0.09 & $\pm$ 0.03 &    0.01 & $\pm$ 0.04 &         & ...        &         & ...        &         & ...        &    0.06 & $\pm$ 0.02 &    0.09 & $\pm$ 0.12 &         & ...        \\
 \object{CoRoT-2} &  9&.8   & $-$0.05 & $\pm$ 0.07 & $-$0.07 & $\pm$ 0.09 & $-$0.13 & $\pm$ 0.15 & $-$0.08 & $\pm$ 0.12 & $-$0.10 & $\pm$ 0.14 & $-$0.07 & $\pm$ 0.05 &         & ...        &    0.11 & $\pm$ 0.12 \\
...               &   & ... &         & ...        &         & ...        &         & ...        &         & ...        &         & ...        &         & ...        &         & ...        &         & ...        \\
    \object{XO-4} & 11&.6   &    0.11 & $\pm$ 0.12 &    0.11 & $\pm$ 0.18 &    0.07 & $\pm$ 0.13 &    0.03 & $\pm$ 0.12 &    0.08 & $\pm$ 0.14 &    0.08 & $\pm$ 0.07 &         & ...        &         & ...        \\
    \object{XO-5} &  3&.5   &    0.09 & $\pm$ 0.05 &    0.08 & $\pm$ 0.06 &    0.11 & $\pm$ 0.12 &    0.10 & $\pm$ 0.12 &    0.17 & $\pm$ 0.14 &    0.09 & $\pm$ 0.04 &    0.04 & $\pm$ 0.12 &    0.07 & $\pm$ 0.11 \\
\hline
\end{tabular}
}
\tablefoot{The first two columns give the target name and the average value of the composite of velocity fields (e.g., rotation and macroturbulent velocity) computed from single estimates. In the columns from 3 to 8, we list the carbon abundance ratios derived from each indicator plus the weighted mean and corresponding standard error, computed as explained in Sect.~\ref{section:c_abundances}. The last two columns show the N and O abundance ratios. The complete table is available at the CDS.}
\end{table*}

Even within the range of effective temperature that we adopt, we see in the upper panels of Fig.~\ref{figure:xfe_teff_feh} that there is a significant correlation of [C/Fe] with \teff\ for some of the carbon indicators. The trends are more pronounced for the two atomic lines, which are in general more susceptible to LTE departures in comparison with molecular features. However, a few studies in the literature \citep{Asplundetal2005,Amarsietal2019} compared the abundances derived in both non-LTE and LTE conditions and the differences are quite small for the atomic lines adopted in the current work. According to the 3D non-LTE calculations performed by \citet{Amarsietal2019}, the corrections to be applied to 1D LTE calculations mostly affects low-metallicity F dwarfs. For most of the stars in our sample, such corrections remains within 0.05~dex, achieving 0.08~dex for only a few cases. Applying these corrections to our sample do not remove neither reduce the trends with the effective temperature. Therefore, the carbon abundances that we provide do not include any correction due to 3D non-LTE effects. The question concerning the origin of the observed trends still remains. In order to avoid misleading conclusions that we may draw from the current data, we decided to correct the [C/Fe] abundance ratios from the linear regressions shown in Fig.~\ref{figure:xfe_teff_feh}. This means that the data plotted in Fig.~\ref{figure:cfe_feh_5} and in the other figures shown in this paper (unless explicitly stated) are all corrected from these trends. For comparison, the upper panels of Fig.~\ref{figure:xfe_feh_teff} shows a different version of Fig.~\ref{figure:cfe_feh_5}, before removing the dependence of [C/Fe] on the effective temperature. Cool and hot stars are clearly split into two sub-samples, specially for the atomic lines, causing a larger scatter.

A particularity of the different indicators plotted in Fig.~\ref{figure:cfe_feh_5} is that, on the one hand, the carbon abundances provided by the two atomic lines are clearly correlated with metallicity in the range of metal-poor stars, as expected from Galactic chemical evolution models (see discussion in the next section). On the other hand, this trend is not clearly seen for the abundances provided by the two molecular bands, in part due to the larger scatter. The most metal-poor star in our sample is definitely C overabundant and supports a negative slope. However, a higher number of stars in this range of metallicity is required to allow a better understanding of how these ${\rm C}_2$ molecular lines behave with decreasing metallicity.

The mean C abundances were calculated as follows: we used three indicators (5128, 5165, and 5380.3~\AA\ from the spectral synthesis method) for stars cooler than 5000~K, and four indicators (5128, 5165, and 5380.3~\AA\ from spectral synthesis, and 5052.2~\AA\ from the equivalent widths method) for stars hotter than 5000~K. For the 5380.3~\AA\ indicator, we see in Fig.~\ref{figure:delta_cfe5380_teff} that the abundances provided by the two methods agree quite well for stars hotter than 5200~K. Moreover, we see in Fig.~\ref{figure:cfe_feh_5} that the [C/Fe] ratios have a smaller scatter when applying spectral synthesis. For all these reasons, we preferred to adopt only the results from the spectral synthesis for this indicator.

We computed the final C abundances as mean values weighted by the uncertainties on the abundances from single indicators. Table~\ref{table:cno_abundances} lists the [C/Fe] abundance ratios, single and mean values, derived for each sample star together with the corresponding uncertainties. These are the values corrected, as described in the previous paragraphs, from the trends with the effective temperature.

\subsection{Nitrogen abundances}
\label{section:n_abundances}

The nitrogen abundances were also derived using pyMOOGi. We applied the spectral synthesis method to molecular lines of the electronic-vibrational band head of the CN blue system centred at 4215~\AA. As in the case of the carbon abundances, the model atmospheres include the elemental abundances derived for each star, in particular, the C abundances obtained as described in the previous section. This spectral region also contains molecular features of CH and, therefore, they were taken into account when computing the synthetic spectrum. The oscillator strengths for atomic lines (from VALD) and molecular features \citep[from][]{Kurucz1992} were also revised to match the solar spectrum, and small changes were applied to the \vbroad\ parameter if needed. The dissociation energy adopted for the CN transitions is ${\rm D}_0$ = 7.63~eV \citep{Reddyetal2003}.

Similarly to [C/Fe], the [N/Fe] abundance ratios also show a dependence on the effective temperature, as seen in the middle panel of Fig.~\ref{figure:xfe_teff_feh}. \citet{Ecuvillonetal2004} found similar trends for their samples of stars with and without detected planets. Their results were afterwards improved by \citet{SuarezAndresetal2016}, by using a larger sample with higher quality spectra. Although the slope that they found was not so steep, remaining within 2$\sigma$, they found a systematic underabundance of nitrogen for cool stars (\teff\ $<$ 5000~K). Possible explanations for the correlation with \teff\ are: $i)$ wrong definition of the continuum due to the presence of too many spectral features in metal-rich or in cool stars); $ii)$ the CN profile of hot stars is quite weak and difficult to fit (difficult to perform accurate measurements); $iii)$ the wavelength region measured is blended with several atomic lines of other elements, requiring accurate atomic parameters for a large number of transitions; $iv)$ or 1D LTE departures, as discussed in \citet{Amarsietal2020,Amarsietal2021} and \citet{Ryabchikovaetal2022}. Non-negligible 3D-1D abundance difference has been found for CN molecules in the Sun, but a complete list of corrections, to be applied to different stellar types, is still missing in the literature. Therefore, as done for carbon, we correct the [N/Fe] abundance ratios from the linear regression plotted in Fig.~\ref{figure:xfe_teff_feh}.

\begin{figure*}
\centering
\begin{minipage}[t]{0.87\textwidth}
\centering
\resizebox{\hsize}{!}{\includegraphics{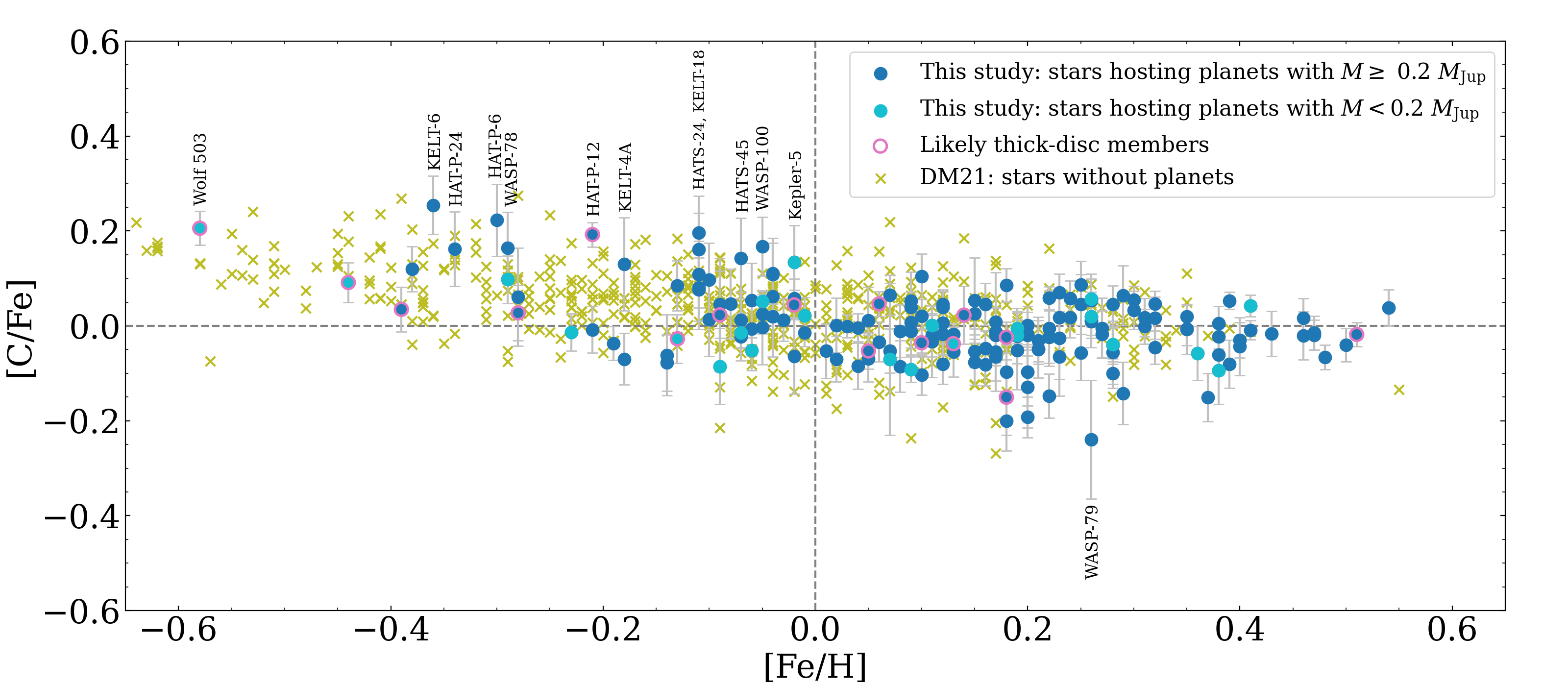}}
\end{minipage}
\caption{Mean [C/Fe] abundance ratios as a function of the stellar metallicity. Our results for planet-hosting stars are compared with the abundances from \citet[][DM21]{DelgadoMenaetal2021} for stars in the Galactic thin disc with no planet detected. The highlighted stars are discussed in the text.}
\label{figure:cfe_feh}
\end{figure*}
\begin{figure*}
\centering
\begin{minipage}[t]{0.87\textwidth}
\centering
\resizebox{\hsize}{!}{\includegraphics{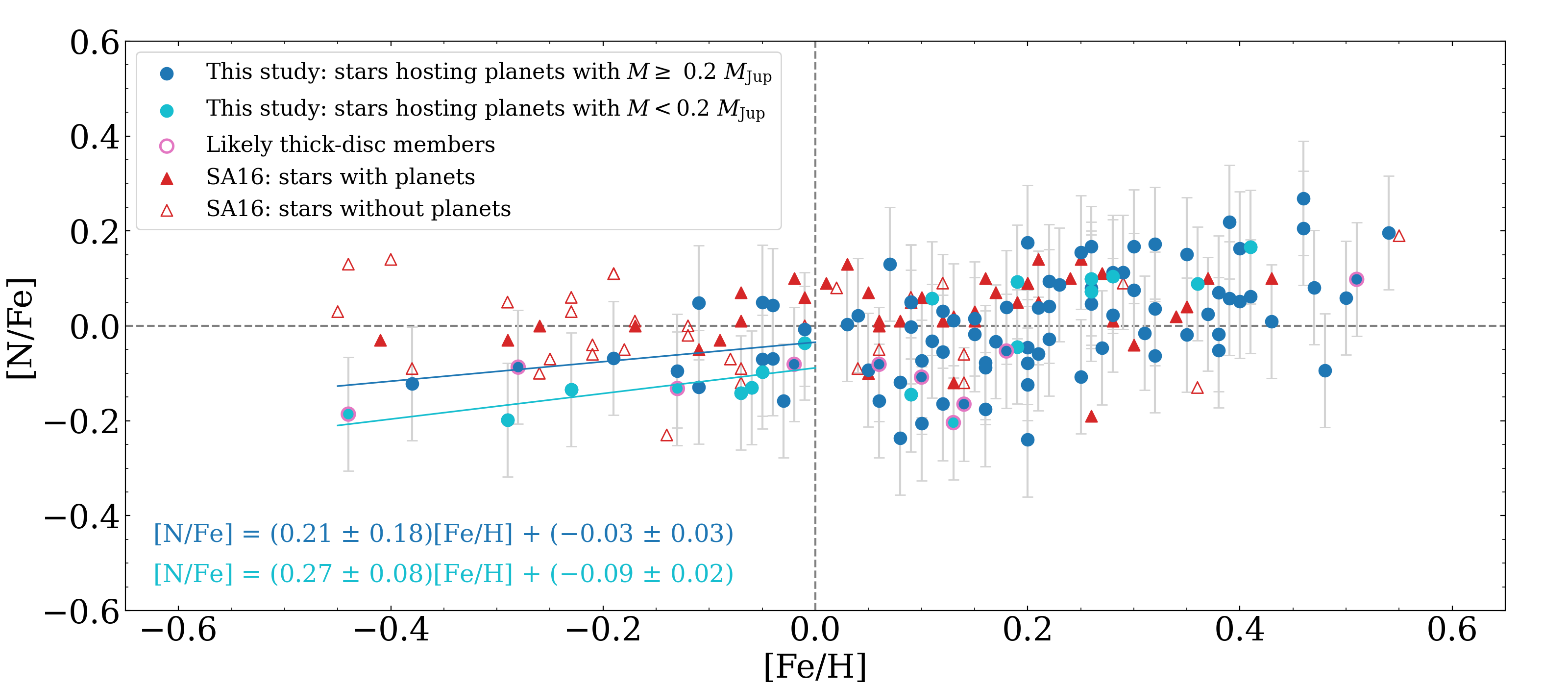}}
\end{minipage}
\caption{[N/Fe] abundance ratios as a function of the stellar metallicity. Same as in Fig.~\ref{figure:cfe_feh} but comparing our results with those from \citet[][SA16]{SuarezAndresetal2016} for stars with and without detected planets.}
\label{figure:nfe_feh}
\end{figure*}
\begin{figure*}
\centering
\begin{minipage}[t]{0.87\textwidth}
\centering
\resizebox{\hsize}{!}{\includegraphics{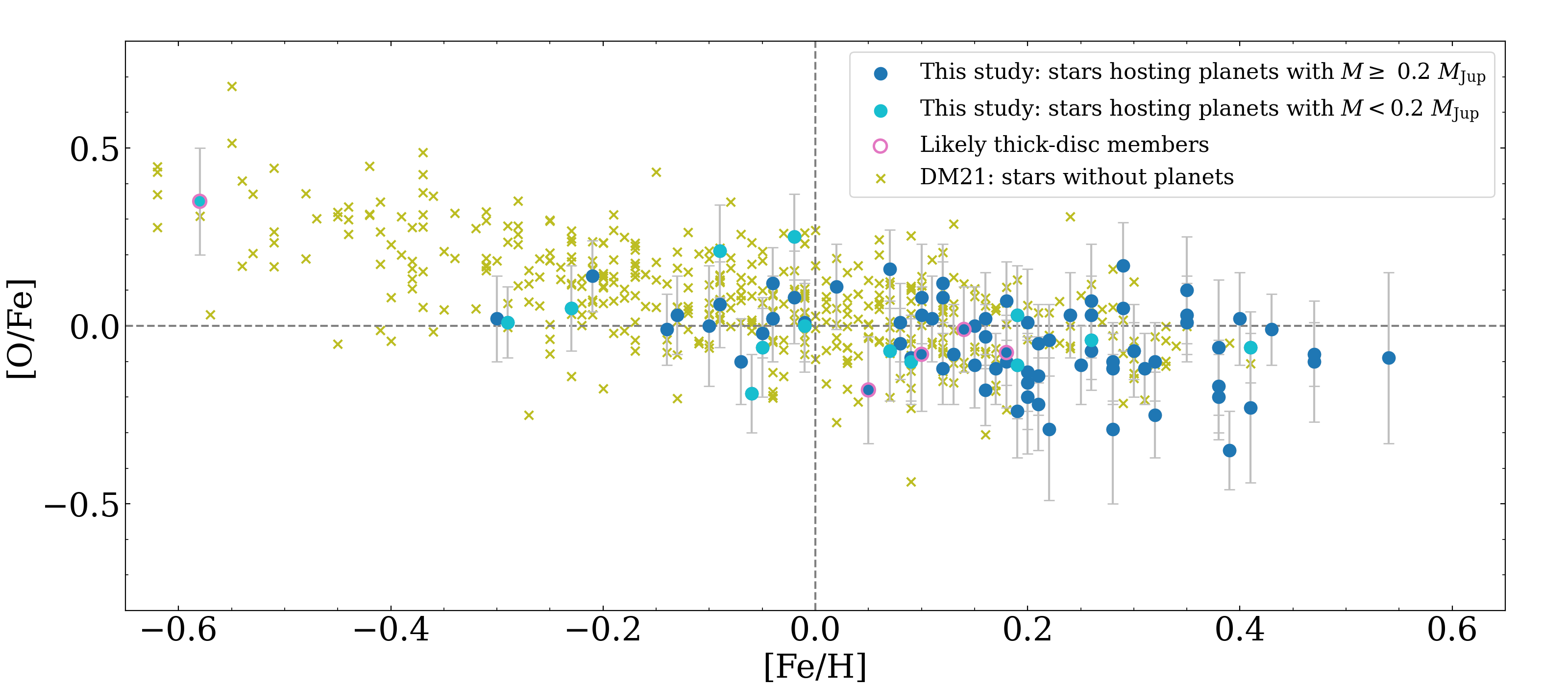}}
\end{minipage}
\caption{[O/Fe] abundance ratios as a function of the stellar metallicity. Same as in Fig.~\ref{figure:cfe_feh}, but comparing our abundance results from the oxygen line at 6300.3~\AA\ with those by \citet{Bertrandelisetal2015} updated by \citet{DelgadoMenaetal2021} and derived using the same line.}
\label{figure:ofe_feh}
\end{figure*}

The uncertainties in the N abundances, if computed in the same way as we did for the individual C indicators (i.e., using modified versions of the model atmospheres and then combining single contributions), provided overestimated values, much larger than the observed dispersion. Therefore, for this element, we preferred to adopt a fixed value of 0.12~dex, which is the standard deviation of the [N/Fe] measurements after removing the dependence on \teff.

Figure~\ref{figure:n_o_synthesis_ex} (left panel) shows an example of spectral synthesis around the CN band head for the same spectrum plotted in Fig.~\ref{figure:c_synthesis_ex} (star \object{K2-24}). The [N/Fe] abundance ratios, corrected from the trends with the effective temperature, and the corresponding uncertainties are listed in Table~\ref{table:cno_abundances}.

\subsection{Oxygen abundances}
\label{section:o_abundances}

The oxygen lines usually adopted to derive its abundance are very often weak. Therefore, very high S/N spectra are required to obtain reliable abundances for this element. In the current work, we applied the spectral synthesis method to the [\ion{O}{i}] forbidden line at 6300.3~\AA, which has a profile relatively weak for equivalent width measurements but strong enough for spectral synthesis. We note that the profile of this line might be affected by the presence of neighbouring atomic or molecular transitions. In the following paragraphs, we comment on the contribution, for instance, of some CN molecular features, of one \ion{Ni}{i} atomic line at 6300.336~\AA\ \citep{Lambert1978,AllendePrietoetal2001}, and of several telluric lines.

As for carbon and nitrogen, we adopted atomic and molecular data from VALD and \citet{Kurucz1992}, respectively. The CN features are relatively weak compared with the oxygen line. Therefore, they do not significantly affect the derived O abundances. The influence of the \ion{Ni}{i} line is taken into account by including the abundance of nickel derived by Delgado Mena (in prep.) to the model atmospheres used as input. We used the spectrum of telluric lines by \citet{Wallaceetal2011} to match our spectra. It was degraded to the resolution of each spectrum and corrected in wavelength according to the radial velocity measured. However, we did not apply any correction from the telluric lines. We just plotted them over the spectrum of each star, aiming at checking whether or not the oxygen line is affected by any of them, as shown in the right panel Fig.~\ref{figure:n_o_synthesis_ex}. If the oxygen and a telluric line are blended (which is not the case in the example shown in this figure), then the derived abundance is not used.

Some of the spectra in our sample show contamination by non-thermal airglow emission lines originating in the Earth's atmosphere \citep[see, e.g.,][]{Nolletal2012}. One of these features is the [\ion{O}{i}]\,6300 emission line, which is close to the oxygen line that we are measuring and can contaminate the stellar oxygen line depending on the radial velocity of the star. Since the intensity of this airglow line depends on the solar activity and the observing time, more than one spectrum of some stars was collected, trying to overcome the problem whenever possible.

Regarding possible departures from 1D LTE assumptions, non-LTE corrections should be negligible for the oxygen line at 6300.3~\AA\ whereas 3D effects are not negligible, but small, mostly affecting high-metallicity F dwarfs \citep{Asplundetal2004,Amarsietal2019,Caffauetal2008}. Therefore, also considering that we see no significant trend with the effective temperature in the bottom panel of Fig.~\ref{figure:xfe_teff_feh}, the [O/Fe] abundance ratios that we derive are not corrected from this trend.

As previously done for carbon, the uncertainties in the O abundances were derived using the modified versions of the model atmospheres (computed for different values of atmospheric parameters and \vbroad) and then calculating the quadratic sum of the single contributions. The derived [O/Fe] abundance ratios and the corresponding uncertainties are listed in Table~\ref{table:cno_abundances}.

\section{Results and discussion}
\label{section:results}

\subsection{Our sample within the Galactic chemical evolution context}

\subsubsection{[X/Fe] as a function of [Fe/H]}

Figure~\ref{figure:cfe_feh} shows the mean [C/Fe] abundance ratios, calculated as explained in Sect.~\ref{section:c_abundances}. For comparison, we also plot in this figure the carbon abundances derived by \citet{DelgadoMenaetal2021} for stars situated in the Galactic thin disc and for which no planet has been detected so far. According to chemical evolution models developed for the Milky Way \citep[see, e.g.,][]{Chiappinietal2003,Akermanetal2004}, at the beginning of the Galactic enrichment history, hence at low metallicities, carbon is synthesised by massive stars. On the other hand, iron is predominantly produced by type Ia supernovae (SNe~Ia) in longer timescales. At solar metallicities, stars having low or intermediate masses start contributing to the production of carbon, which causes a flattening of [C/Fe] as a function of [Fe/H]. Apart from some dispersion, we see that most of the stars in our sample follow this correlation with metallicity expected for stars belonging to the thin disc. We note that a population membership analysis was already performed by \citet{Magrinietal2022} for the current sample. A few particular cases are discussed in the Appendix~\ref{appendix:peculiar_stars}. Concerning the observed dispersion, it might also be related to differences in the age of the systems according to the theoretical predictions discussed in \citet[][and references therein]{Romano2022}.

Still regarding the [C/Fe] vs. [Fe/H] plot, a few studies in the literature suggested an opposite trend in the range of super-solar metallicity stars, compared to what we see in the metal-poor domain, in the sense that the C abundances seems to increase with increasing metallicity \citep{daSilvaetal2015,SuarezAndresetal2017}. This positive slope was not confirmed by \citet{DelgadoMenaetal2021} and it is not seen in the current work, at least not for the whole sample. \citet{SuarezAndresetal2017} found different slopes for different planetary masses, which seems to be the case for our results as well regardless the small number of lower-mass planet hosts in our sample of metal-rich stars.

The [N/Fe] abundance ratio is normally observed to increase with metallicity in the metal-rich regime \citep{Ecuvillonetal2004,daSilvaetal2015,SuarezAndresetal2016,Magrinietal2018}, where the so-called secondary production of nitrogen, based on pre-existing C and O in the star, starts dominating the primary production, in which nitrogen is synthesised from the material produced in the star itself \citep[see, e.g.,][and references therein]{Liangetal2006,Vincenzoetal2016}. This is also the behaviour that we see for our data in Fig.~\ref{figure:nfe_feh}, despite the larger scatter compared with literature data. Assuming that the dependence of the N abundances on the effective temperature (see Fig.~\ref{figure:xfe_feh_teff} and \ref{figure:xfe_teff_feh}) was properly removed, as explained in Sect.~\ref{section:n_abundances}, the distribution of points in this figure is worth some discussion.

An interesting behaviour of our data is that the decreasing of [N/Fe] with decreasing [Fe/H] seems to continue also for stars with sub-solar metallicity, which is not seen in the results by \citet{SuarezAndresetal2016}. However, these authors adopt the NH molecules to derive the nitrogen abundances, which could explain the disagreement. We see a similar trend in \citet[][also using the NH lines]{Ecuvillonetal2004}, though in the limit of 2$\sigma$, and in \citet[][using CN molecular bands]{daSilvaetal2015}, but the number of sub-solar metallicity stars is small. This trend is more clearly seen in \citet[][using the CN bands]{Magrinietal2018}, at least for stars belonging to the thin-disc population, and in \citet[][using the NH lines]{Takeda2023} for his sample of solar analogues.

Another peculiarity of our results for nitrogen is an apparent offset, in the range of low metallicities, between the abundances derived for gas-giant and lower-mass ($M<0.2~M_{\rm Jup}$) planet hosts. We see in Fig.~\ref{figure:nfe_feh} that nearly half of the stars contributing to the trend for [Fe/H] $<$ 0 (see the corresponding linear regressions) are orbited by a gas-giant planet and have [N/Fe] abundances systematically higher than stars hosting lower-mass planets. Nevertheless, though the difference seems systematic, a higher number of stars is required to support this outcome.

\begin{figure*}
\centering
\begin{minipage}[t]{0.7\textwidth}
\centering
\resizebox{\hsize}{!}{\includegraphics{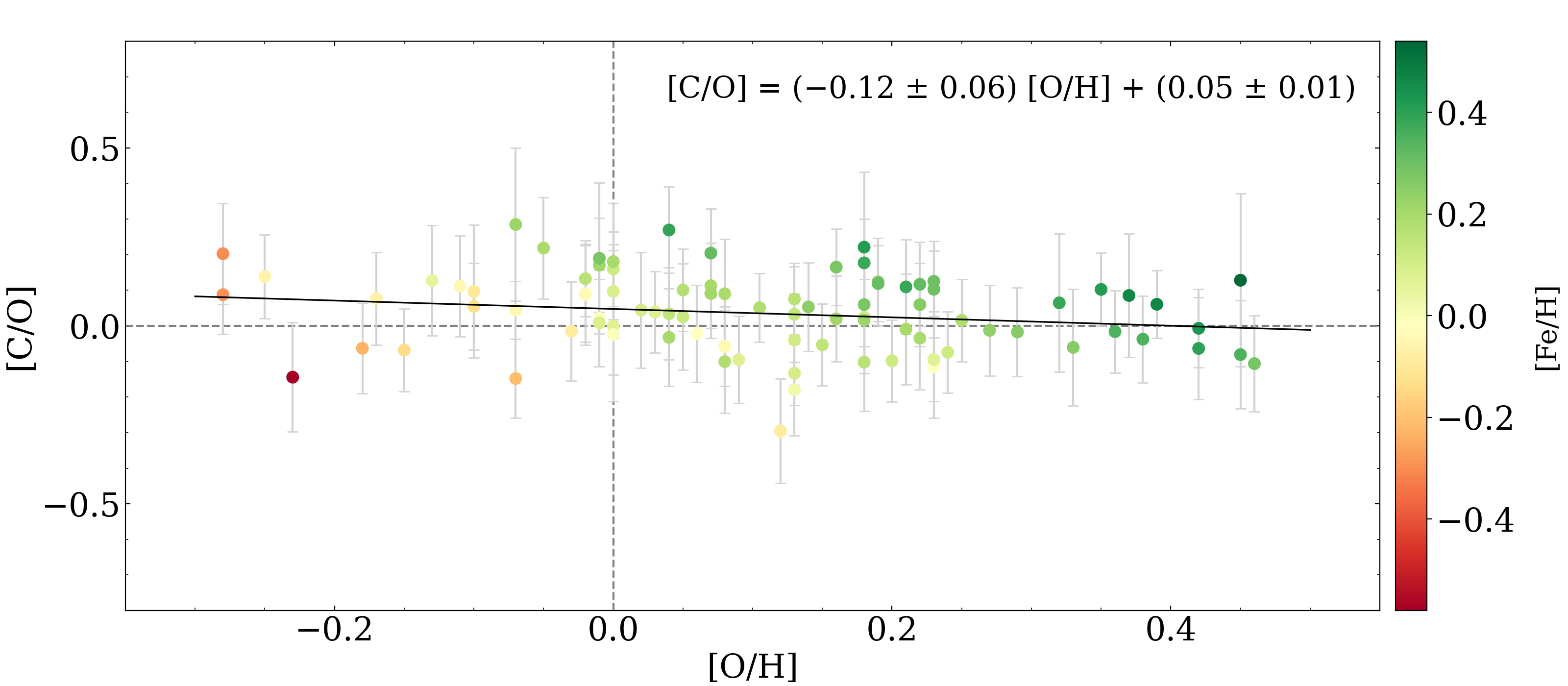}}
\end{minipage} \\
\begin{minipage}[t]{0.7\textwidth}
\centering
\resizebox{\hsize}{!}{\includegraphics{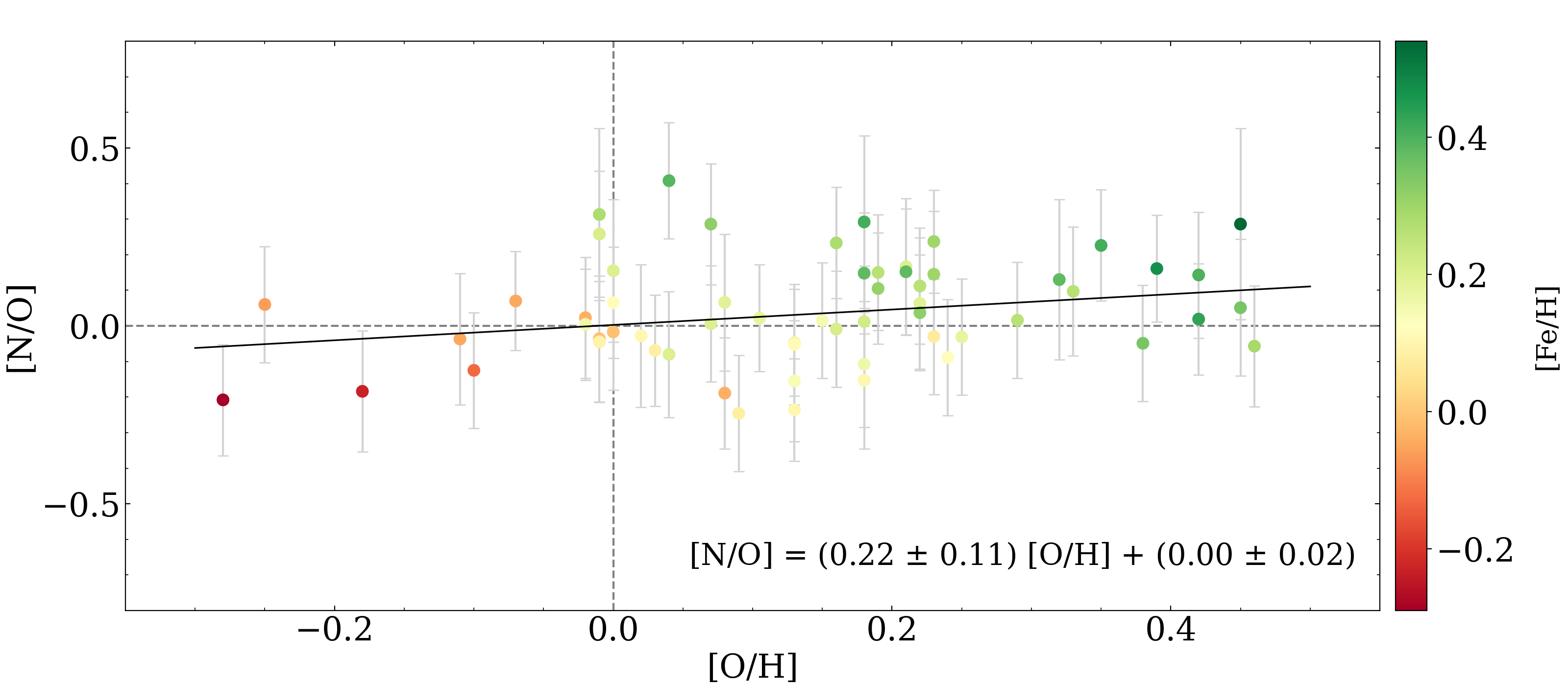}}
\end{minipage}
\caption{Abundance ratios as a function of [O/H] colour coded according to the stellar metallicity. The panels show the [X1/X2] ratios calculated correcting the abundances from the trends with the stellar effective temperature (see discussion in the text). The black dashed lines indicate the solar values.}
\label{figure:x1x2_ratios_x2}
\end{figure*}

Figure~\ref{figure:ofe_feh} shows the [O/Fe] abundance ratios and how they decrease with increasing metallicity. On the one hand, the trend in the range of low metallicities is more clearly seen in the data by \citet{Bertrandelisetal2015} or by \citet{DelgadoMenaetal2021} given the small number of metal-poor stars in our sample for which we were able to derive the oxygen abundance. On the other hand, despite the large scatter, a clear negative slope is seen the range of higher metallicities. The overall trend is in line with the predictions from chemical evolution models \citep[see, e.g.,][and references therein]{Chiappinietal2003,Spitonietal2015,Romano2022} and with several other results found in the literature, for instance, for stars with and without detected planets \citep[e.g.,][]{Ecuvillonetal2006,daSilvaetal2015}, for solar analogues compared with FGK dwarf stars \citep{Takeda2023}, and for the stellar abundance compilation of the Hypatia catalogue \citep{Hinkeletal2014}, which includes the results, though not homogeneous, of more than three thousand stars of almost one hundred studies. Again, as for the [C/Fe] ratios, the [O/Fe] abundance dispersion might be related to different ages of our stars \citep[see][and references therein]{Romano2022}.

\subsubsection{[X1/X2] as a function of [O/H]}

The usual way to follow the evolution of abundance ratios determined from stellar spectra is to use the Fe abundance as a reference. We do this because Fe is the element most easily measured in the spectra of FGK-type stars. A few examples are shown in Figs.~\ref{figure:cfe_feh}, \ref{figure:nfe_feh}, and \ref{figure:ofe_feh}. However, from the point of view of chemical evolution, it is more appropriate to refer to an element of primary nature, such as oxygen, instead of iron, which is produced with a time delay due to the evolution of the lower-mass component in the precursor of SNe~Ia. Still regarding the chemical evolution, on the one hand, we expect the ratio between two primary elements to remain constant as a function of time, here represented by the abundance of the primary element. This ratio will only be driven by the ratio of the stellar yields of the two elements. On the other hand, the ratio of a secondary element to a primary element should increase with time, i.e., with the abundance of the primary element \citep[see][]{PagelBernard1997}.

In Fig.~\ref{figure:x1x2_ratios_x2} we plot [C/O] and [N/O] vs. [O/H]. All abundances are corrected for the trends with the effective temperature. We consider O as a primary element, and we investigate the behaviour of C and N. The linear regression of [C/O] vs. [O/H] (on 89 stars) provides a slightly negative slope $-$0.12 $\pm$ 0.06, which is, within errors, consistent with a constant ratio of [C/O]. The intercept is 0.05 $\pm$ 0.01, consistent with the solar value of [C/O]. In the sampled metallicity range, the two elements appear to have a dominant primary production and their ratio remains constant. The linear regression of [N/O] vs. [O/H] (computed with 60 stars) has instead a positive slope 0.22 $\pm$ 0.11, consistent with a secondary production of N \citep[see, e.g.,][]{Romano2022}, whose abundances increase with [O/H]. The intercept of the fit is 0.00 $\pm$ 0.02. So, since the slope is positive, we have positive [N/O] at super-solar [O/H], usually larger than the solar value. From the point of view of the formation of planets and their atmospheres, at high metallicity we are provided with more N than O, while [C/O] is less affected and remains more or less constant.

\begin{table}[t]
    \caption{Coefficients of the linear regressions of abundance ratios as a function of [Fe/H] fitted to the data in Fig.~\ref{fig:2d_ratios_feh}.}
    \label{tab:CNO_vs_FeH_coefficients}
    \centering
    \begin{tabular}{c r@{.}l r@{.}l r@{}l c}
       \hline\hline\noalign{\smallskip}
       [X1/X2] &
       \multicolumn{2}{c}{$a$} &
       \multicolumn{2}{c}{$b$} &
       \multicolumn{2}{c}{p-value} &
       $N_{\rm stars}$ \\
       \noalign{\smallskip}\hline\noalign{\smallskip}
       $[$C/O$]$ &    0&012 &    0&175 & 2.636&$\cdot 10^{-3}$  &  88 \\
       $[$C/N$]$ &    0&076 & $-$0&43  & $<$ 2&$\cdot 10^{-16}$ & 105 \\
       $[$N/O$]$ & $-$0&068 &    0&578 & 4.071&$\cdot 10^{-9}$  &  60 \\
       \hline
    \end{tabular}
    \tablefoot{The columns show the coefficients of [X1/X2] = $a + b\cdot$[Fe/H], the corresponding p-values, and number of stars used for each linear regression. This number varies depending on whether it was possible to determine both X1 and X2 elements in the ratio.
    }
\end{table}

\subsection{Relations with planetary properties}

We explored possible correlations between the stellar elemental ratios [C/O], [C/N], and [N/O] and their planetary mass and radius, retrieved from the NASA Exoplanet Archive. For this part of the analysis, we only selected the sample of stars for which all three elemental abundances, together with the planetary properties (radius and mass), were determined, corresponding to a total of 58 systems. Before removing the dependence on \teff\ (see Sect.~\ref{section:abundances}) we found correlation between [C/N] and both planetary radius (p-value = 0.01) and planetary mass (p-value = 0.008), as well as correlation between [N/O] and the planetary radius (p-value = 0.041). Yet, after correcting the data for the \teff\ dependence, no significant correlations are present (p-value $>$ 0.05) among any elemental ratio and the planetary radii and masses (see Fig.~\ref{figure:x1x2_elem_ratios_planetary_mass_radius}, which shows the elemental ratios not normalised to the solar values).

\subsection{About the assumption of stellar properties when inferring the planetary composition.}
\label{sec:wrongassumption}

\begin{figure*}
    \centering
        \includegraphics[trim=3.5cm 1cm 0cm 0.5cm,
      clip,width=.42\textwidth]{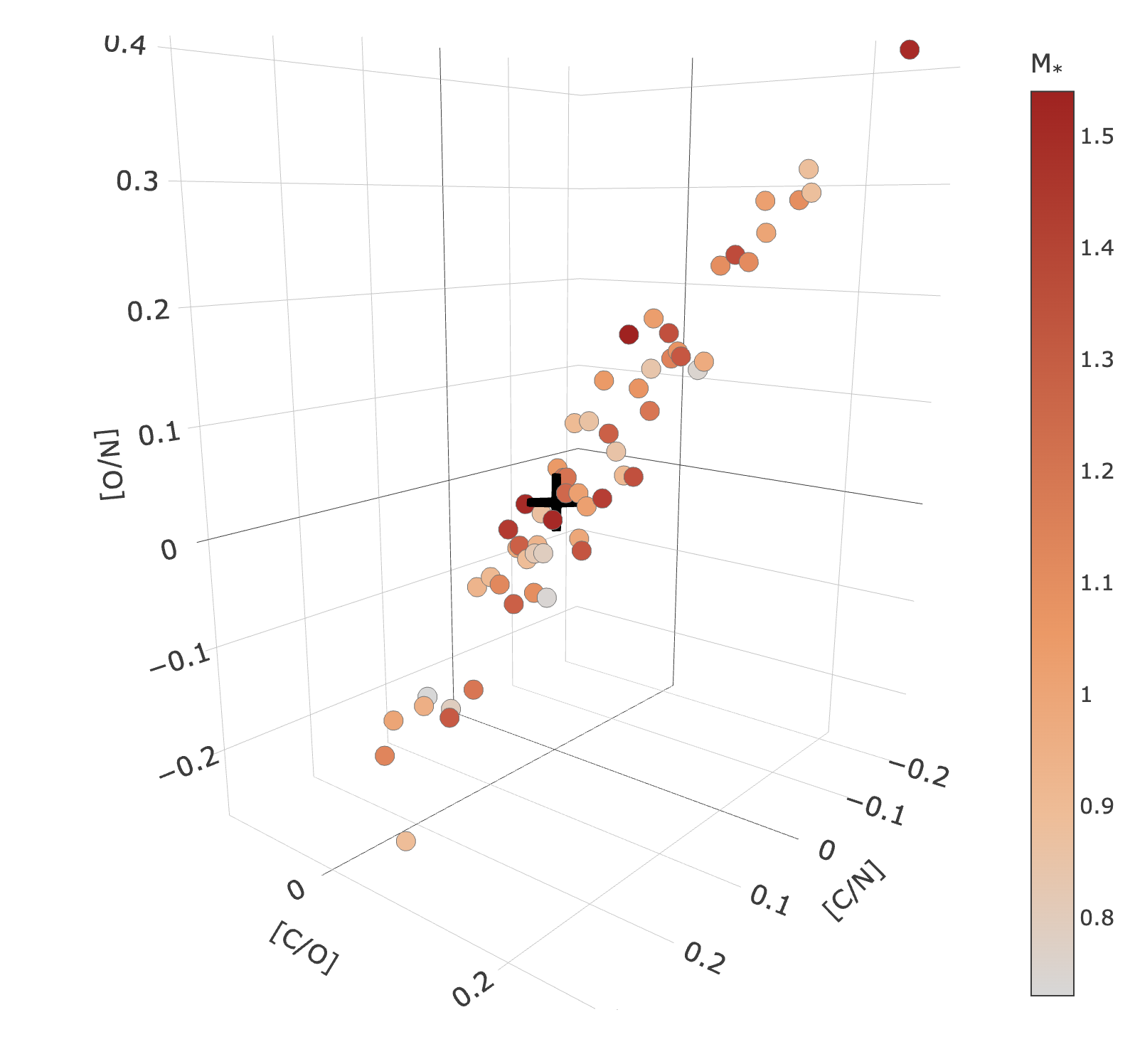}\quad \quad
    \includegraphics[trim=3.5cm 1cm 0cm 1cm,
      clip,width=.42\textwidth]{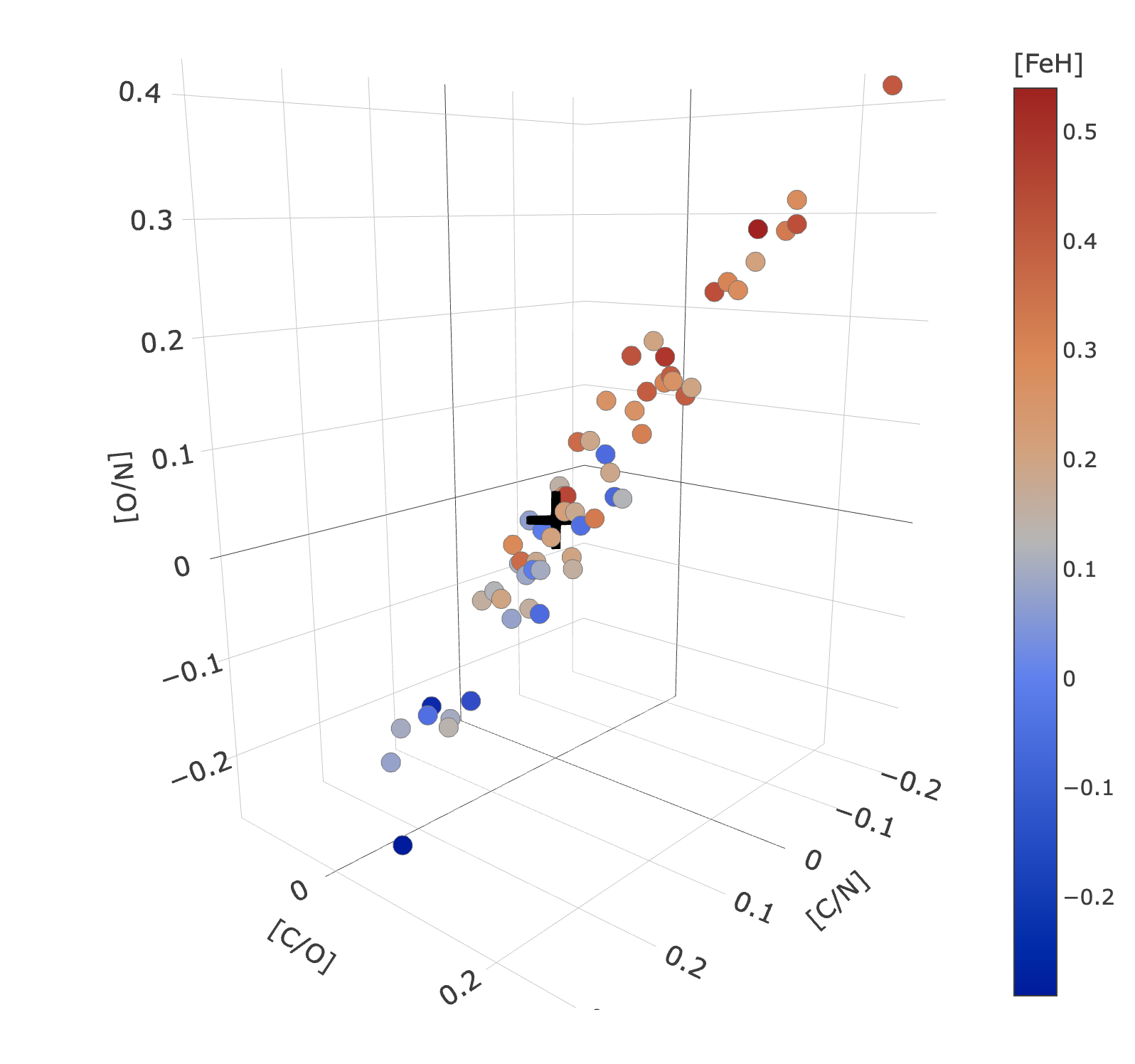}\quad \quad
    \caption{Relation among the three stellar elemental ratios, corrected from the trends with the effective temperature and normalised to the solar values, for the sample of 60 stars with all three C, N, and O determined (see also Table~\ref{tab:CNO_vs_FeH_coefficients}). Markers are colour-coded based on the stellar mass \textit{(left)} and \feh\ \textit{(right)}. The black cross represents the Sun.
    }
    \label{fig:ratios_feh_sun}
\end{figure*}

\begin{figure*}
    \centering
    \includegraphics[width=.46\textwidth]{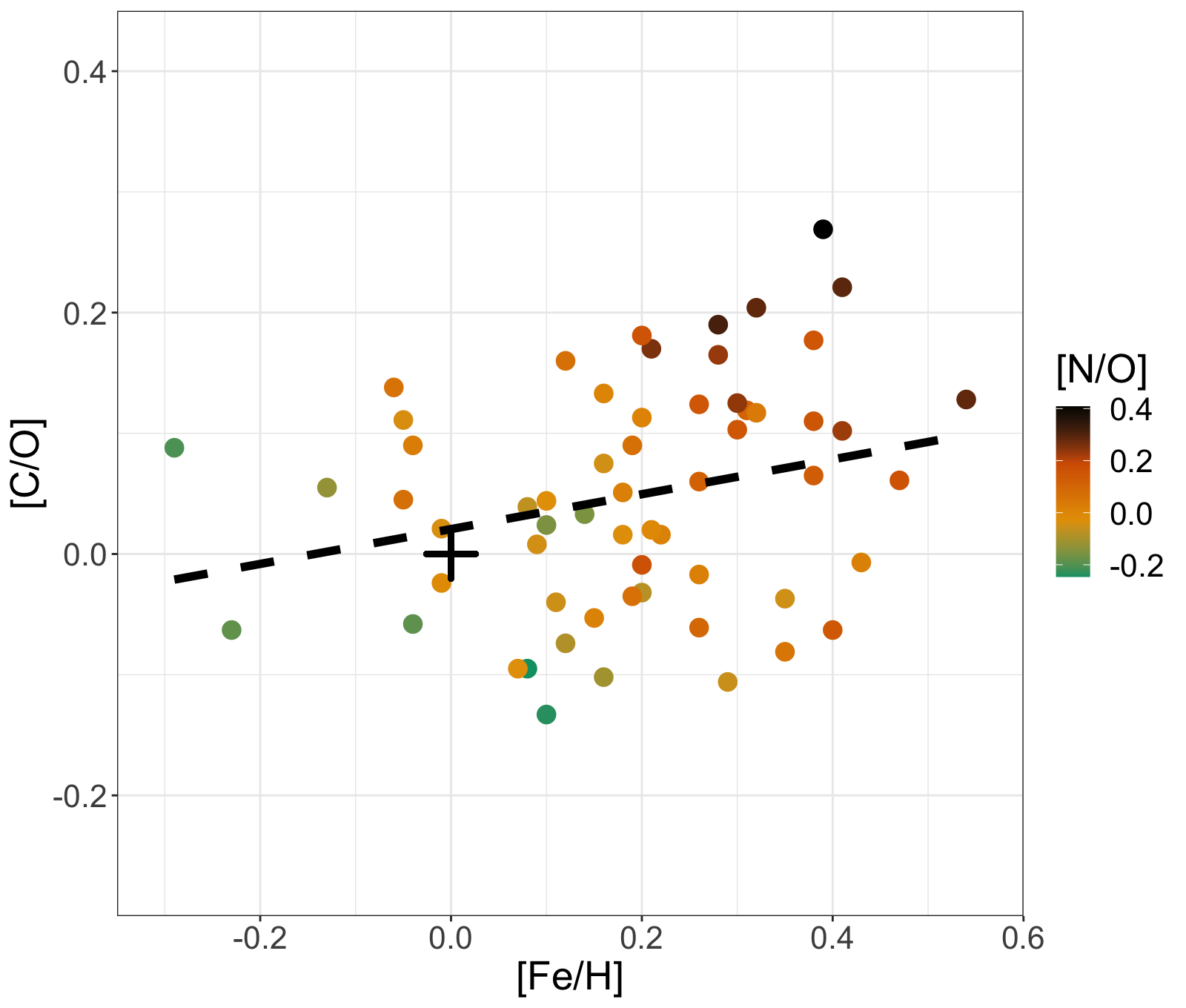}
    \includegraphics[width=.46\textwidth]{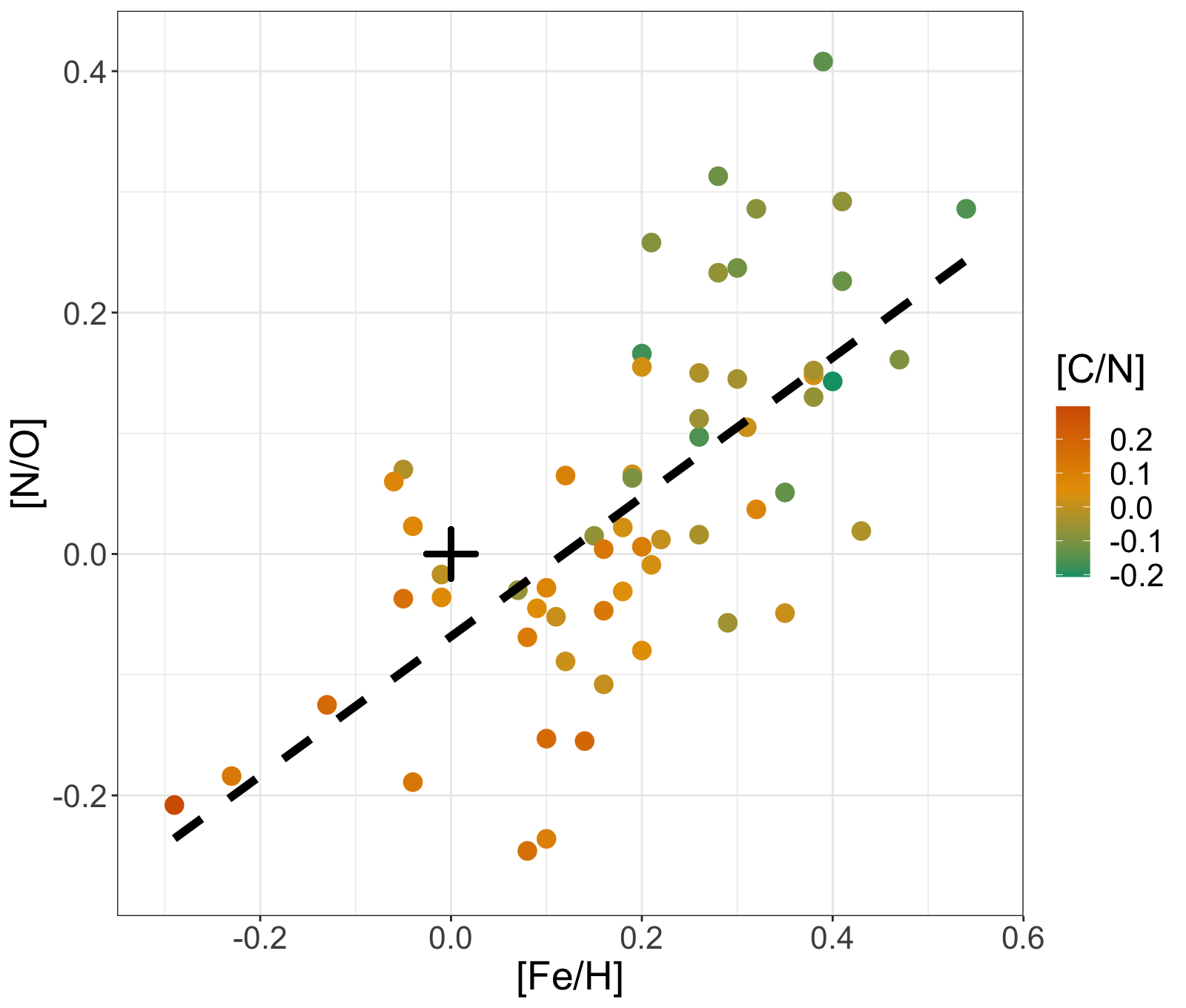}
    \caption{Stellar [C/O] \textit{(left)} and [N/O] \textit{(right)} as function of the stellar \feh~ for the sample of 60 stars with all three C, N and O determined. Markers are colour-coded with [N/O] and [C/N] values, respectively. All elemental ratios are normalised to the solar values. The black cross represents the Sun. Dashed line mark the linear regression line. Y-axis scale was kept the same in both plots for comparison purposes.
    }
    \label{fig:2d_ratios_feh}
\end{figure*}

We present in Figs.~\ref{fig:ratios_feh_sun} and \ref{fig:2d_ratios_feh} the 4D and 3D relations, respectively, between the homogeneously derived [C/O], [C/N], [N/O] ratios and both stellar mass and metallicity \citep[both from][]{Magrinietal2022}. We note that, for the purpose and clarity of the upcoming discussion, the elemental ratios have been normalised to the solar abundances from \citet{Asplundetal2009} (see Appendix~\ref{appendix:x1x2_elem_ratios}). While there is no evident correlation with the first parameter, a clear trend appear with \feh\ (see also Figs.~\ref{figure:x1x2_elem_ratios_feh} and \ref{figure:x1x2_elem_ratios_stellar_mass}). Moving to sub-solar metallicities both [C/O] and [N/O] decrease, while the stellar [C/N] increases. Vice-versa, going towards super-solar metallicities, where the majority of our sample lies, both [N/O] and [C/O] increase, while [C/N] decreases. The 3D trends can be better visualised in Fig.~\ref{fig:2d_ratios_feh}. The largest spread of values is given by [N/O], which covers a range of $\sim$0.654~dex, followed by [C/N] (0.586~dex), and finally by [C/O] (0.58~dex).

We assessed possible correlations between the stellar ratios and \teff, \feh, and \Mstar\ by applying a linear regression. We tested various linear combinations with either \feh, \teff, or \Mstar\ alone, and a combination of all them. In all cases, while \feh\ was always a significant variable, neither \teff\ or \Mstar\ appeared to be of any weight in the fit. For such, and following Occam's razor principle, we applied a linear fit only as a function of \feh. We report in Table~\ref{tab:CNO_vs_FeH_coefficients} the coefficients of the linear regression applied to the stellar [C/O], [C/N], and [N/O] ratios as a function of the stellar metallicity \feh. These relations are valid for FGK dwarf stars with planetary companions.

The observed trends among the abundance ratios of C, O, and N and the stellar metallicity (Fig.~\ref{fig:ratios_feh_sun}) have an important impact on the science of exoplanets. Generally, when studying the planetary composition, the stellar composition is used as a reference for the chemistry of the disc, i.e., the planet-forming environment. Often, it is assumed that the stellar host has solar metallicity and composition, all stellar elemental abundances being scaled accordingly. These assumed stellar abundances are then used to determine whether the exoplanet is enriched or not in the observed elements. As a result, the planetary metallicity or a given elemental abundance can be estimated as 10x solar when the relevant value in the actual host star is 3x solar, meaning that the real planetary enrichment is actually 3.3x. This inaccurate approach can be responsible for introducing biases in the interpretation of the planetary compositions and formation histories.

Specifically, the assumption of a solar composition of the host star biases the interpretation of the planetary elemental ratios and the information they carry on the planet formation process. As an illustrative example, let's imagine a giant planet is observed to have [C/O] = 0.2, meaning that its C/O $\approx$ 0.9 if we adopt the solar C/O$_\odot$ = 0.55 as reference \citep{Asplundetal2009}. The observed planetary value would suggest the accretion of C and O from the gas in the disc (being C/O > C/O$_\odot$), with limited contribution from the accretion of planetesimals \citep[see][and references therein]{Madhusudhan2019}. However, if the planet orbits one of the stars with [C/O] $\sim$ 0.2 in our sample (see Fig. \ref{fig:2d_ratios_feh}, left), the observed planetary C/O value could very well be stellar (i.e., C/O$^{*}$ $\approx$ 1), which would imply instead that the main source of C and O is the accretion of planetesimals \citep{Turrinietal2021}. The unverified adoption of the solar reference therefore leads to wrong inferences on the formation process of the planet. Another example focused on the C/O value is linked to the possibility of identifying giant planets that formed in discs that underwent global heating events due to stellar flares. Astrochemical models \citep{Eistrupetal2016,Pacettietal2022} indicate that giant planets formed in such environments could be characterised by a normalised C/O$^{*}$ < 1 (i.e., substellar) when the main source of their C and O is the accretion of disc gas. A giant planet with C/O = 0.55 orbiting a supersolar metallicity star with C/O = 0.7 ([C/O] = 0.1) would have C/O$^{*}$ < 1, i.e., substellar, suggesting that its native disc underwent such global thermal events. We refer readers to \citet{Biazzoetal2022,Guilluyetal2022,Carleoetal2022} for observational applications of these considerations.

A more subtle bias is linked to the different correlations between the stellar C/N, C/O and N/O ratios and the stellar metallicity. The relative deviations of these ratios in planetary atmospheres with respect to their respective stellar values have been shown to increase with the distance of their native region from the host star \citep{Turrinietal2021,Pacettietal2022}, meaning that the farther the planets start their formation process, the more their C/N$^{*}$, C/O$^{*}$ and N/O$^{*}$ ratios are expected to deviate from 1, marking the stellar value in the normalised scale. Since the stellar C/N systematically decreases with respect to the solar one for increasing stellar metallicities (Fig.~\ref{fig:2d_ratios_feh}, right), while N/O and C/O systematically increase (Fig.~\ref{fig:2d_ratios_feh}, left), the adoption of solar values means that we systematically underestimate (overestimate) the planetary C/N (N/O and C/O) values for stars of supersolar metallicity, the opposite being true for stars of subsolar metallicity. This leads to incorrect constraints on the formation region of the giant planets from their deviations from the stellar values, e.g., causing giant planets around supersolar metallicity stars to appear to have formed closer to their host stars than they actually have (see, e.g., Figs.~7-8 in \citealt{Turrinietal2021} or Figs.~4-7 in \citealt{Pacettietal2022}).

Finally, as the C/O ratio increases with \feh\ (as seen in Fig.~\ref{fig:2d_ratios_feh} with [C/O]) while the [O/Fe] decreases for increasing \feh\ (see Fig.~\ref{figure:ofe_feh}), adopting the solar abundances leads to incorrect interpretation of the atmospheric chemistry of giant planets, particularly at equilibrium temperatures below 1100~K. In the atmospheres of these planets O also binds with Si, Fe, and Mg to form oxides, as their bonds are chemically favoured. This leads to decreased abundances of H$_2$O and the overestimation of the planetary C/O ratio when the O sequestration by oxides is not quantified and corrected for \citep{Fonteetal2023}, and the element is measured using only the observed CO, H$_2$O, and CH$_4$. As the magnitude of this effect depends on the relative abundance of Fe, Mg, and Si with respect to O, using the solar values instead of the real stellar ones can lead to the wrong estimation of the correction factor and, consequently, the planetary C/O value, once again introducing biases in the interpretation of the formation history of the planets. When we take into account this effect and the changes in the stellar C/O with the stellar metallicity, it is immediate to see that the temperature at which CO and H$_2$O have equal abundances, marking the transition between CH$_4$-dominated to CO-dominated atmospheres \citep[e.g.][]{Madhusudhanetal2016}, will be different in the atmospheres of giant planets with the same equilibrium temperature orbiting stars with different composition (e.g., different [O/Fe]). The adoption of solar values as references can then bias our understanding of the atmospheric chemical environment of planetary atmospheres. As illustrate by these examples, referencing the real stellar abundances is therefore critical for the interpretation of the atmospheric composition of exoplanets.

Therefore, we strongly recommend to characterise at least the stellar metallicity, which is a parameter that can be more easily determined than stellar abundances, and use the relations in Table~\ref{tab:CNO_vs_FeH_coefficients} to infer the host star elemental ratios, for a proper interpretation of the planetary composition.

One more consideration to make is related to the history of a planetary system, and its migration across the Galaxy. As reported by the kinematics of the star measured by \citet{Magrinietal2022}, the sample of stars we have analysed came from different regions of the Milky Way, e.g., from either the inner disc, the outer disc, or thin disc versus thick disc. It is known that the chemical properties of the stars depend on the galactic environment they originated within, and such aspect is also true for their planet(s). A planet formed in a region of overdensity will present different characteristics from a planet formed within an isolated region. To best interpret the planetary atmospheric data, it is consequently pivotal to trace the migration history of their host stars. In Fig.~\ref{figure:cfe_feh_teff_vpec} (bottom panel) we show our stellar sample as a function of their [C/Fe] and metallicity, highlighting the role of galactic migration: stars with orbits having a higher eccentricity might come from the inner or outer regions compared to their current position. Some of these stars are located at the extremes of the metallicity distribution of our sample. It is precisely because of the effect of stellar migration that we can, therefore, extend our sample in metallicity and in abundance ratios, thus investigating the relation between host star and planetary system in chemically different environments.

\section{Summary and conclusions}
\label{section:conclusions}

In this work, we have presented a homogeneous determination of the abundances of the volatile elements C, N, and O for a sample of 181 stars belonging to the Ariel Mission Candidate Sample, based on the homogeneously derived stellar parameters (\feh, \teff, \logg, $\xi$, and \Mstar) presented in \cite{Magrinietal2022} and for which we were able to obtain reliable abundances. By using the same grid of model atmospheres and the same radiative transfer code, we applied the spectral synthesis method to atomic lines (the \ion{C}{i} line at 5380.3~\AA, and the [\ion{O}{i}] forbidden line at 6300.3~\AA) and molecular features (the ${\rm C}_2$ bands at 5128 and 5165~\AA, and the CN band at 4215~\AA), and the equivalent width method to the \ion{C}{i} atomic lines at 5052.2 and 5380.3~\AA.

In order to avoid misleading conclusions regarding the chemical enrichment of the Milky Way or the relations between stellar abundances and planetary parameters, we corrected the [C/Fe] and [N/Fe] abundance ratios from the trends with the effective temperature. The same procedure was not applied to the [O/Fe] ratios considering the absence of a significant trend. We checked for possible departures from 1D LTE but, even if they are taken into account, they do not reduce the observed trends.

The final [C/Fe] abundance ratios were calculated as a mean of several indicators, after removing the dependence on \teff. The derived mean C abundances were then used to derive the N abundances from the CN indicator. The influence of the \ion{Ni}{i} atomic line at 6300.336~\AA\ was taken into account when deriving the O abundances. Moreover, the presence of telluric or airglow emission lines was also considered and hence contaminated lines were not included in our analysis. The main results of the current study are listed below.

\begin{enumerate}

\item Most of the stars in our sample follow the Galactic chemical evolution (global) trends expected for the [C/Fe], [N/Fe], and [O/Fe] abundance ratios as a function of [Fe/H]. A possible increasing trend of [C/Fe] with increasing [Fe/H], at least for stars hosting planets with $M<0.2~M_{\rm Jup}$, would confirm previous findings available in the literature, but it should be investigated using more stars. For carbon, there are some chemically particular cases, which are discussed in the Appendix~\ref{appendix:peculiar_stars}. The expected decreasing of [N/Fe] with decreasing [Fe/H] seems to continue also for stars with sub-solar metallicity. Moreover, there is an apparent offset of the [N/Fe] ratios between gas-giant and lower-mass planet hosts. However, a higher number of metal-poor stars are required to confirm these findings.

\item The [C/O] abundances as a function of [O/H] show, within errors, a constant ratio, with a slope of $-$0.12 $\pm$ 0.06 that is in line with chemical evolution predictions for the ratio between two primary elements. The [N/O] ratios as a function of [O/H] have instead a positive slope of 0.22 $\pm$ 0.11, which is consistent with a secondary production of N. Regarding the formation of planets and their atmospheres, both these results indicate that, as we move to higher metallicities, we are provided with more N than O, whereas the relative amount of C and O tends to remain constant.

\item For what it concerns the science of planetary systems, we show that the [C/N], [C/O], and [N/O] ratios are correlated with [Fe/H]. The observed trends among the abundance ratios of C, O, and N and the stellar metallicity prove once and for all that the commonly adopted approach of using the solar abundances as reference for modelling and interpreting the atmospheric composition of giant planets around host stars of different metallicity is incorrect and introduces biases in the interpretation of the planetary compositions and formation histories. In the absence of stellar abundances, we provide relations as function of [Fe/H] that can be used to infer reference values to qualitatively estimate whether the atmospheric composition of planets is enriched or not with respect to the host stars, providing a more physically-justified framework than the use of solar abundances.

\end{enumerate}

The current investigation presents the first results of a work in progress. We plan to increase the number of abundance indicators, and to apply a similar analysis to larger sample of stars belonging to the Ariel Tier 1. At the time of writing this paper, new spectra have been collected and new observing programmes have been proposed aiming at improving the current results and complementing the wavelength ranges not yet covered by the spectra already available.

Finally, all the C, N, and O abundance ratios presented here (Table~\ref{table:cno_abundances}) are available via both the CDS\footnote{http://cdsportal.u-strasbg.fr} and the Ariel Stellar Catalogue\footnote{Publicly available via e-mail to \href{}{scwgariel@gmail.com}}. The latter also includes atmospheric parameters (\feh, \teff, \logg, and $\xi$), kinematic properties and for 187 planet-host FGK dwarf stars.

\begin{acknowledgements}
We thank the reviewer for their positive words and constructive suggestions concerning an earlier version of the present paper, which improved its content and readability.
This work has been developed within the framework of the Ariel ``Stellar Characterisation'' working group of the ESA Ariel space mission Consortium.
The team is very grateful to the service astronomers that performed our observations at ESO (with UVES during P105 and P106), with the TNG (using HARPS-N during A41 and A42), with the LBT (using PEPSI during 2021 and 2022), and at the SAAO (with SALT during 2023).
This work has made use of the VALD database, operated at Uppsala University, the Institute of Astronomy RAS in Moscow, and the University of Vienna. 
We acknowledge financial support from the ASI-INAF agreement n. 2022-14-HH.0.
E.D.M acknowledges the support from Funda\c{c}\~ao para a Ci\^encia e a Tecnologia (FCT) through national funds and from FEDER through COMPETE2020 by the following grants: UIDB/04434/2020 \& UIDP/04434/2020 and 2022.04416.PTDC. E.D.M. further acknowledges the support from FCT through Stimulus FCT contract 2021.01294.CEECIND.
C.D. acknowledges financial support from the INAF initiative ``IAF Astronomy Fellowships in Italy'', grant name \textit{GExoLife}.
Polish participation in SALT is funded by MEiN grant No. 2021/WK/01.

The LBT is an international collaboration among institutions in the United States, Italy and Germany. LBT Corporation partners are: Istituto Nazionale di Astrofisica, Italy; The University of Arizona on behalf of the Arizona Board of Regents; LBT Beteiligungsgesellschaft, Germany, representing the Max-Planck Society, The Leibniz Institute for Astrophysics Potsdam, and Heidelberg University; The Ohio State University, representing OSU, University of Notre Dame, University of Minnesota and University of Virginia.
The TNG is operated by the Fundación Galileo Galilei (FGG) of the Istituto Nazionale di Astrofisica (INAF) at the Observatorio del Roque de los Muchachos (La Palma, Canary Islands, Spain).
Some of the observations reported in this paper were obtained with the Southern African Large Telescope (SALT).
This research has made use of the NASA Exoplanet Archive, which is operated by the California Institute of Technology, under contract with the National Aeronautics and Space Administration under the Exoplanet Exploration Program.
\end{acknowledgements}

\newpage
\clearpage

\bibliographystyle{aa}
\bibliography{aanda.bib}


\begin{appendix}

\section{Abundance dependence on \teff}
\label{appendix:atm_par_dependence}

\begin{figure}
\centering
\begin{minipage}[t]{0.49\textwidth}
\centering
\resizebox{\hsize}{!}{\includegraphics{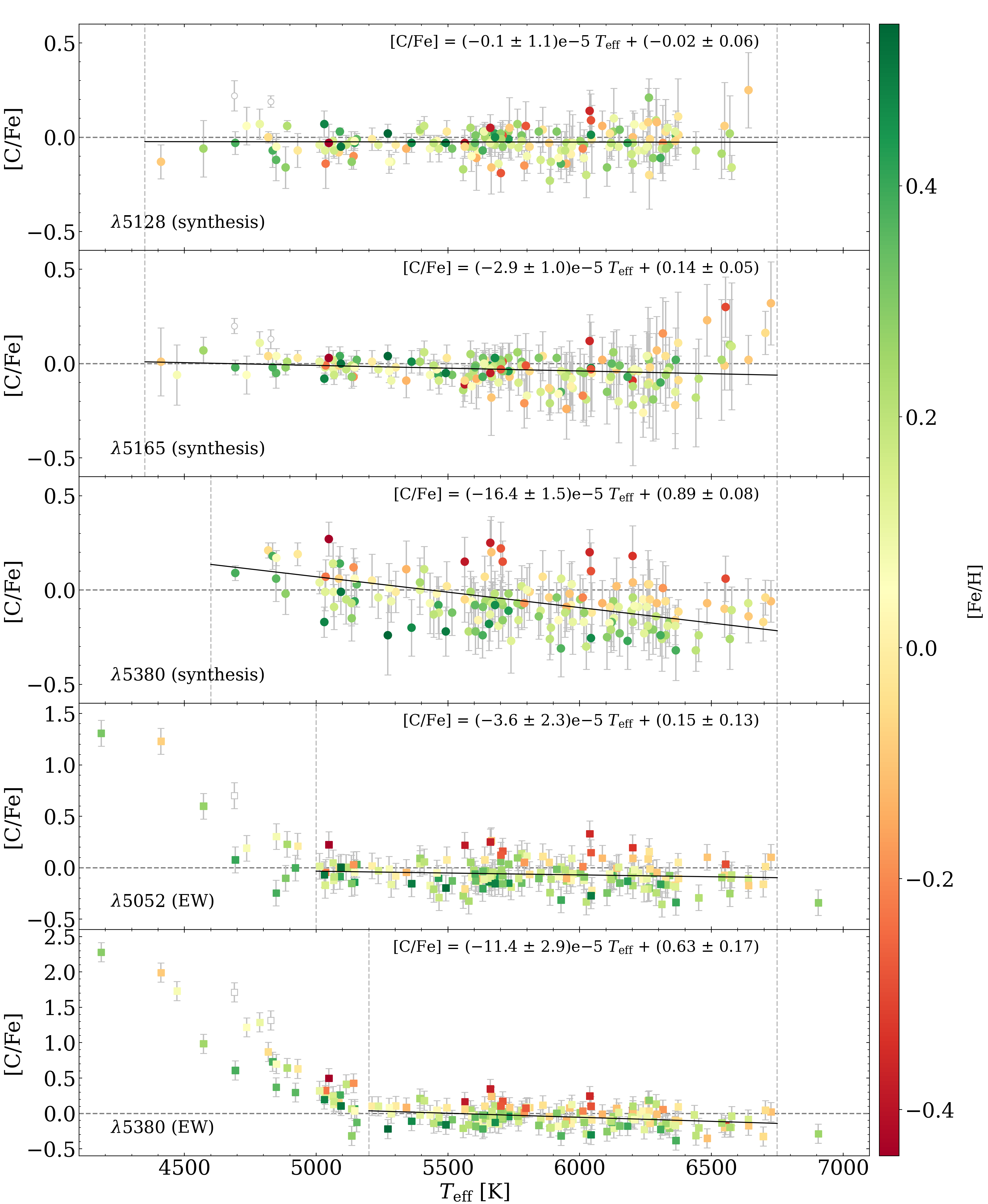}}
\end{minipage} \\
\begin{minipage}[t]{0.49\textwidth}
\centering
\resizebox{\hsize}{!}{\includegraphics{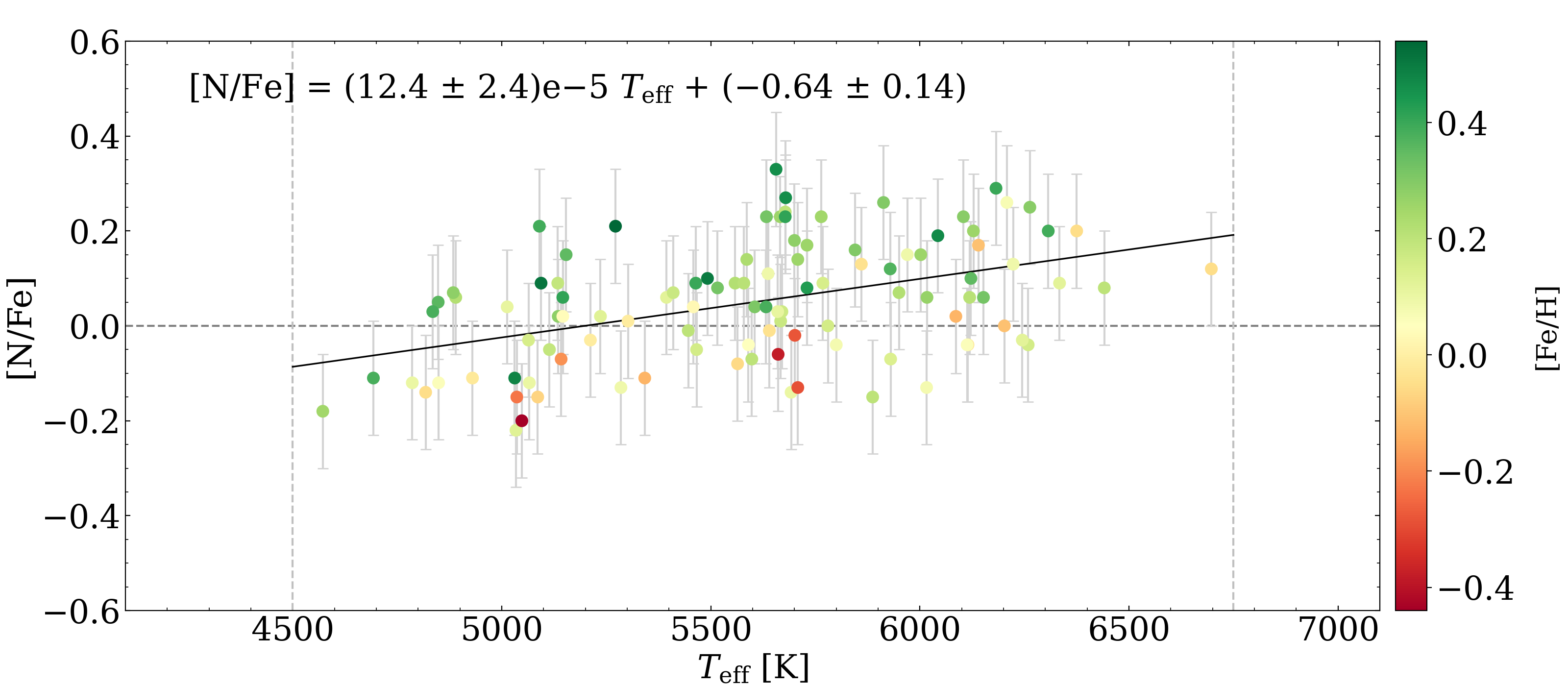}}
\end{minipage} \\
\begin{minipage}[t]{0.49\textwidth}
\centering
\resizebox{\hsize}{!}{\includegraphics{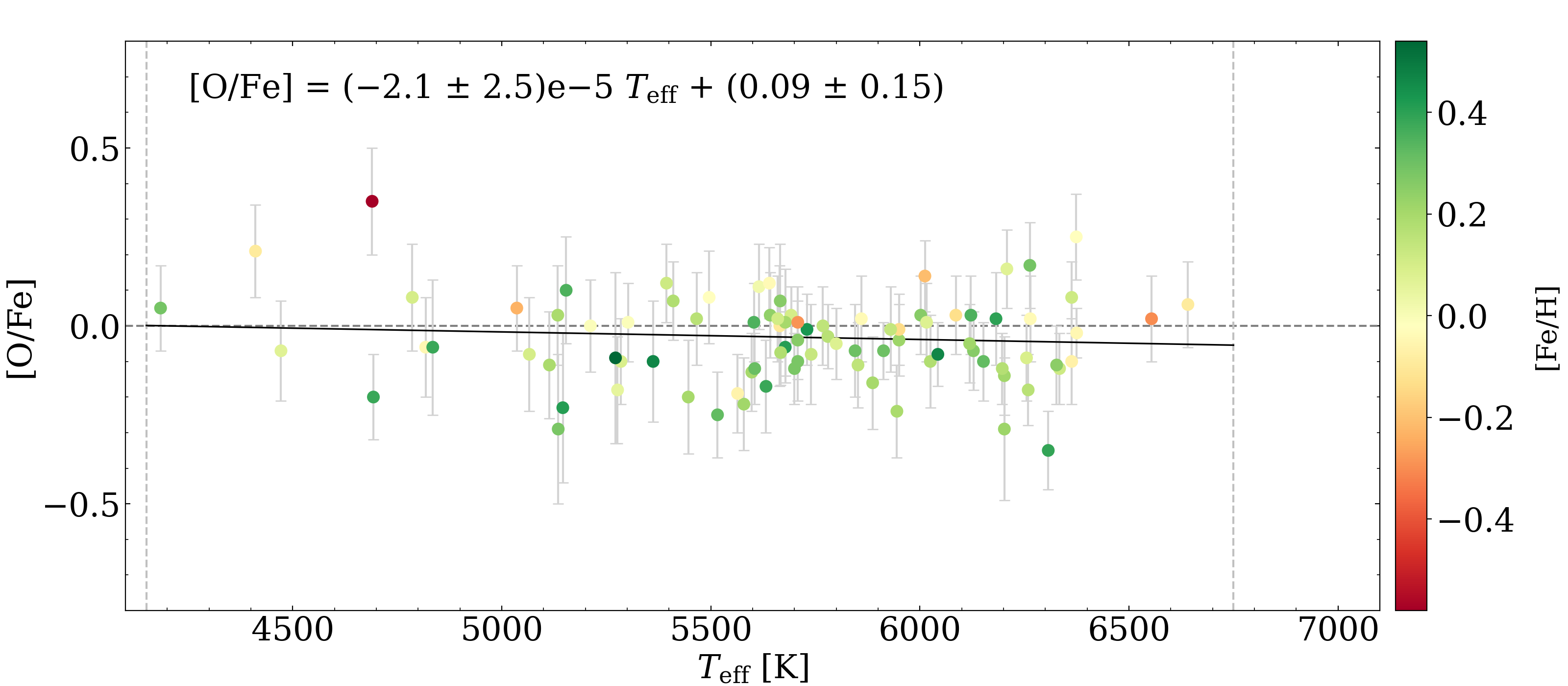}}
\end{minipage}
\caption{Abundance ratios as a function of the effective temperature colour coded according to the stellar metallicity. The white symbols indicate two stars that most likely belong to the thick-disc population. The black lines display linear regressions fitted to stars within a given \teff\ range (delimited by the vertical dashed lines). The corresponding equations are also shown.}
\label{figure:xfe_teff_feh}
\end{figure}

\begin{figure}
\centering
\resizebox{\hsize}{!}{\includegraphics{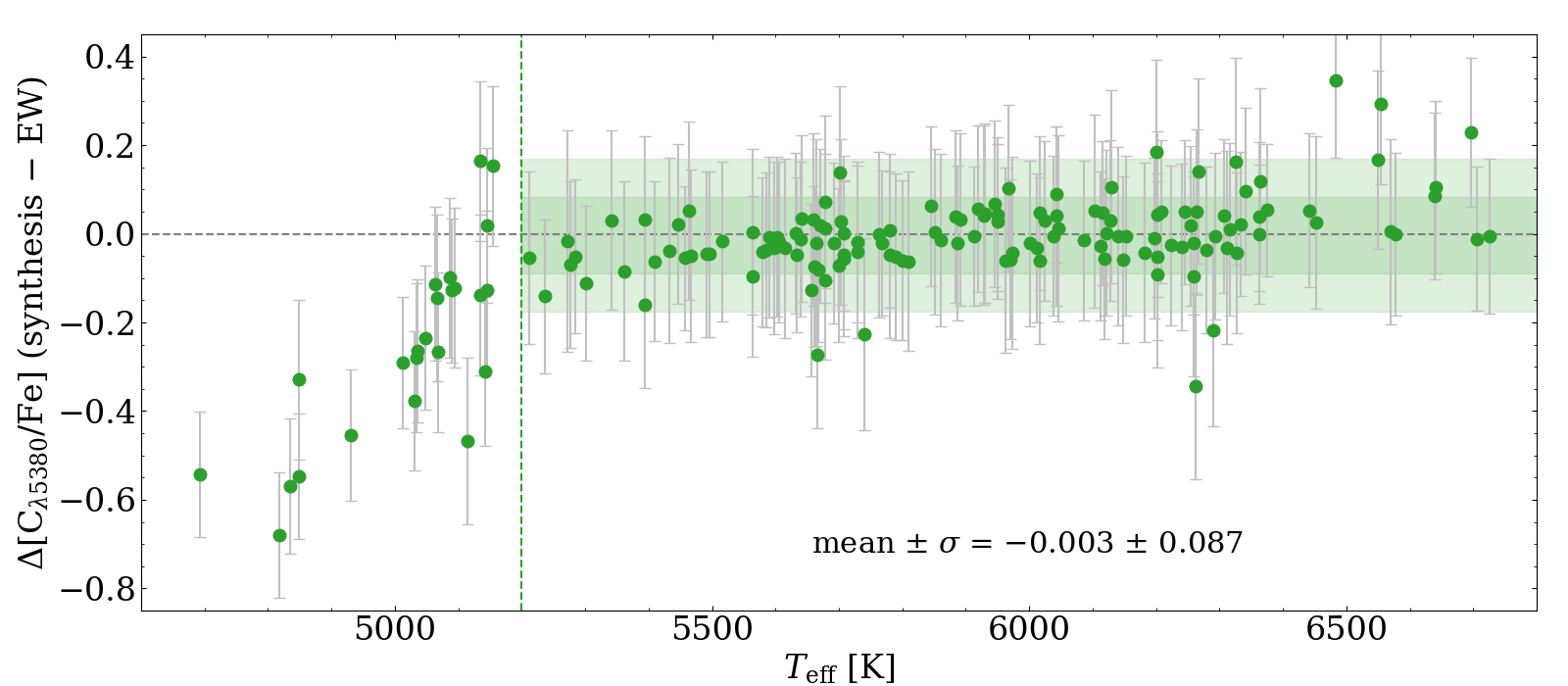}}
\caption{Abundance differences between [C/Fe] derived from spectral synthesis and from equivalent widths.}
\label{figure:delta_cfe5380_teff}
\end{figure}

The carbon abundances derived using the EW method applied to cool stars are not always reliable for the two \ion{C}{i} atomic lines under investigation \citep[see, e.g.,][]{DelgadoMenaetal2021,Biazzoetal2022}. The smaller the effective temperature, the weaker these lines. Therefore, it is more difficult to measure their EWs. Our results clearly show the same effect for the line at 5052.2~\AA\ in stars with \teff\ $\lesssim$ 5000~K, and for the line at 5380.3~\AA\ in stars with \teff\ $\lesssim$ 5200~K, as seen in Fig.~\ref{figure:xfe_teff_feh}. In the same figure, we see that the abundances derived for the line at 5380.3~\AA\ using the spectral synthesis method do not show the same systematic overabundance. The difference between the two methods is more clearly seen in Figure~\ref{figure:delta_cfe5380_teff}, which shows a high agreement for stars hotter than 5200~K and an increasing disagreement for cooler stars.

\begin{figure}
\centering
\begin{minipage}[t]{0.49\textwidth}
\centering
\resizebox{\hsize}{!}{\includegraphics{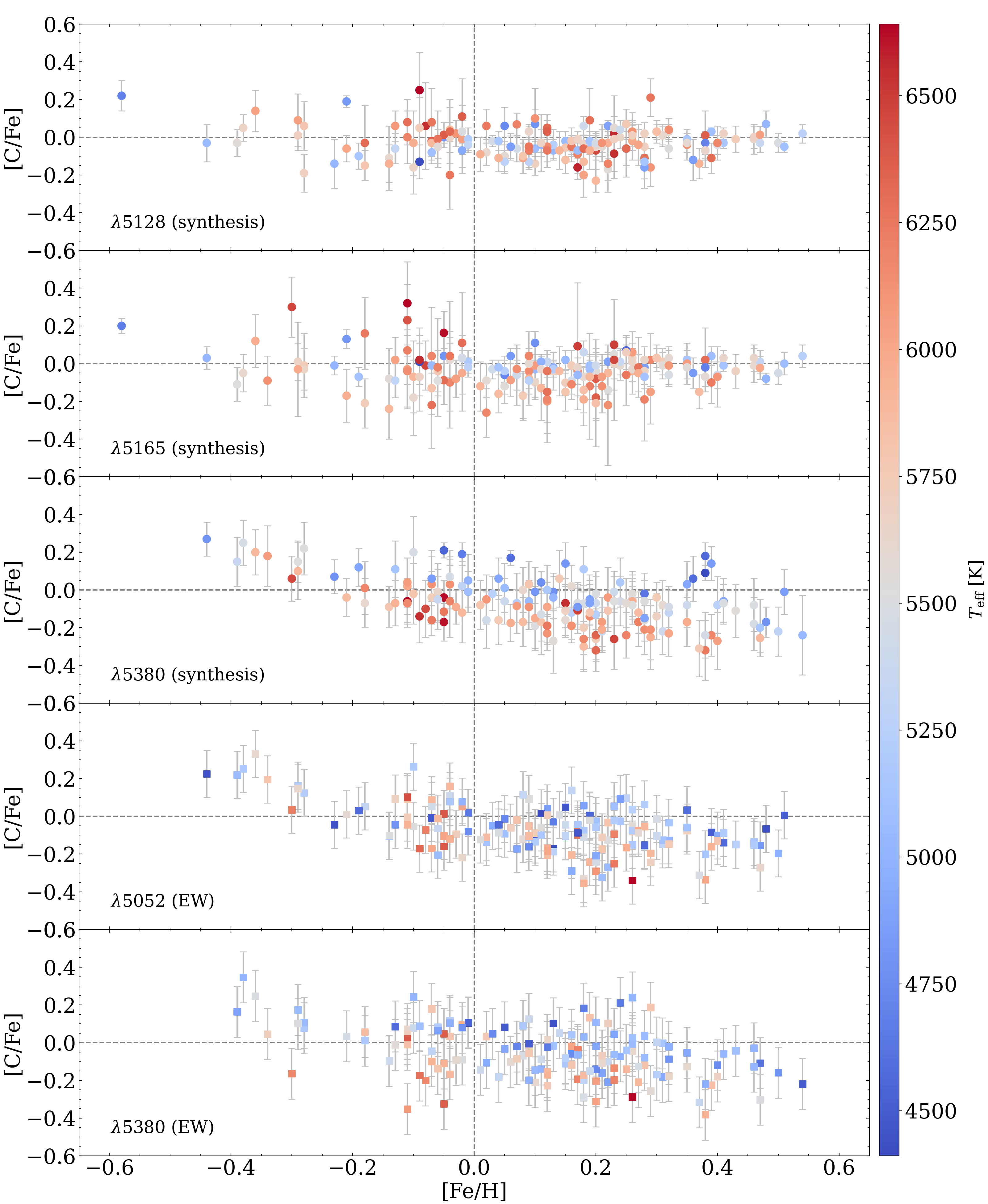}}
\end{minipage} \\
\begin{minipage}[t]{0.49\textwidth}
\centering
\resizebox{\hsize}{!}{\includegraphics{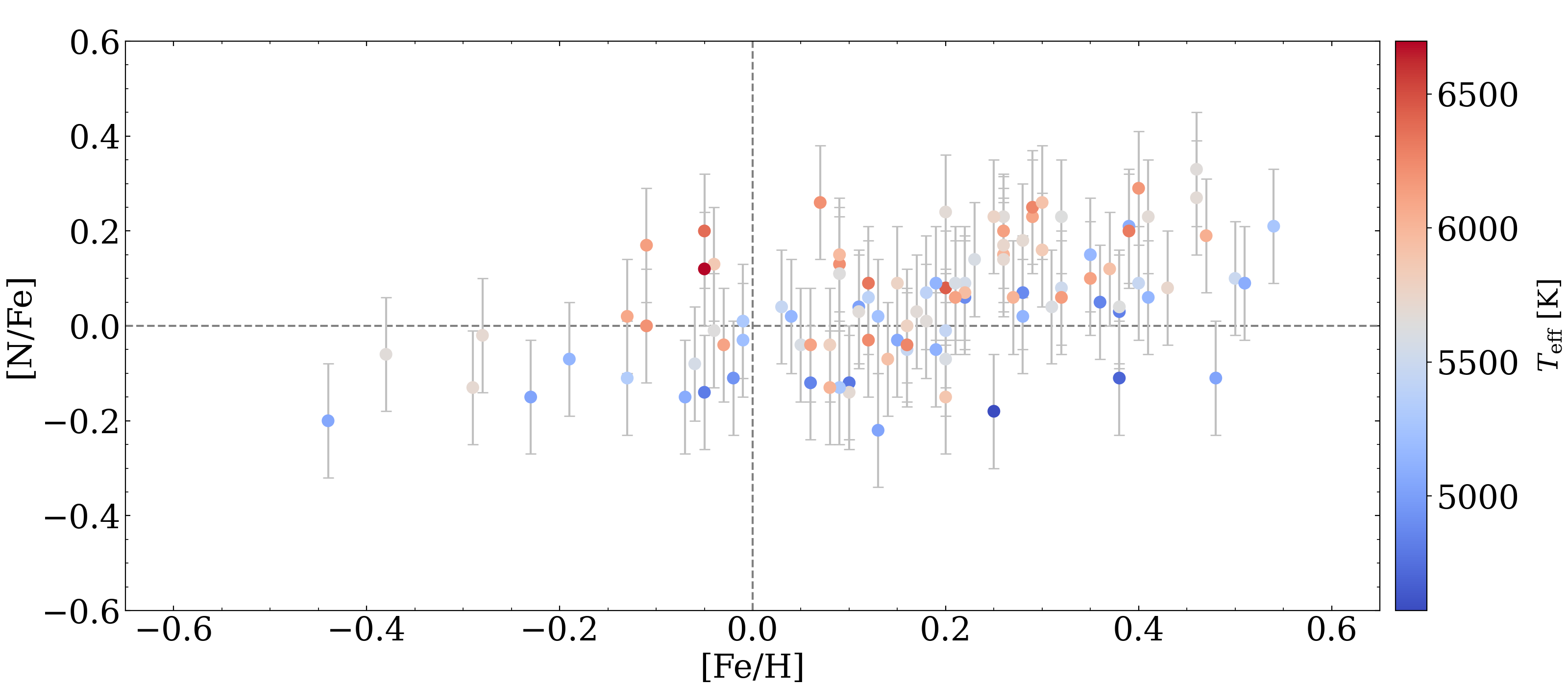}}
\end{minipage} \\
\begin{minipage}[t]{0.49\textwidth}
\centering
\resizebox{\hsize}{!}{\includegraphics{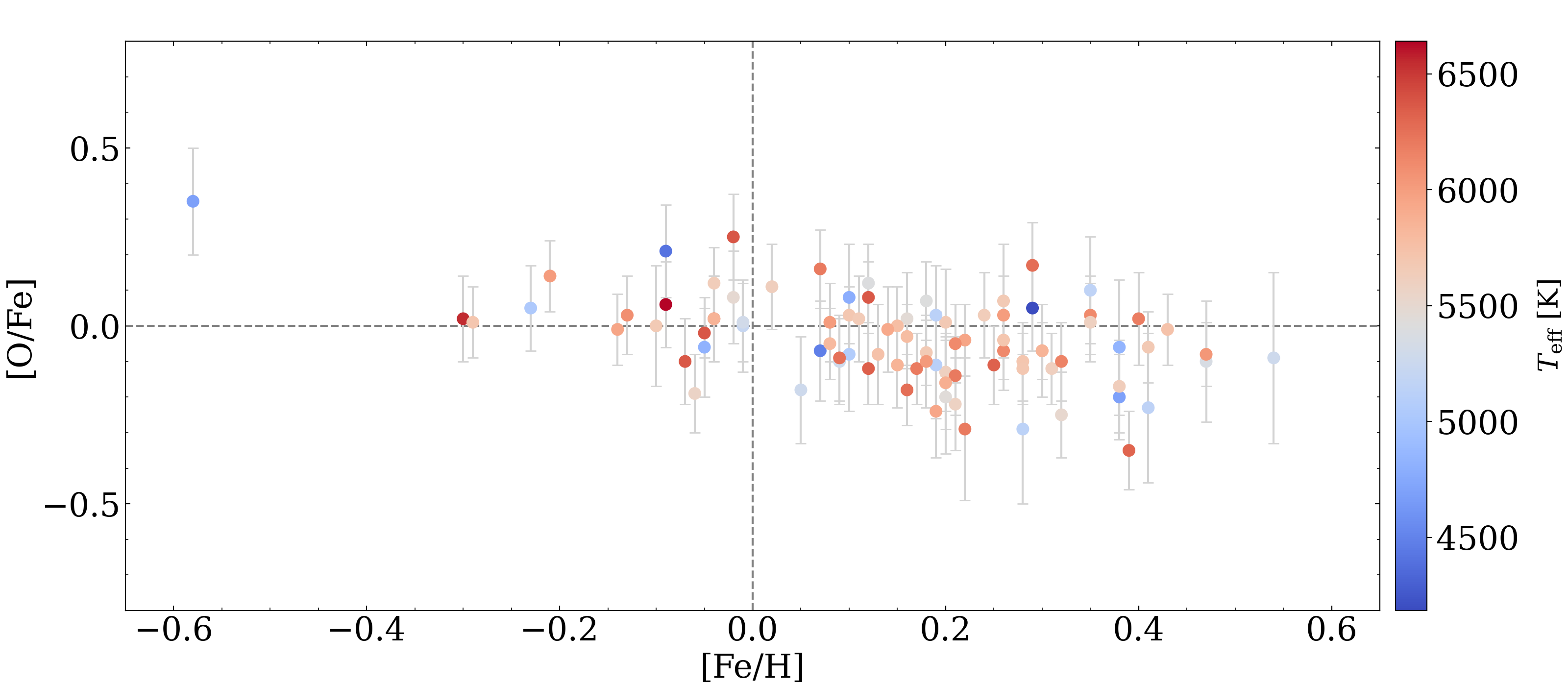}}
\end{minipage}
\caption{Abundance ratios as a function of the stellar metallicity colour coded according to the effective temperature. The panels show the stars in our sample as in Fig.~\ref{figure:cfe_feh_5} for carbon, Fig.~\ref{figure:nfe_feh} for nitrogen, and Fig.~\ref{figure:ofe_feh} for oxygen but before correcting the abundances from the trends with \teff.}
\label{figure:xfe_feh_teff}
\end{figure}

Figure~\ref{figure:xfe_teff_feh} also shows that each of the indicators provide abundance ratios that have some dependence on the effective temperature. For carbon, the slope is negative for all the five indicators, though within 2$\sigma$ for 5128 and 5052.2~\AA. A significant trend is seen for nitrogen as well, for which the [N/Fe] ratios increase with increasing \teff. For oxygen, the slope has no statistical significance. A consequence of these trends is that the [X/Fe] abundance ratios as a function of [Fe/H] show different distributions for cool and hot stars, as observed in Fig.~\ref{figure:xfe_feh_teff}, increasing the scatter of the stars in those diagrams. In addition, since the derived carbon abundances are used to derive the nitrogen composition from CN molecules, any trend in [C/Fe] is propagated to the [N/Fe] ratios. The presence of these trends, therefore, has an impact on the conclusions that we may draw concerning, for instance, the Galactic chemical evolution or the relations between stellar abundances and planetary parameters. These are the reasons why we prefer to correct the abundances from the observed slopes, removing any dependence on the effective temperature. As described in Sect.~\ref{section:abundances}, this procedure was applied to the carbon and nitrogen indicators, but not to oxygen considering the absence of a significant trend.

\section{C-enhanced and other particular cases}
\label{appendix:peculiar_stars}

\begin{figure}
\centering
\begin{minipage}[t]{0.49\textwidth}
\centering
\resizebox{\hsize}{!}{\includegraphics{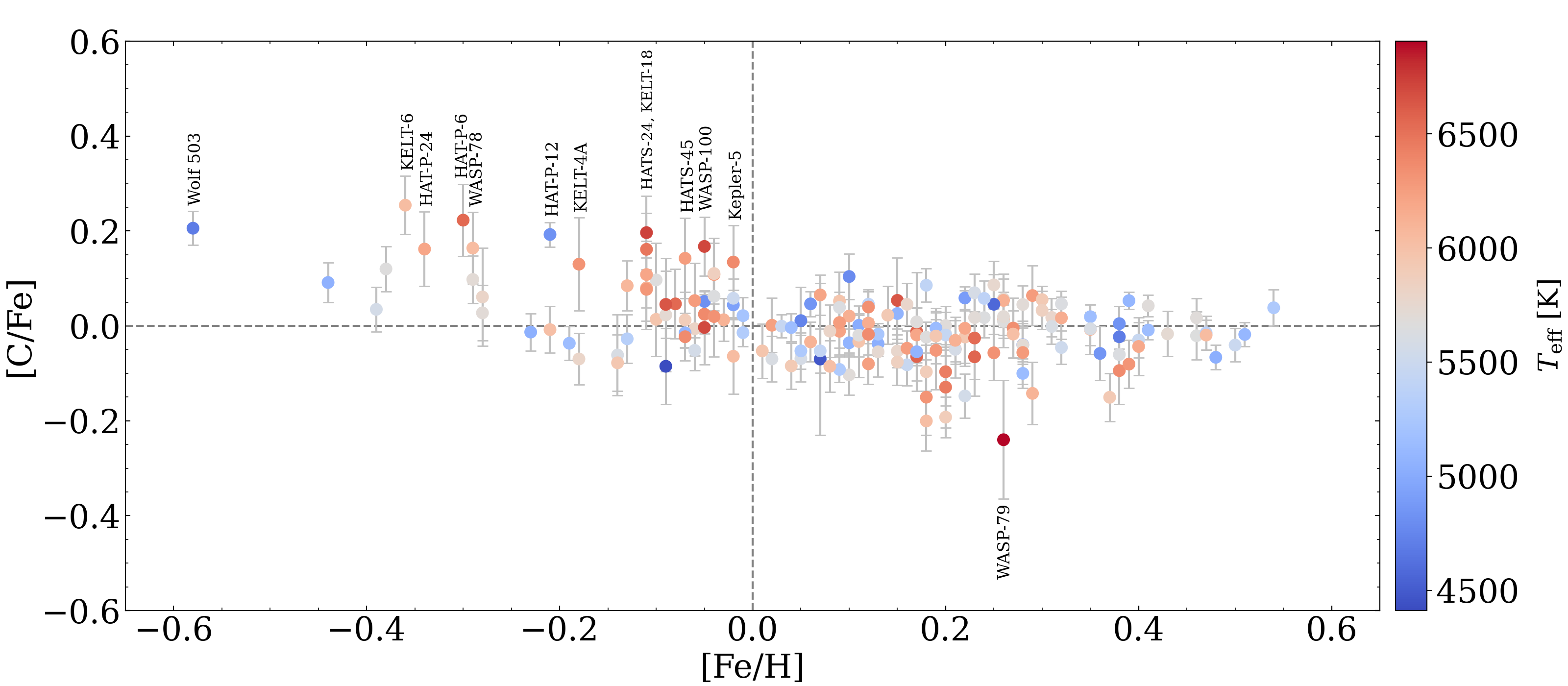}}
\end{minipage} \\
\begin{minipage}[t]{0.49\textwidth}
\centering
\resizebox{\hsize}{!}{\includegraphics{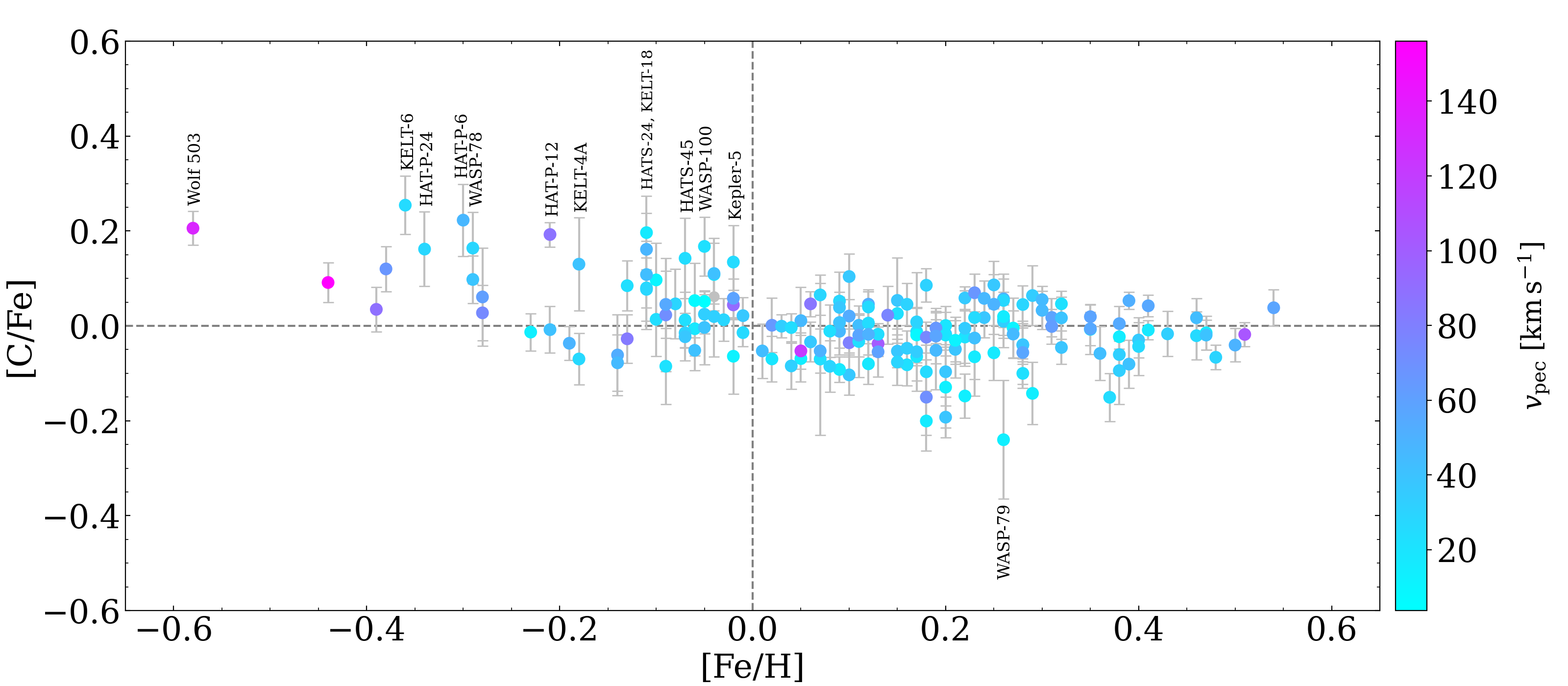}}
\end{minipage} \\
\begin{minipage}[t]{0.49\textwidth}
\centering
\resizebox{\hsize}{!}{\includegraphics{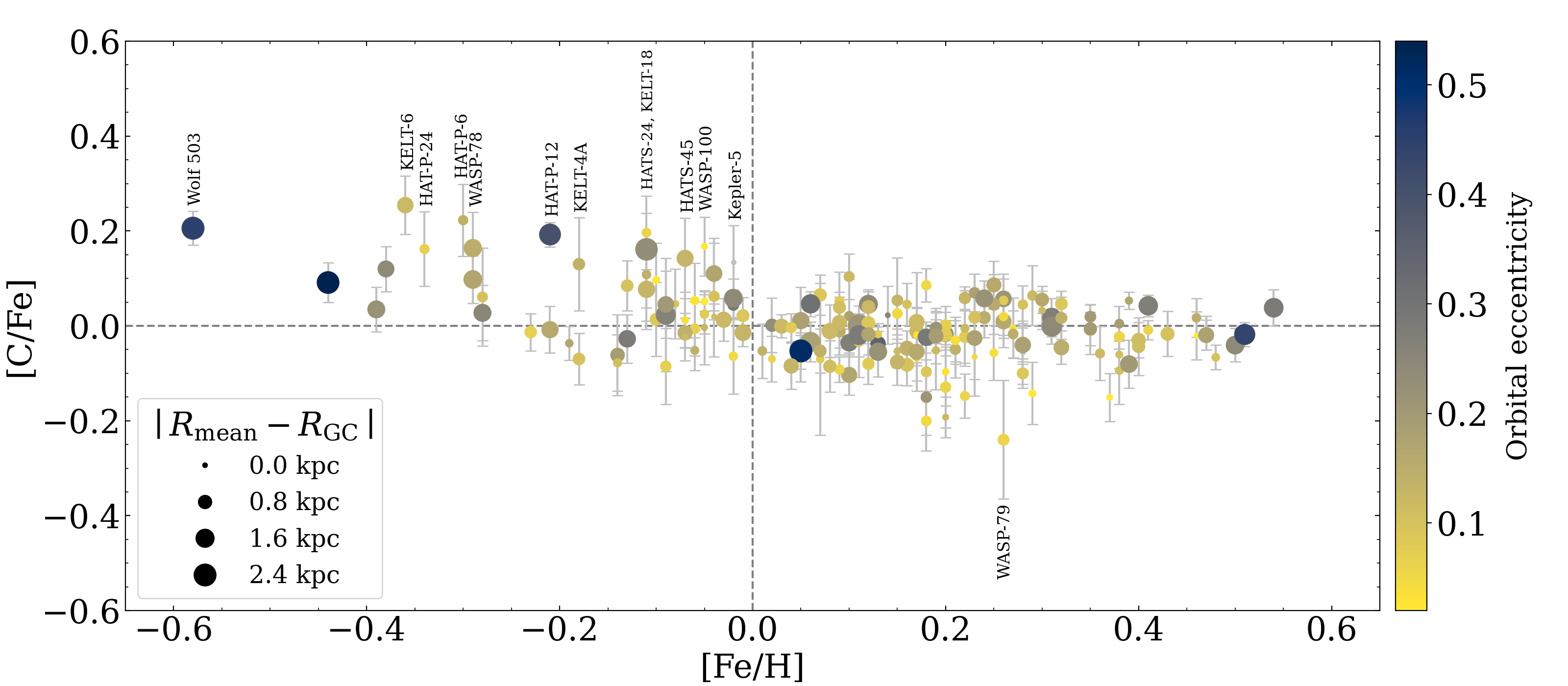}}
\end{minipage}
\caption{Mean [C/Fe] abundance ratios as a function of the stellar metallicity. Same as in Fig.~\ref{figure:cfe_feh} but showing our stars colour-coded according to their effective temperature (top panel), peculiar space velocity (middle panel), and orbital eccentricity (bottom panel) from \citet{Magrinietal2022}. In the bottom panel, the symbol sizes are proportional to the difference between the mean distance of the stars in their Galactic orbit ($R_{\rm mean}$) and their current galactocentric distance ($R_{\rm GC}$).} 
\label{figure:cfe_feh_teff_vpec}
\end{figure}

 We see in Fig.~\ref{figure:cfe_feh} that some stars in our sample are situated in a particular position (slightly over or underabundant in carbon) compared with the typical distribution of thin-disc members. Here we highlight a few of theses cases and we comment on possible explanations for the atypical abundances. To help with the discussion, Fig.~\ref{figure:cfe_feh_teff_vpec} shows the same stars with symbols colour-coded according to some stellar parameters taken from \citet{Magrinietal2022}. In the top panel, the stars are colour-coded based on their effective temperature. In the middle panel, the colour code follows the so-called peculiar space velocity ($v_{\rm pec}$), calculated using the $U$, $V$, and $W$ radial velocities. In the bottom panel, the symbols are colour-coded according to the stellar orbital eccentricity and have sizes proportional to the difference between the mean distance of each star in its Galactic orbit and its current galactocentric distance.

The star \object{Wolf\,503} ([Fe/H] = $-$0.58 $\pm$ 0.17 and [C/Fe] = 0.21 $\pm$ 0.04), the most metal-poor in our sample, is very probably a member of thick disc according to the population membership analysis performed by \citet{Magrinietal2022}. The same classification is also valid for \object{HAT-P-12} ([Fe/H] = $-$0.21 $\pm$ 0.18 and [C/Fe] = 0.19 $\pm$ 0.03). Moreover, both these stars are situated among thick-disc members according to the classification performed by \citet{DelgadoMenaetal2021} using an independent approach (chemical instead of dynamical) to distinguish the different populations. Last but not least, these two stars have a large orbital eccentricity and might have migrated from a different region of the Milky Way. Close to \object{HAT-P-12} we see the star \object{KELT-4A} ([Fe/H] = $-$0.18 $\pm$ 0.12 and [C/Fe] = 0.13 $\pm$ 0.10). It seems to be C-richer than the average of surrounding stars, but the uncertainty is relatively large, which might be due to its relatively high effective temperature (\teff\ = 6316~K) that makes the spectral features weaker (specially the molecular bands) and more difficult to measure. Though this star is probably a member of the thin disc, it is also situated close to the distribution of the $\alpha$-enhanced stars discussed in \citet{DelgadoMenaetal2021}, for which an overabundance of carbon is derived.

Another group of stars with [Fe/H] and [C/Fe] values that place them among the $\alpha$-enhanced stars shown in \citet{DelgadoMenaetal2021} are the following: \object{HATS-24}, \object{KELT-18}, \object{HATS-45}, \object{WASP-100}, and \object{Kepler-5}. Though they most likely belong to the thin disc population, they are all C-enhanced compared with other stars in our sample in the same metallicity range. On the other hand, they are relatively hot (\teff\ $>$ 6200~K), which may also be affecting our abundance measurements. A particular case of hot star is \object{WASP-79}, the hottest in our sample (\teff\ = 6906~K). This quite high value for the effective temperature may explain the large error bar and its peculiar position in Figure~\ref{figure:cfe_feh}. It is also the star for which we obtain the smallest value for the carbon abundance ([C/Fe] = $-$0.24 $\pm$ 0.13). Another explanation for this relatively low abundance value could be the age of the system. \object{WASP-79} is situated in a position that agrees with the evolutionary stage of young stars according to the theoretical predictions shown in \citet{Romano2022}.

The star \object{KELT-6} ([Fe/H] = $-$0.36 $\pm$ 0.10 and [C/Fe] = 0.25 $\pm$ 0.06) has a high probability of belonging to the thin disc according to the classifications described above. In spite of that, it is situated significantly above the main distribution typical of thin-disc members in this range of metallicity. An explanation for this peculiarity could be related to the presence of planets. This is a system composed by two Jupiter-like planets \citep{Damassoetal2015}. According to the results from \citet{DelgadoMenaetal2021}, planet-host stars in the metal-poor range ([Fe/H] $\lesssim$ $-$0.2) seem to be overabundant in carbon compared to stars for which no planet has been detected. In that paper, the authors remark that this overabundance is typical of thick-disc stars. However, even for a selection of stars in the thick-disc population, they also find a difference in carbon abundances between single stars and planet hosts, specially for stars hosting low-mass planets. Other three stars in our sample are in a similar situation: \object{HAT-P-24}, \object{HAT-P-6}, and \object{WASP-78}. They are all probably members of the thin disc and host a Jupiter-like planet. The number of stars in this range of metallicity (from $\sim$$-0.3$ to $\sim$$-0.35$) is small, but their systematic overabundance in carbon (likely related to the presence of planets) supports the results obtained by \citet{DelgadoMenaetal2021}.

\section{X1/X2 elemental ratios}
\label{appendix:x1x2_elem_ratios}

The absolute elemental ratios X1/X2, differently from the [X1/X2] ratios normally used, which are defined with respect to the Sun, are defined as:
\begin{equation}
{\rm X1/X2} = 10^{\log{\epsilon({\rm X1})}} / 10^{\log{\epsilon({\rm X2})}},
\end{equation}
where $\log{\epsilon({\rm X1})}$ and $\log{\epsilon({\rm X2})}$ are absolute abundances. For the Sun, by adopting the absolute abundances from \citet{Asplundetal2009}, we have C/O$_\odot$ = 0.550, N/O$_\odot$ = 0.138, C/N$_\odot$ = 3.981.

\begin{figure*}
\centering
\begin{minipage}[t]{0.49\textwidth}
\centering
\resizebox{\hsize}{!}{\includegraphics{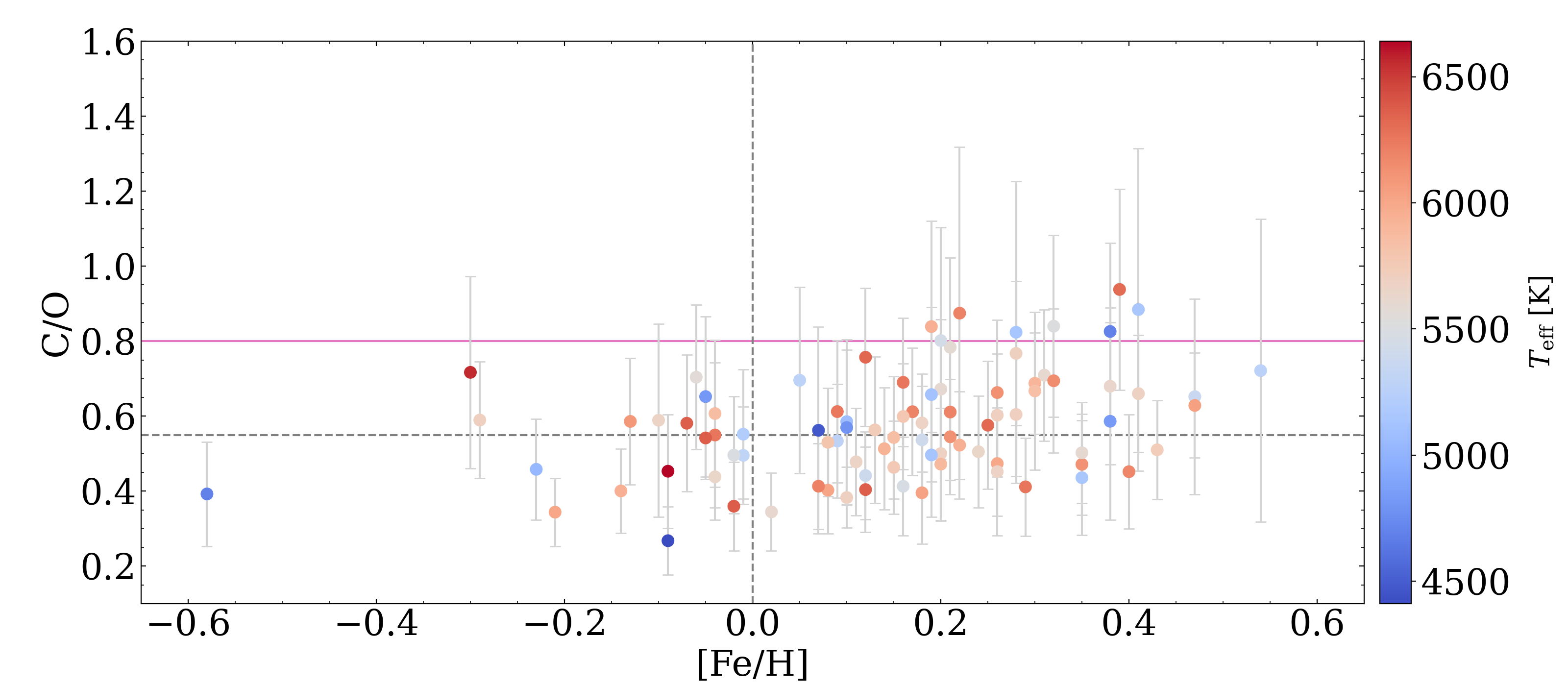}}
\end{minipage}
\begin{minipage}[t]{0.49\textwidth}
\centering
\resizebox{\hsize}{!}{\includegraphics{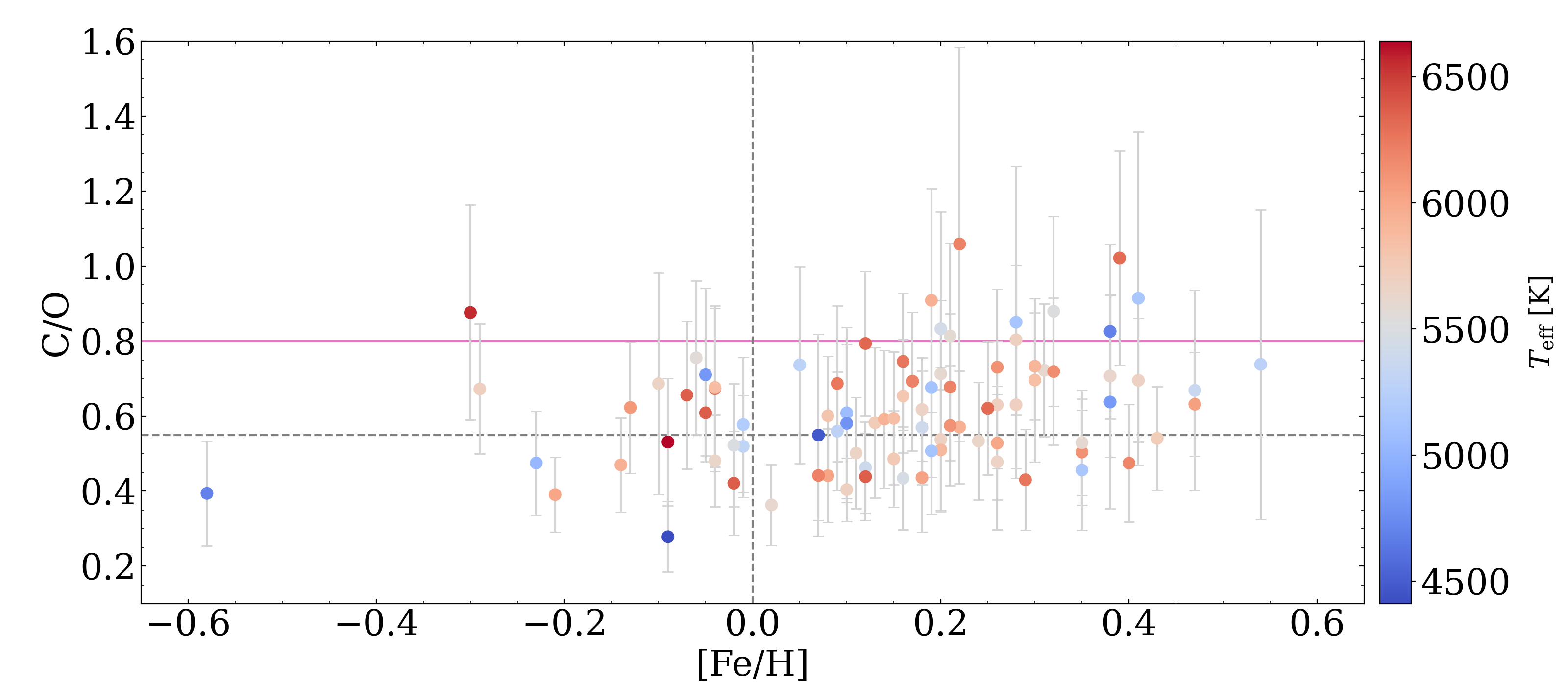}}
\end{minipage} \\
\begin{minipage}[t]{0.49\textwidth}
\centering
\resizebox{\hsize}{!}{\includegraphics{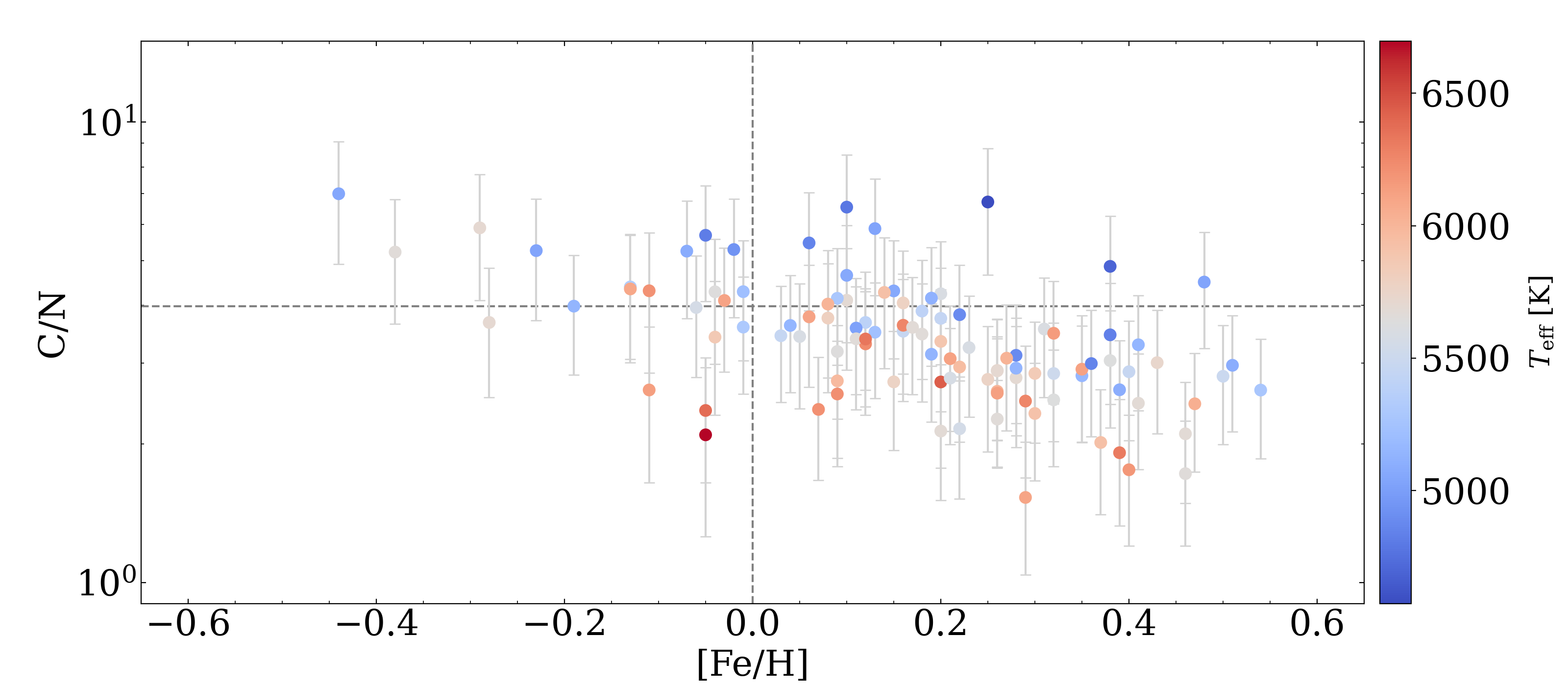}}
\end{minipage}
\begin{minipage}[t]{0.49\textwidth}
\centering
\resizebox{\hsize}{!}{\includegraphics{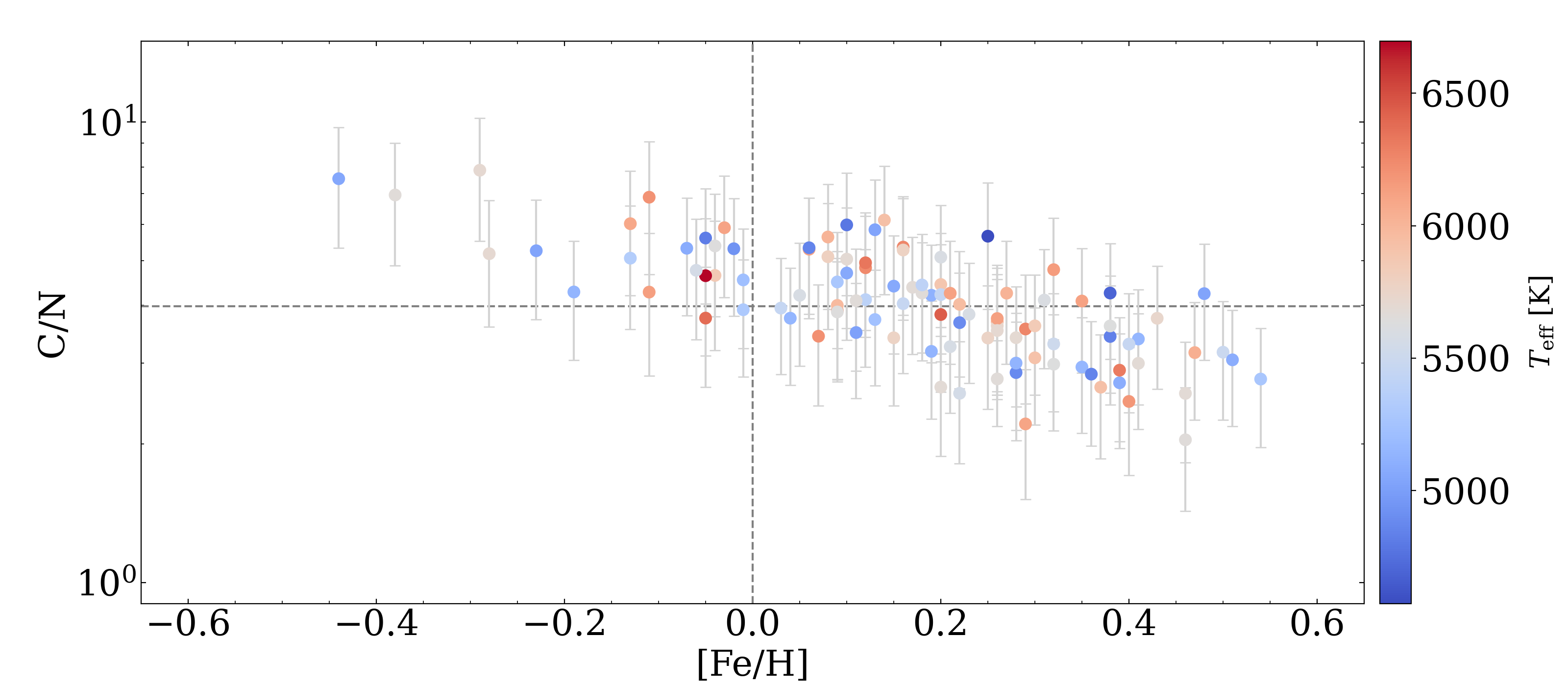}}
\end{minipage} \\
\begin{minipage}[t]{0.49\textwidth}
\centering
\resizebox{\hsize}{!}{\includegraphics{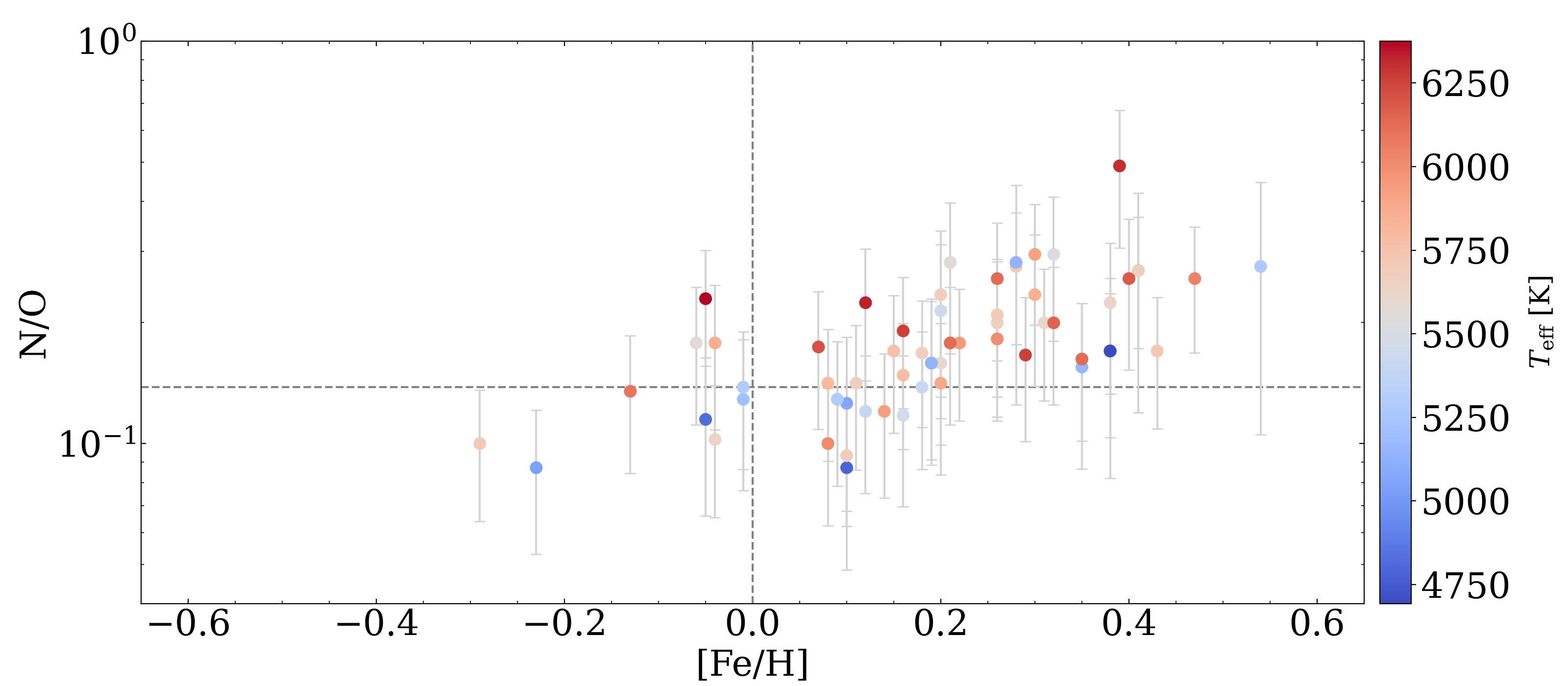}}
\end{minipage}
\begin{minipage}[t]{0.49\textwidth}
\centering
\resizebox{\hsize}{!}{\includegraphics{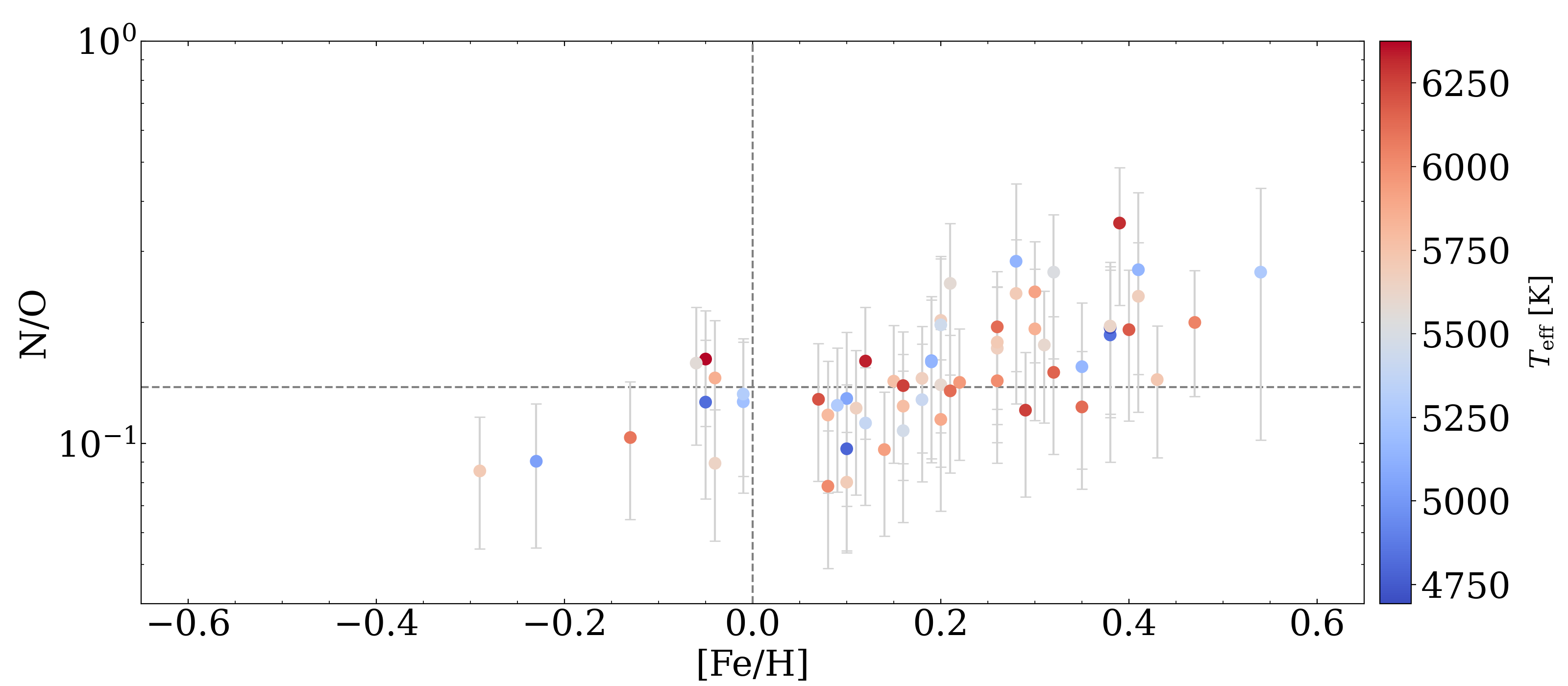}}
\end{minipage}
\caption{Abundance ratios as a function of the stellar metallicity colour coded according to the stellar effective temperature. The panels show the X1/X2 elemental ratios calculated before (left) and after (right) correcting the abundances from the trends with \teff\ (see discussion in the text). The black dashed lines indicate the solar values and the magenta solid line represents the lower limit of C/O = 0.8 for carbon-rich stars.}
\label{figure:x1x2_elem_ratios_feh}
\end{figure*}

\begin{figure*}
\centering
\begin{minipage}[t]{0.49\textwidth}
\centering
\resizebox{\hsize}{!}{\includegraphics{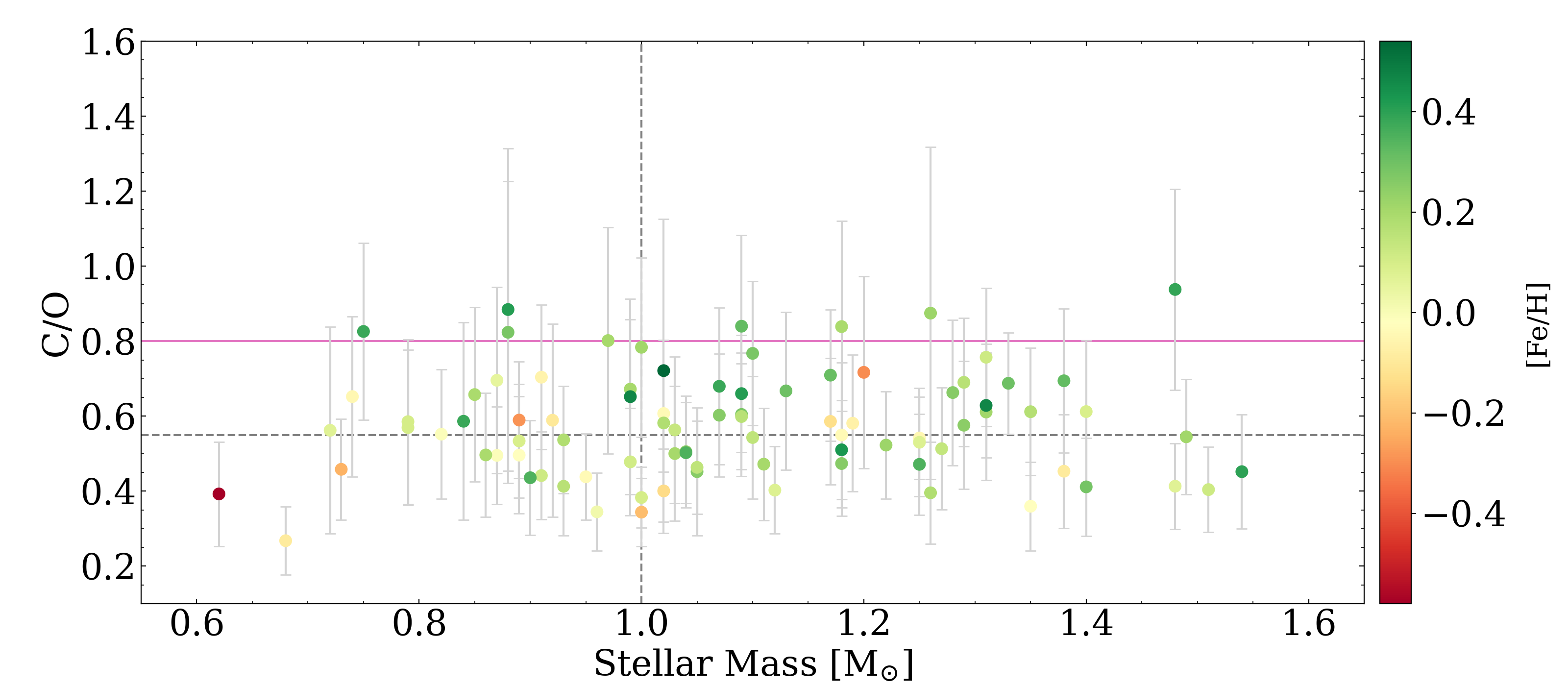}}
\end{minipage}
\begin{minipage}[t]{0.49\textwidth}
\centering
\resizebox{\hsize}{!}{\includegraphics{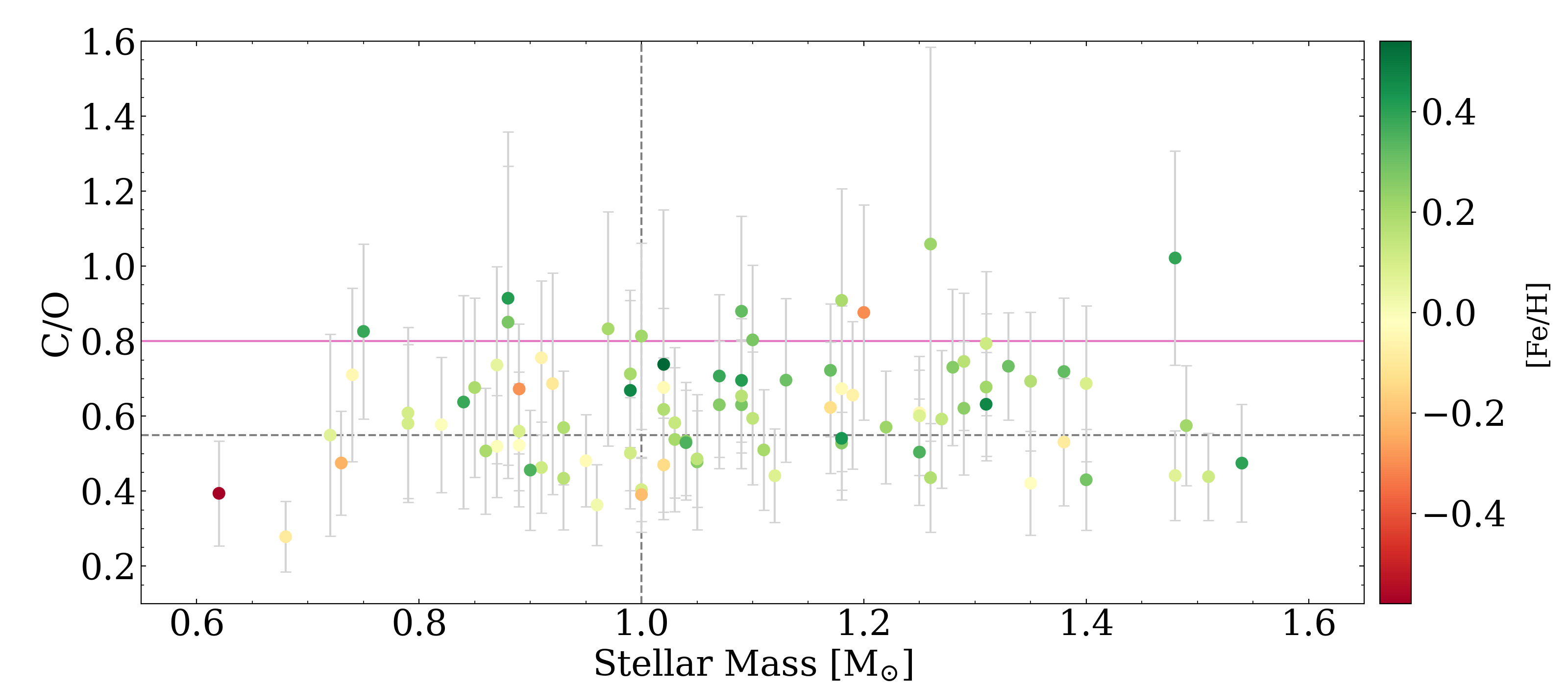}}
\end{minipage} \\
\begin{minipage}[t]{0.49\textwidth}
\centering
\resizebox{\hsize}{!}{\includegraphics{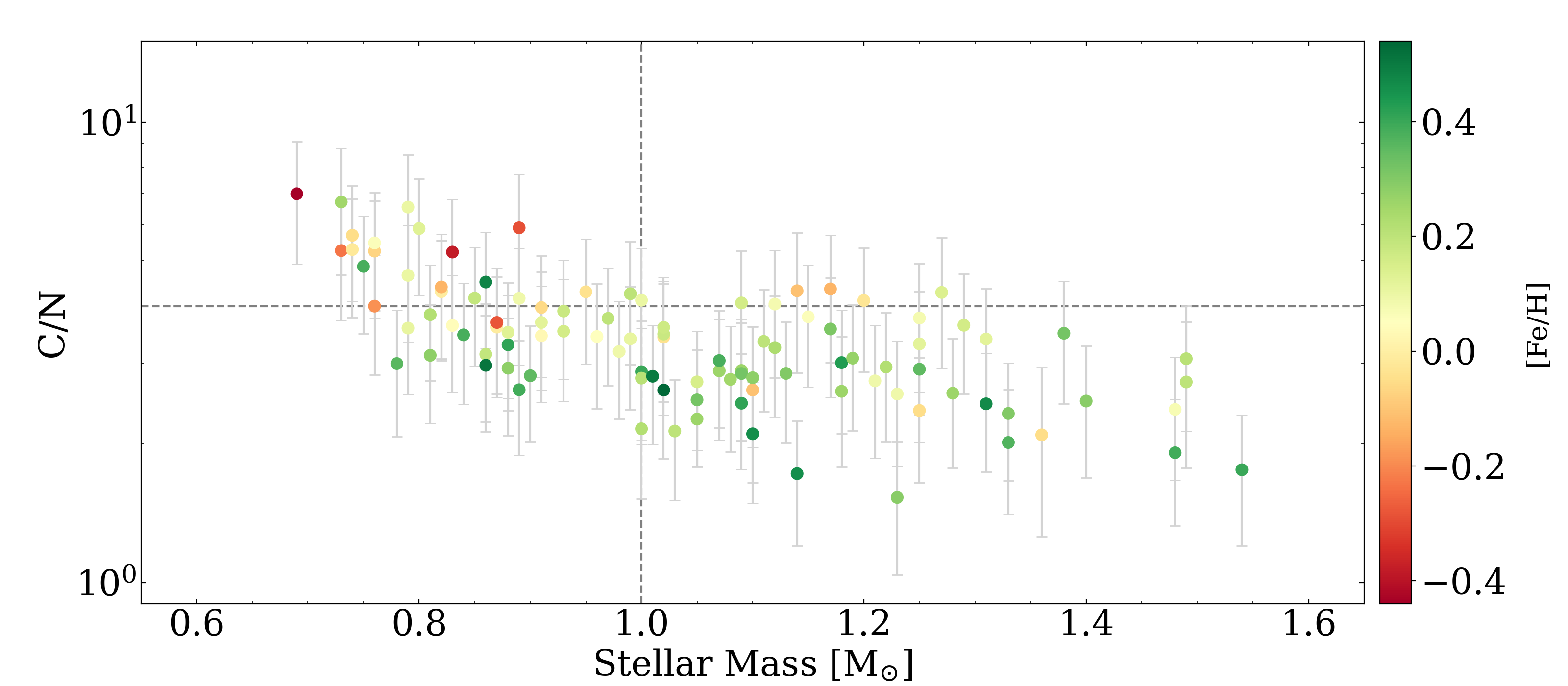}}
\end{minipage}
\begin{minipage}[t]{0.49\textwidth}
\centering
\resizebox{\hsize}{!}{\includegraphics{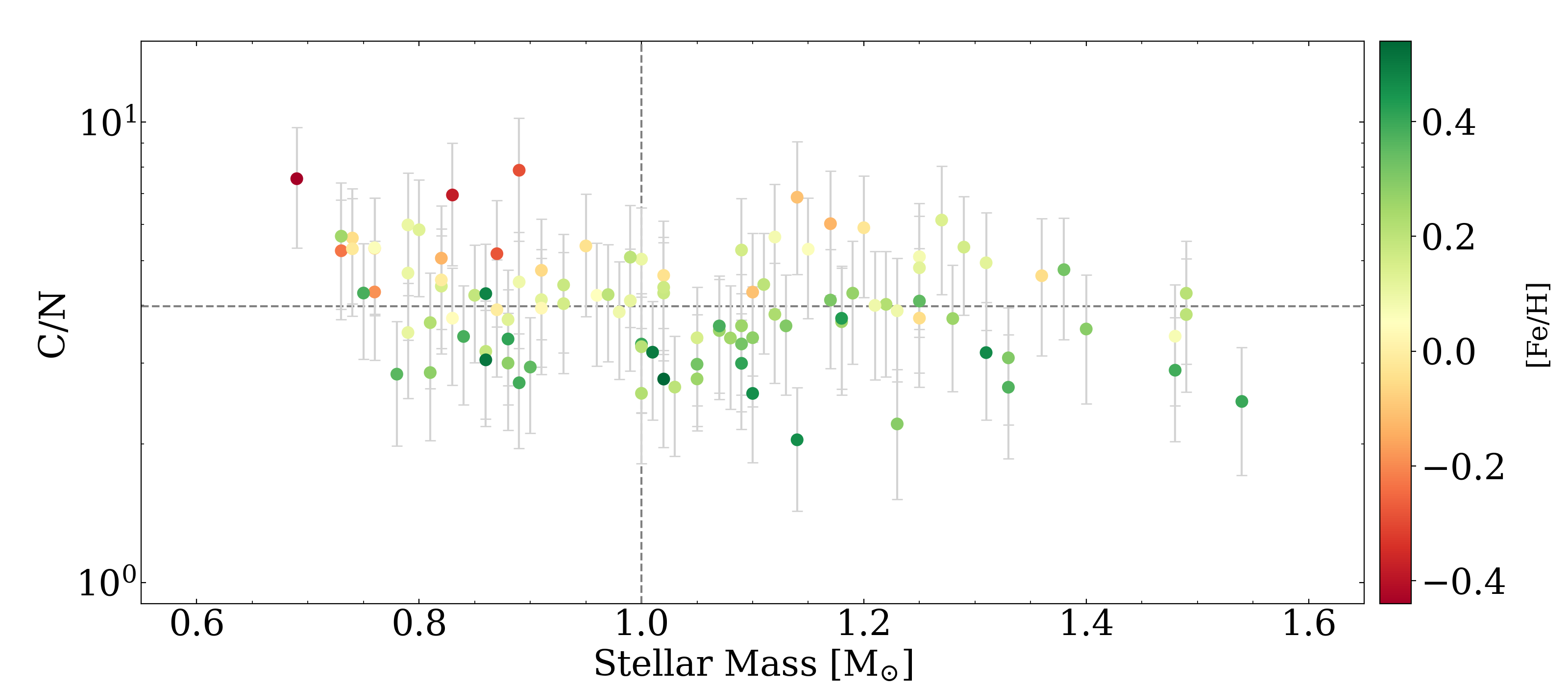}}
\end{minipage} \\
\begin{minipage}[t]{0.49\textwidth}
\centering
\resizebox{\hsize}{!}{\includegraphics{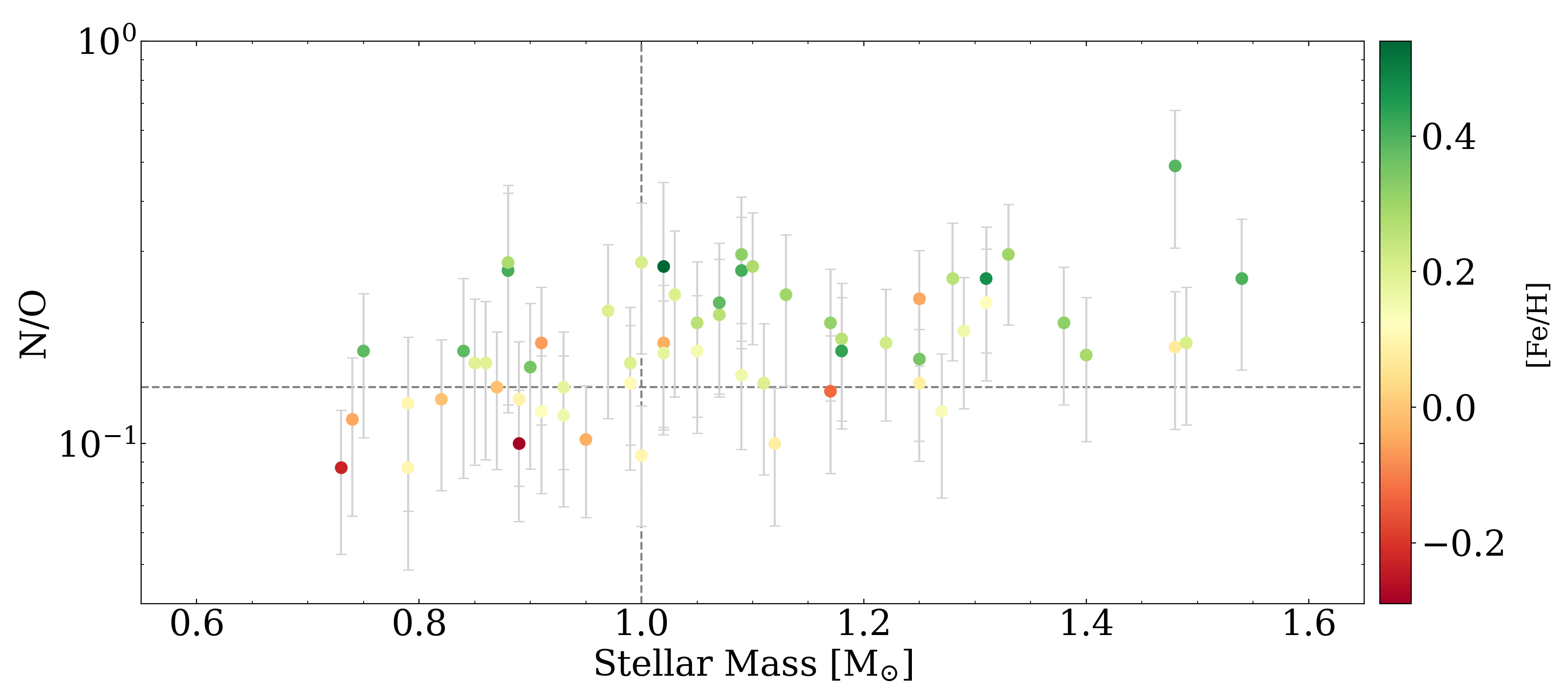}}
\end{minipage}
\begin{minipage}[t]{0.49\textwidth}
\centering
\resizebox{\hsize}{!}{\includegraphics{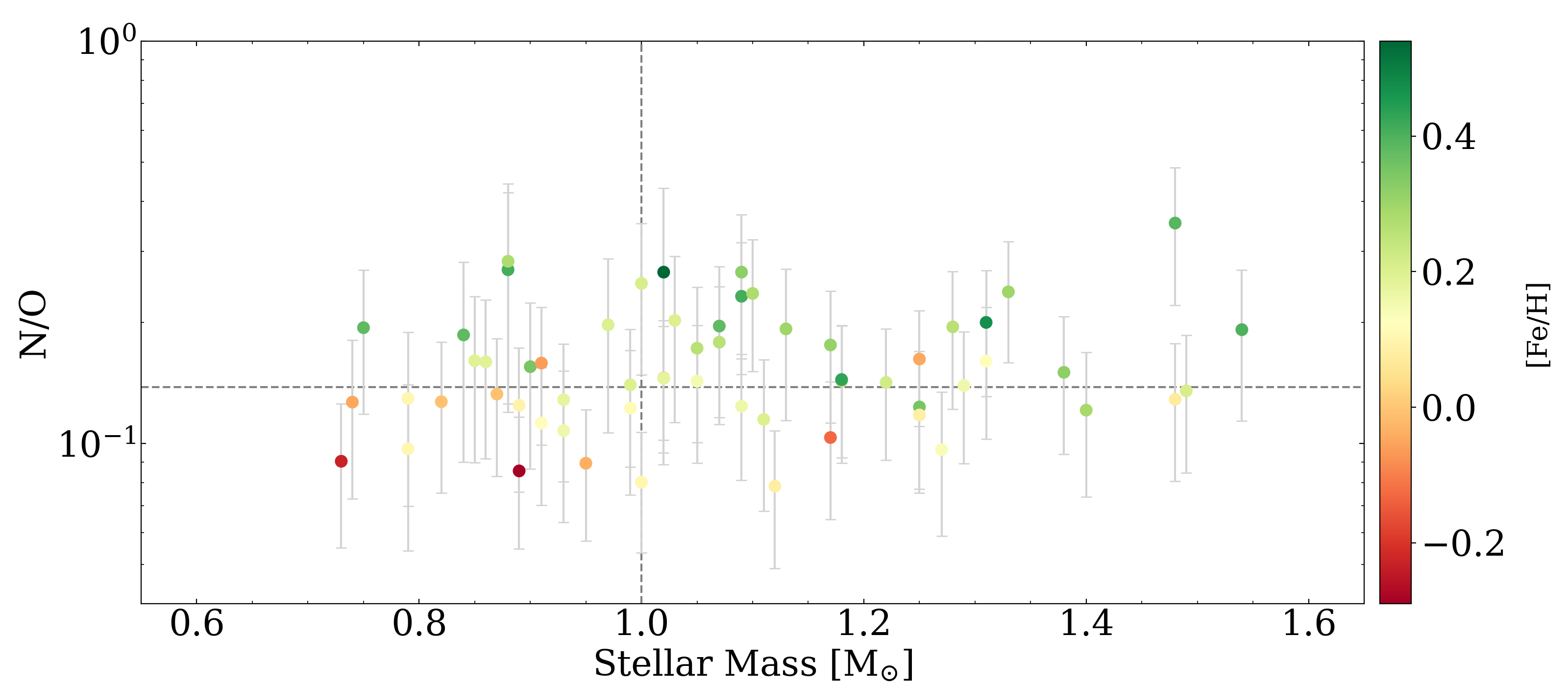}}
\end{minipage}
\caption{Same as in Fig.~\ref{figure:x1x2_elem_ratios_feh} but showing the X1/X2 ratios as a function of the stellar mass colour coded according to the stellar metallicity.}
\label{figure:x1x2_elem_ratios_stellar_mass}
\end{figure*}

\begin{figure*}
\centering
\begin{minipage}[t]{0.49\textwidth}
\centering
\resizebox{\hsize}{!}{\includegraphics{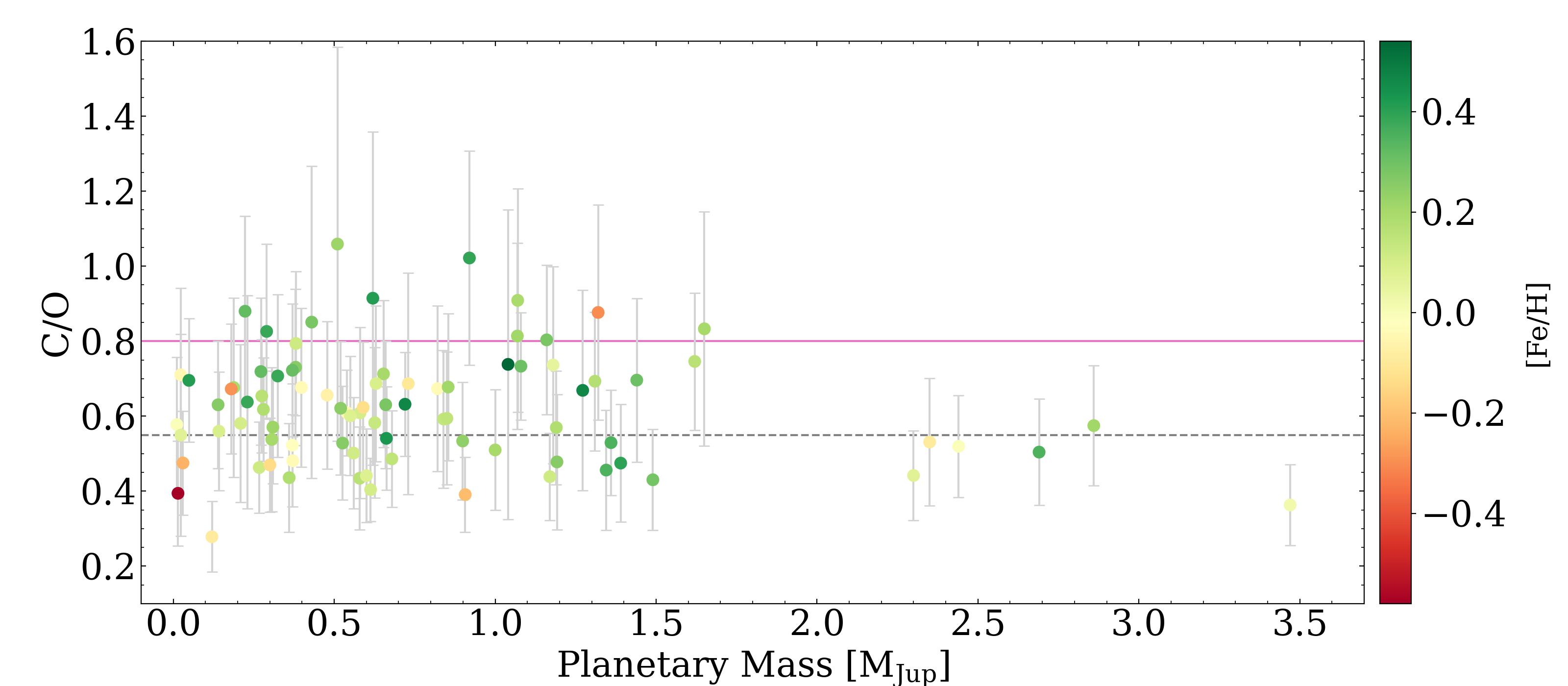}}
\end{minipage}
\begin{minipage}[t]{0.49\textwidth}
\centering
\resizebox{\hsize}{!}{\includegraphics{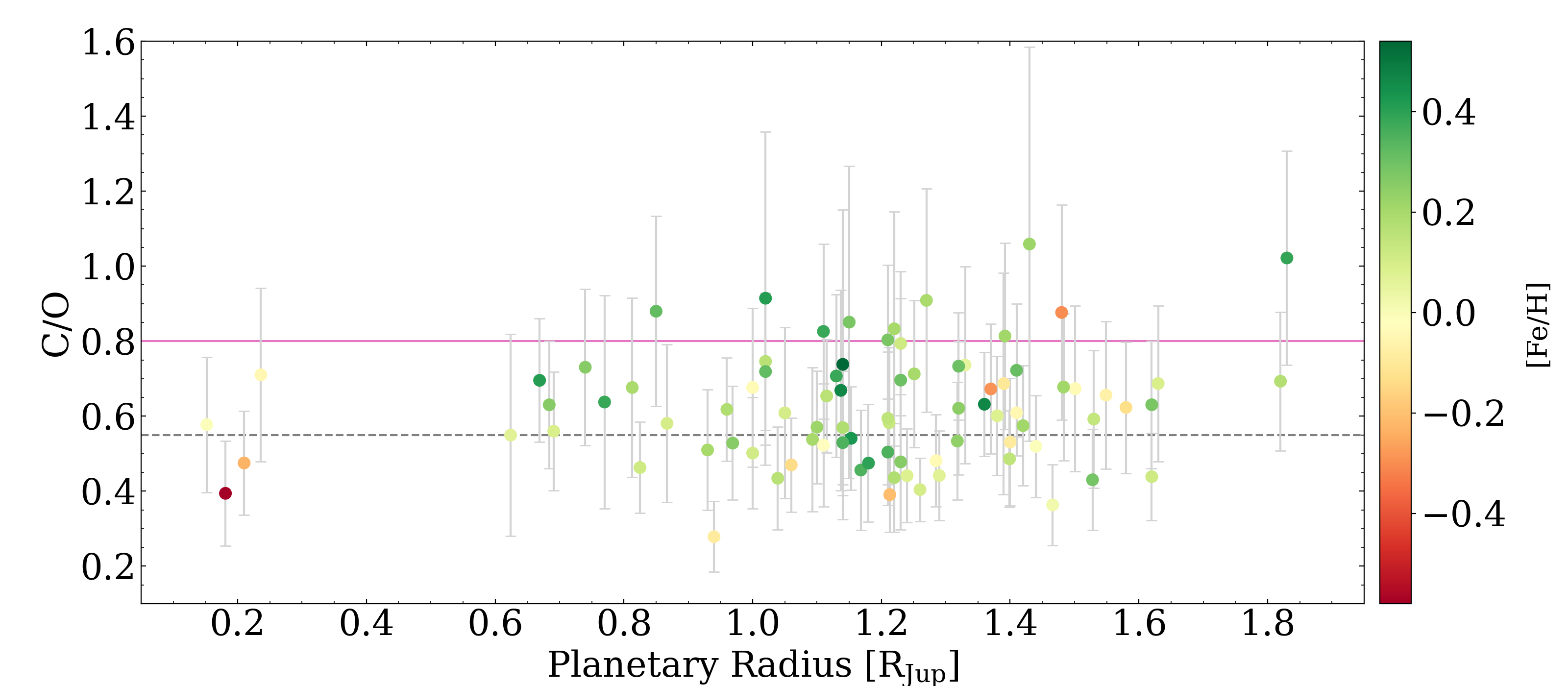}}
\end{minipage} \\
\begin{minipage}[t]{0.49\textwidth}
\centering
\resizebox{\hsize}{!}{\includegraphics{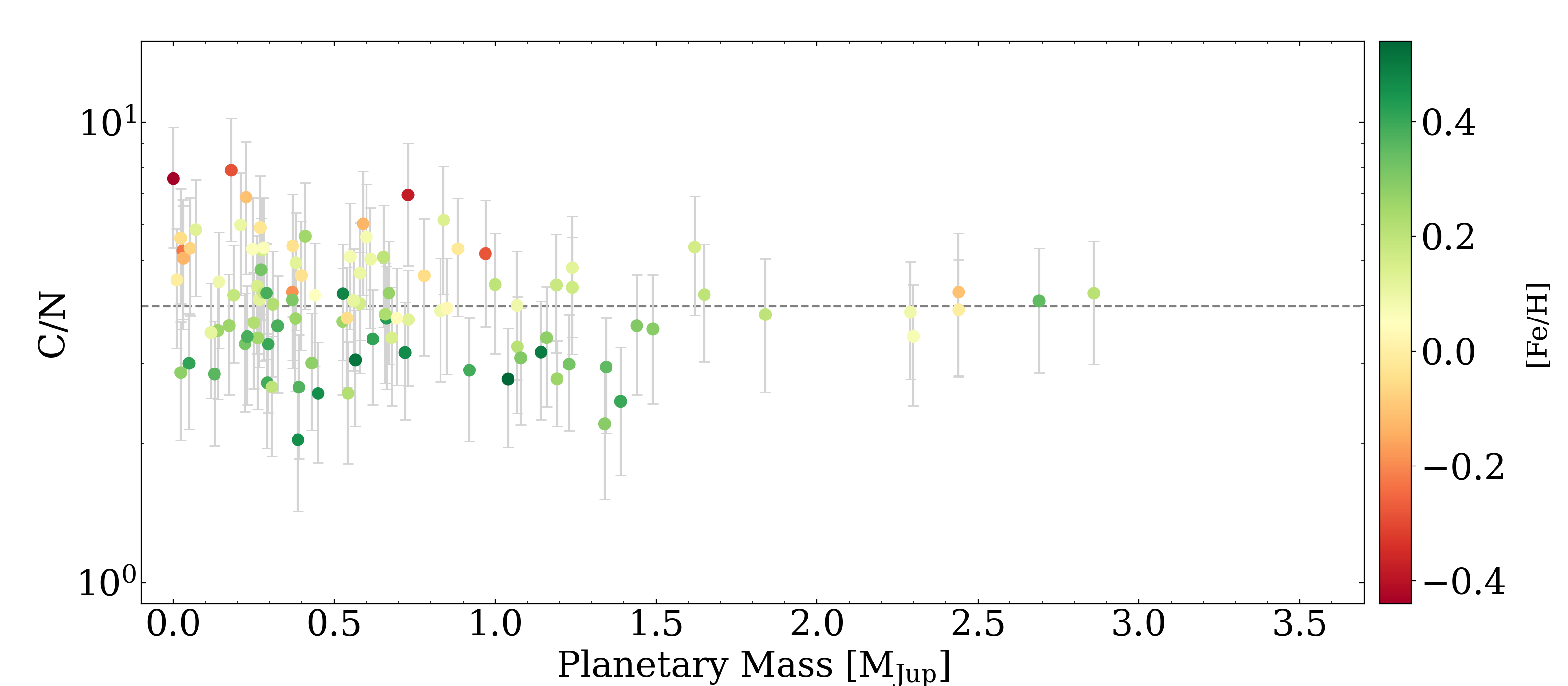}}
\end{minipage}
\begin{minipage}[t]{0.49\textwidth}
\centering
\resizebox{\hsize}{!}{\includegraphics{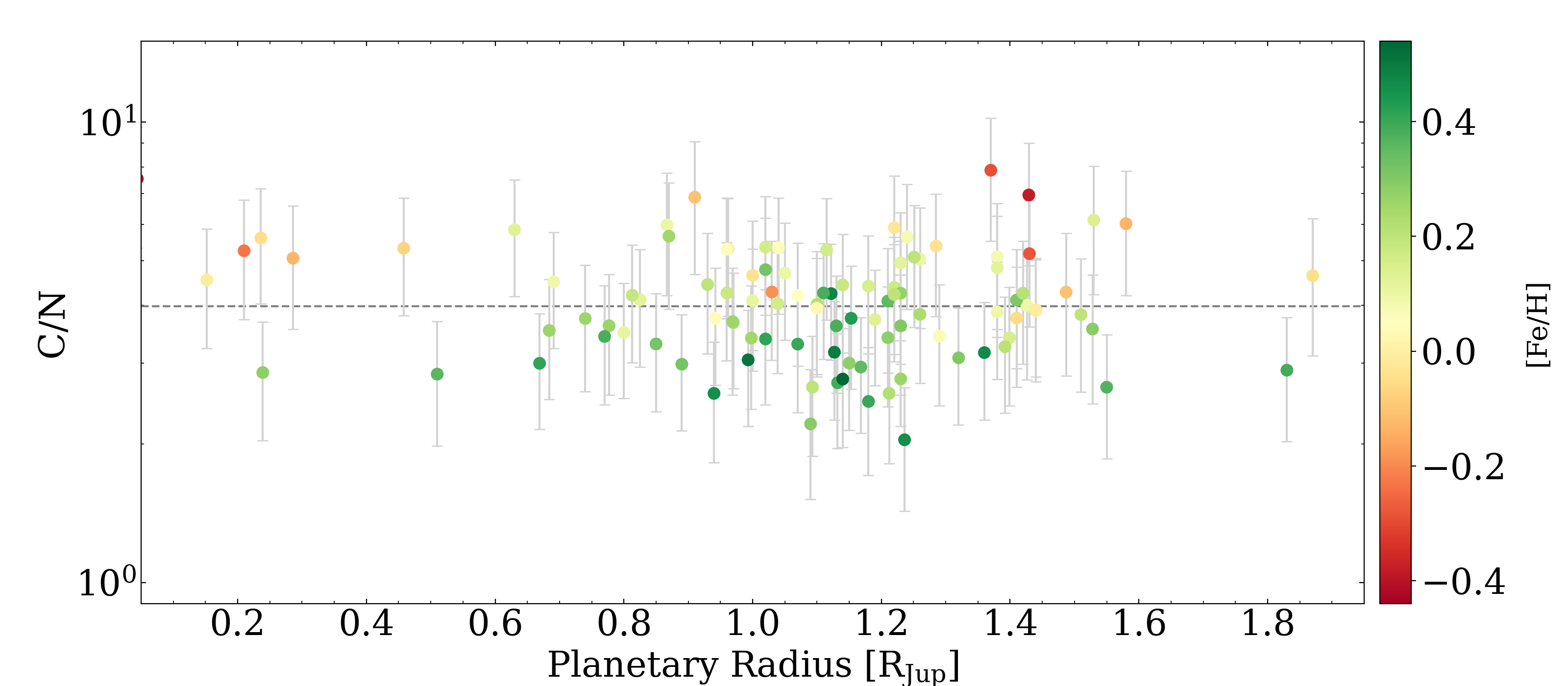}}
\end{minipage} \\
\begin{minipage}[t]{0.49\textwidth}
\centering
\resizebox{\hsize}{!}{\includegraphics{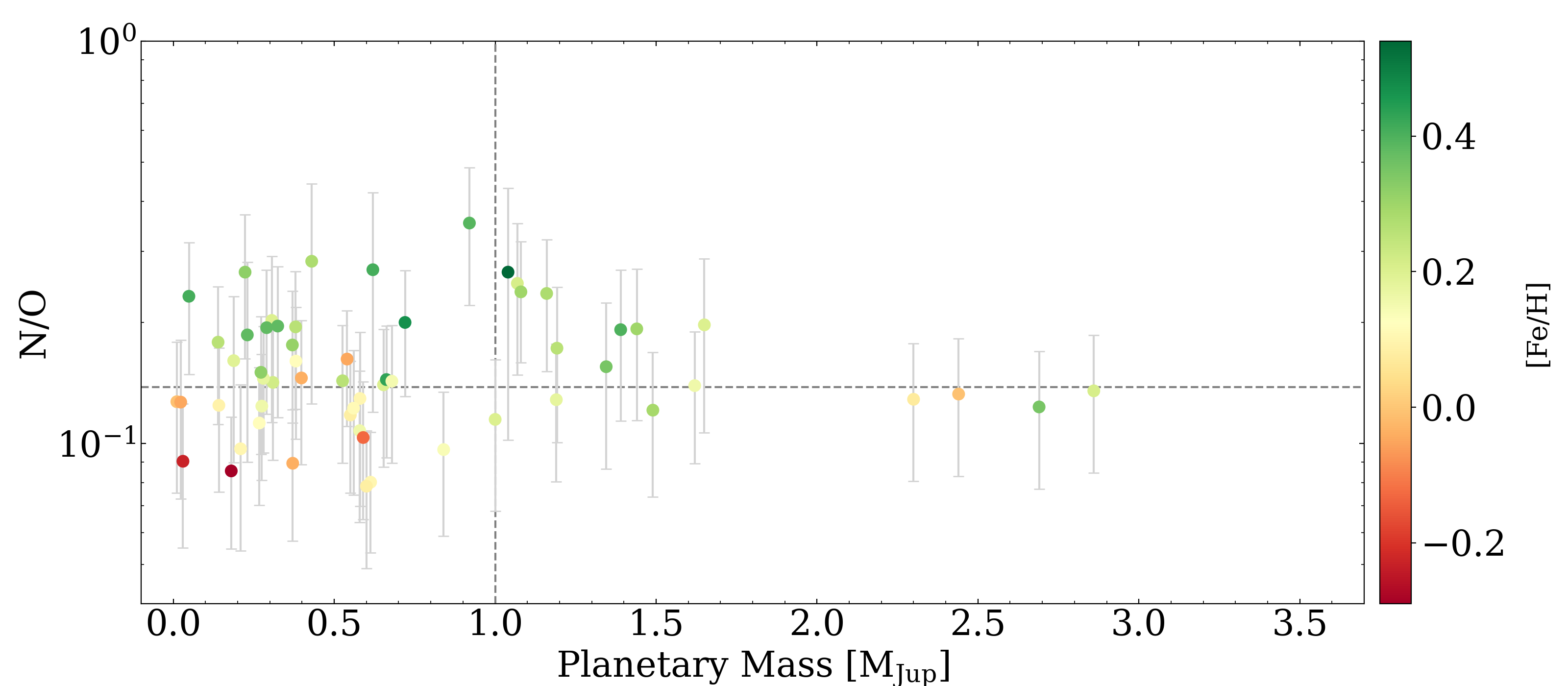}}
\end{minipage}
\begin{minipage}[t]{0.49\textwidth}
\centering
\resizebox{\hsize}{!}{\includegraphics{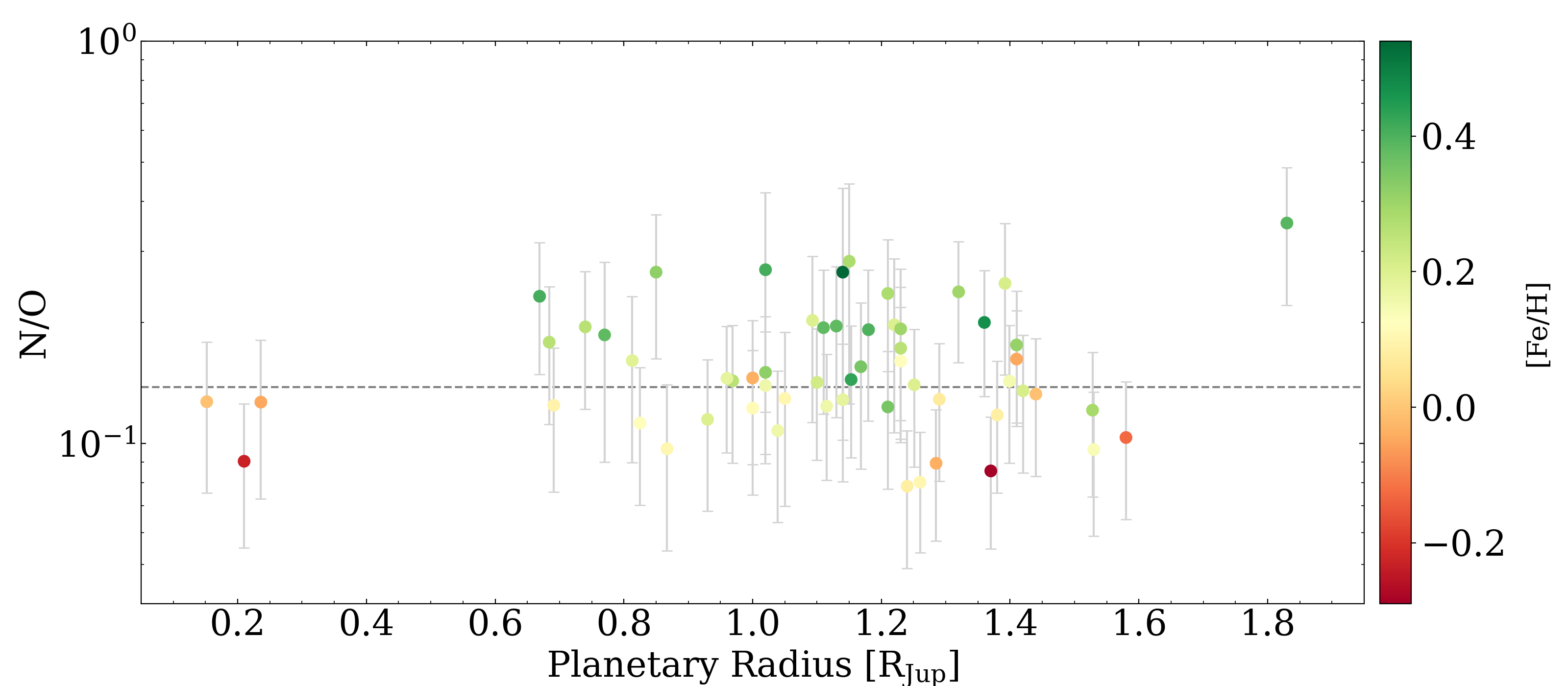}}
\end{minipage}
\caption{X1/X2 elemental ratios as a function of the planetary mass (left panels) and planetary radius (right panels) colour coded according to the stellar metallicity. The dependence on \teff\ of the C and N abundances were removed, as discussed in the text. The black dashed lines indicate the solar values and the magenta solid line represents the lower limit of C/O = 0.8 for carbon-rich stars.} 
\label{figure:x1x2_elem_ratios_planetary_mass_radius}
\end{figure*}

Figures~\ref{figure:x1x2_elem_ratios_feh}, \ref{figure:x1x2_elem_ratios_stellar_mass}, and \ref{figure:x1x2_elem_ratios_planetary_mass_radius} show the relations of the C/O, C/N, and N/O elemental ratios with some stellar ([Fe/H], \teff, mass) and planetary (mass, radius) parameters. We note that these are only complementary figures that might be of interest to the reader. Possible correlations among these parameters are discussed in the main text regarding the [X1/X2] ratios.

The presence of some stars above the limit of C/O = 0.8 for which a star is considered enriched in carbon shall be confirmed. As discussed in \citet{Nissen2013}, high-metallicity stars may have their oxygen abundances underestimated due to a wrong estimation of the contribution that the \ion{Ni}{i} line might give to the formation of the [\ion{O}{i}] line. To overcome this problem in future works, we plan to include other oxygen indicators in our analysis, such as the \ion{O}{i} triplet at 7774~\AA, which is one of the oxygen abundance indicators quite often used in the literature.

\end{appendix}

\end{document}